\documentclass{article}

\usepackage{arxiv}

\usepackage[utf8]{inputenc} 
\usepackage[T1]{fontenc}    
\usepackage{hyperref}       
\usepackage{url}            
\usepackage{booktabs}       
\usepackage{amsfonts}       
\usepackage{nicefrac}       
\usepackage{microtype}      
\usepackage{lipsum}
\usepackage{graphicx}
\graphicspath{ {./images/} }

\usepackage{amsmath, amssymb, bm, epsfig,natbib}
\usepackage{graphicx,subfigure}
\usepackage[dvipsnames]{xcolor}
\usepackage{srcltx}
\usepackage{pdfpages}
\usepackage{rotating}
\usepackage{epstopdf}
\usepackage{lscape}
\usepackage{subfigure}
\usepackage{verbatim, setspace}
\usepackage{calc, color}
\usepackage{tabularx}
\usepackage{longtable}
\usepackage{multirow}
\usepackage{booktabs}
\usepackage{indentfirst}
\usepackage{ragged2e}
\usepackage{subfigure}
\usepackage{tikz}
\usepackage{ragged2e}

\usepackage{times}

\usepackage{url}
\usepackage{rotating}

\newcommand{\Y}{\mathbf{Y}}
\newcommand{\bmu}{\mbox{\boldmath $\mu$}}
\newcommand{\bSigma}{\mbox{\boldmath $\Sigma$}}
\newcommand{\blambda}{\mbox{\boldmath $\lambda$}}
\newcommand{\yp}{\mathbf{y}}
\newcommand{\bOmega}{\mbox{\boldmath $\Omega$}}
\newcommand{\bDelta}{\mbox{\boldmath $\Delta$}}
\newcommand{\bvarphi}{\mbox{\boldmath $\varphi$}}
\newcommand{\y}{\mathbf{y}}
\newcommand{\by}{\mathbf{y}}

\newcommand{\SN}{\textrm{SN}}

\newcommand{\TESN}{\textrm{TESN}}
\newcommand{\TN}{\textrm{TN}}
\newcommand{\ap}{\mathbf{a}}
\newcommand{\bp}{\mathbf{b}}
\newcommand{\w}{\mathbf{w}}
\newcommand{\W}{\mathbf{W}}
\newcommand{\LL}{\mathcal{L}}
\newcommand{\zero}{\mathbf{0}}
\newcommand{\dr}[1]{{\mathrm d}#1}
\newcommand{\bGamma}{\mbox{\boldmath $\Gamma$}}
\newcommand{\tautil}{\tilde\tau}
\newcommand{\bW}{\mathbf{W}}
\newcommand{\bw}{\mathbf{w}}
\newcommand{\balpha}{\mbox{\boldmath $\alpha$}}
\newcommand{\bV}{\mathbf{V}}
\newcommand{\bC}{\mathbf{C}}
\newcommand{\CG}[1]{\textcolor{black}{#1}}
\newcommand{\tr}{\textrm{tr}}
\newcommand{\ESN}{\textrm{ESN}}
\newcommand{\TSN}{\textrm{TSN}}
\newcommand{\btheta}{\mbox{\boldmath $\theta$}}

\newcommand{\sumas}{\sum^n_{i=1}}
\newcommand{\C}{\mathbf{C}}
\newcommand{\ii}{i\in\{1,\ldots,n\}}
\newcommand{\jj}{j\in \{1,\ldots,G\}}

\newcommand{\bR}{\mathbf{R}}
\newcommand{\sumasj}{\sum_{j=1}^G}
\newcommand{\bI}{\mathbf{I}}
\newcommand{\se}{\mathbf{s}}
\newcommand{\bsigma}{\mbox{\boldmath $\sigma$}}
\newcommand{\E}{\textrm{E}}
\newcommand{\bF}{\mathbf{F}}
\newcommand{\bdelta}{\mbox{\boldmath $\delta$}}
\newtheorem{prop}{Proposition}
\newtheorem{remark}{Remark}

\title{Finite mixture modeling of censored and missing data using the multivariate skew-normal distribution}

\author{
Francisco H. C. de Alencar  \\
Departamento de Estat\'{\i}stica\\
Universidade Estadual de Campinas\\
Campinas, Brazil\\
\texttt{hildemardealencar@gmail.com} \\
\And
Christian E. Galarza \\
Departamento de Estad\'{\i}stica\\
Escuela Superior Politecnica del Litoral\\
Guayaquil, Ecuador \\
\texttt{chedgala@espol.edu.ec} \\
\And
Larissa A. Matos \\
Departamento de Estat\'{\i}stica\\
Universidade Estadual de Campinas\\
Campinas, Brazil\\
\texttt{larissam@unicamp.br} \\
\And
Victor H. Lachos \\
Department of Statistics\\
University of Connecticut\\
Storrs CT  06269, U.S.A. \\
\texttt{hlachos@uconn.edu} \\
}

\begin{document}
\maketitle
\begin{abstract}
Finite mixture models have been widely used to model and 	analyze data from a heterogeneous populations. Moreover, data of this kind can be missing or subject to some upper and/or lower detection limits because of the restriction of experimental apparatuses. Another complication arises when measures of each population depart significantly from normality, for instance, asymmetric behavior. For such data structures, we propose a robust model for censored and/or missing data based on finite mixtures of multivariate skew-normal distributions. This approach allows us to model data with great flexibility, accommodating multimodality and skewness, simultaneously, depending on the structure of the mixture components.  We develop an analytically simple, yet efficient, EM-type algorithm for conducting maximum likelihood estimation of the parameters.  The algorithm has closed-form expressions at the E-step that rely on formulas for the mean and variance of the  truncated multivariate skew-normal distributions.
Furthermore, a general information-based method for approximating the asymptotic covariance matrix of the estimators is also presented. Results obtained from the analysis of both simulated and real datasets are reported to demonstrate the effectiveness of the proposed method. The proposed algorithm and method are implemented in the new \verb|R| package \verb|CensMFM|.
\end{abstract}


\section{Introduction}
\label{intro} 
Modeling based on finite mixture distributions is a rapidly developing area with a wide range of applications. Finite mixture models are now applied in such diverse areas as biology, biometrics, genetics, medicine and marketing, among others. There are various features of finite mixture distributions that make them useful in statistical modeling. For instance, statistical models which are based on finite mixture distributions capture many specific properties of real data such as multimodality, skewness, kurtosis, and unobserved heterogeneity. The importance of mixture distributions can be noted from the large number of books on mixtures, including \cite{Peel2000}, \cite{fruhwirth2006finite}, \cite{mcnicholas2016mixture}, \cite{davila2018finite} and \cite{recent2019mixture}.

In many research areas, such as  environmental pollution and infectious diseases measurements often exhibit complex features such as censored responses and  missing values \citep{lin2018multivariate,lin2019multivariate}. Moreover, the proportion of censoring  in these studies may be substantial, so the use of crude/ad hoc methods, such as substituting a threshold value or some arbitrary point like a midpoint between zero and cutoff for detection, might lead to biased estimates of the model parameters. Furthermore, multivariate data are commonly seen with simultaneous occurrence of multimodality and skewness and inferential procedures become complicated when the data exhibit these features. The mixture distribution can be used quite effectively to analyze this kind of data. \cite{lin2009maximum} proposed a flexible mixture modeling framework using the multivariate skew-normal distribution, where a feasible EM algorithm is developed for finding the maximum likelihood (ML) estimates. In the context of finite mixtures for correlated censored data, \cite{he2013mixture} proposed a Gaussian mixture model to  flexibly approximate the underlying distribution of the observed data, where an EM algorithm in a multivariate setting was developed to cope with the censored data. More recently,  \cite{lachos2017finite} proposed a robust model for censored data based on finite mixtures of multivariate Student-t distributions (FM-MtC model), including the implementation of an exact EM algorithm for ML estimation. This approach allows modeling data with great flexibility, accommodating multimodality, and kurtosis depending on the structure of the mixture components. These methods are undoubtedly very flexible, but the problems related to the simultaneous occurrence of skewness, anomaly observations and multimodality remain. Even when modeling using Student-t mixtures, overestimation of the number of components necessary to capture the asymmetric nature of each subpopulation can occur \citep{Cabral2012}. So far, to the best of our knowledge there are no studies simultaneously accounting for multivariate censored responses, missing values, heterogeneity and skewness.

In this article, we propose a robust mixture model for censored data based on the multivariate skew-normal distribution so that the FM-MSNC model is defined and a fully likelihood-based approach is carried out, including the implementation of an exact EM-type algorithm for the ML estimation. The interval censoring mechanism of the proposed model allows us to handle missing  and censored values simultaneously. We show that the E-step reduces to computing the first two moments of a truncated multivariate skew-normal distribution. The general formulas for these moments were derived efficiently by \cite{GalarzaTrunSN2019}, for which we use the \verb|MomTrunc| package in \verb|R|. The likelihood function is easily computed as a byproduct of the E-step and is used for monitoring convergence and for model selection. Furthermore, we consider a general information-based method for obtaining the asymptotic covariance matrix of the ML estimate. The method proposed in this paper is implemented in the \verb|R| package \verb|CensMFM|, which is available for download from the CRAN repository.

The remainder of the paper is organized as follows. In Section~\ref{model}, we briefly discuss some preliminary results related to the multivariate extended skew-normal (ESN) and related truncated extended skew-normal (TESN) distributions, in addition, to some of their key properties are presented. In section 3, we present the multivariate skew-normal censored (MSNC) model  and the related ML estimation. In Section~\ref{sectionFM-tMC}, we introduce the robust FM-MSNC model, including the EM algorithm for ML estimation, and derive the empirical information matrix  analytically to obtain the standard errors. In Sections~\ref{sec-simStudy} and \ref{app}, numerical examples using both simulated and real data, respectively, are given to illustrate the performance of the proposed method. Finally, some concluding remarks are presented in Section~\ref{sec:6}.

\section{Background}
\label{model}

\subsection{The multivariate skew-normal distribution}\label{prelim}

In this subsection we present the skew-normal distribution and some of its properties. We say that a {\small $p\times 1$} random vector {\small ${\Y}$} follows a multivariate SN distribution with {\small $p\times 1$} location vector {\small $\bmu$}, {\small $p\times p$} positive definite dispersion matrix {\small $\bSigma$} and {\small $p\times 1$} skewness parameter vector {\small $\blambda\in\mathbb{R}^p,$} and we write {\small $\textbf{Y}\sim \textrm{SN}_p(\bmu,\bSigma,\blambda),$} if its pdf is given by
{\small \begin{equation}\label{denSN}
SN_p(\mathbf{y};\bmu,\bSigma,\blambda)= 2{\phi_p(\mathbf{y};\bmu,\bSigma)
\Phi_1(\blambda^{\top}\bSigma^{-1/2}(\mathbf{y}-\bmu))},
\end{equation}}
where {\small $\Phi_1(\cdot)$} represents the cumulative distribution function (cdf) of the standard univariate normal distribution. If  {\small $\blambda=\bf 0$}  then (\ref{denSN}) reduces to the symmetric {\small $\textrm{N}_p(\bmu,\bSigma)$}  pdf which is denoted by {\small $\phi_p(\mathbf{y};\bmu,\bSigma)$}.  Except by a straightforward difference in the parameterization considered in (\ref{denSN}), this model corresponds to that introduced by \cite{AzzaliniDV1996}, whose properties  were extensively studied in\cite{AzzaliniC1999} \citep[see also,][]{ArellanoG2005}.

\begin{prop}\label{propdenscond} 
If {\small $\Y\sim SN_{p}(\bmu,\bSigma,\blambda)$}, then for any {\small $\yp\in \mathbb{\mathbb{R}}^p$}
{\small \begin{equation}\label{denscond}
F_{\Y}(\yp)=P(\Y\leq\yp)=2{\Phi_{p+1}\hspace{-0.5mm}\big((\mathbf{z}^
{\scriptscriptstyle\top},0)^{\scriptscriptstyle\top};\mathbf{0},\bOmega
\big)},\,\,
\end{equation}}
where {\small $\mathbf{z}=\yp-\bmu$} and {\small$\bOmega=\left(\begin{array}{cc} \bSigma& -\bDelta \\
-\bDelta^{\top} & 1 \end{array}
\right),$} with {\small $\bDelta=\bSigma^{1/2}\blambda/{(1+\blambda^{\top}\blambda)^{1/2}}.$}
\end{prop}

It is worth mentioning that the multivariate skew-normal distribution is not closed to marginalization and conditioning. Next we present its extended version which has these properties, called the multivariate ESN distribution.

\subsection{The extended multivariate skew-normal distribution (ESN)}

We say that a {\small $p\times 1$} random vector {\small ${\Y}$} follows an ESN distribution with {\small $p\times 1$} location vector {\small $\bmu$}, {\small $p\times p$} positive definite dispersion matrix {\small $\bSigma$}, a {\small $p\times 1$} skewness parameter vector {\small $\blambda\in\mathbb{R}^p,$} and shift parameter {\small $\tau\in\mathbb{R}$}, denoted by {\small $\textbf{Y}\sim \textrm{ESN}_p(\bmu,\bSigma,\blambda,\tau),$} if its pdf is given by
{\small\begin{equation}\label{denESN}
ESN_p(\mathbf{y};\bmu,\bSigma,\blambda,\tau)= \xi^{-1}{\phi_p(\mathbf{y};\bmu,\bSigma)
\Phi_1(\tau+\blambda^{\top}\bSigma^{-1/2}(\mathbf{y}-\bmu))},
\end{equation}}
with {\small $\xi=\Phi_1(\tau/(1+\blambda^{\top}\blambda)^{1/2})$}. Note that when {\small $\tau=0$}, we retrieve the skew-normal distribution defined in (\ref{denSN}), that is, {\small $ ESN_p(\mathbf{y};\bmu,\bSigma,\blambda,0)\equiv SN_p(\mathbf{y};\bmu,\bSigma,\blambda)$}. It is also interesting to note that
{\small$$  ESN_p(\mathbf{y};\bmu,\bSigma,\blambda,\tau) {\longrightarrow} \phi_p(\mathbf{y};\bmu,\bSigma),\,\,{\text as }\,\,\,\tau\rightarrow +\infty.$$}
The following  propositions are crucial to develop our methods. The proofs are given in \cite{arellano2010multivariate}.	

\begin{prop}\label{proposition2}
Let  {\small $\Y\sim ESN_{p}(\bmu,\bSigma,\blambda,\tau)$} and {\small $\Y$} is partitioned as {\small $\Y=(\Y^{\top}_1,\Y^{\top}_2)^{\top}$} of dimensions {\small $p_1$} and {\small $p_2$ ($p_1+p_2=p$)}, respectively. Let
{\small $$\bSigma=\left(\begin{array}{cc}
\bSigma_{11} & \bSigma_{12} \\
\bSigma_{21} & \bSigma_{22}
\end{array}
\right),\ \ \bmu=(\bmu^{\top}_1,\bmu^{\top}_2)^{\top}, \ \ \blambda=(\blambda^{\top}_1,\blambda^{\top}_2)^{\top}\quad\text{and}\quad \bvarphi=(\bvarphi^{\top}_1,\bvarphi^{\top}_2)^{\top}$$} 
be the corresponding partitions of {\small $\bSigma$, $\bmu$, $\blambda$} and {\small $\bvarphi=\bSigma^{-1/2}\blambda$}. Then,
{\small\begin{eqnarray*}
&\Y_1& \sim ESN_{p_1}(\bmu_1,\bSigma_{11}, c_{12}\bSigma_{11}^{1/2}\tilde{\bvarphi}_1,c_{12}\tau),\\
&\Y_2&|\Y_1=\y_1\sim ESN_{p_2}(\bmu_{2.1},\bSigma_{22.1},\bSigma^{1/2}_{22.1}\bvarphi_2, \tau_{2.1})
\end{eqnarray*}}
where {\small $c_{12}=(1+\bvarphi^{\top}_2\bSigma_{22.1}\bvarphi_2)^{-1/2}$, $\tilde{\bvarphi}_1=\bvarphi_1+\bSigma_{11}^{-1}\bSigma_{12}\bvarphi_2$, $\bSigma_{22.1}=\bSigma_{22}-\bSigma_{21}\bSigma^{-1}_{11}\bSigma_{12}$,  $\bmu_{2.1}=\bmu_2+\bSigma_{21}\bSigma^{-1}_{11}(\by_1-\bmu_1)$} and {\small $\tau_{2.1}=\tau+\tilde{\bvarphi}^{\top}_1(\by_1-\bmu_1)$.}
\end{prop}

\begin{prop}\label{propdenscond3} 
If {\small $\Y\sim ESN_{p}(\bmu,\bSigma,\blambda,\tau)$}, then for any {\small $\yp\in \mathbb{\mathbb{R}}^p$}
{\small	\begin{equation}\label{denscond}
F_{\Y}(\yp)=P(\Y\leq\yp)={\frac{				\Phi_{p+1}\hspace{-0.5mm}\big((
\mathbf{z}^{\scriptscriptstyle\top},\tilde\tau)^{\scriptscriptstyle\top};
\mathbf{0},\bOmega
\big)}{\Phi_1(\tilde{\tau})}},
\end{equation}}
with {\small $\mathbf{z}$} and {\small $\bOmega$} as defined in Proposition \ref{propdenscond}, and {\small $\tilde{\tau}=\tau/(1+\blambda^{\top}\blambda)^{1/2}$.}
\end{prop}

Hereafter, for {\small $\Y\sim ESN_{p}(\bmu,\bSigma,\blambda,\tau)$}, we will denote its \emph{cdf} as {\small $F_{\Y}(\yp)\equiv \tilde{\Phi}_p(\yp;\bmu,\bSigma,\blambda,\\\tau)$} for simplicity.

Let {\small $\mathbb{A}$} be a Borel set in {\small $\mathbb{R}^p$}. We say that the random vector {\small $\Y$} has a truncated extended skew-normal distribution on {\small $\mathbb{A}$} when {\small $\Y$} has the same distribution as {\small $\Y | (\Y \in \mathbb{A})$}. In this case, the pdf of {\small $\Y$} is given by
{\small $$f(\y\mid\bmu,\bSigma,\nu;\mathbb{A})=\displaystyle\frac{ESN_p( \y;\bmu,\bSigma,\blambda,\tau)}{P(\Y \in  \mathbb{A})}\mathbf{1}_{\mathbb{A}}(\y),$$}
where  {\small $\mathbf{1}_{ \mathbb{A}}$} is the indicator function of {\small $ \mathbb{A}$}. We use the notation {\small $\Y \sim  {\TESN}_{p}(\bmu,\bSigma,\blambda,\tau;\mathbb{A})$}. If {\small $\mathbb{A}$} has the form
{\small\begin{eqnarray} \label{hyper}
\mathbb{A} &=& \{(x_1,\ldots,x_p)\in \mathbb{R}^p:\,\,\, a_1\leq x_1 \leq b_1,\ldots, a_p\leq x_p \leq b_p \} \nonumber\\ &=&\{\mathbf{x}\in\mathbb{R}^p:\mathbf{a}\leq\mathbf{x}\leq\mathbf{b}\},
\end{eqnarray}}
then we use the notation {\small $\{\Y \in  \mathbb{A}\}=\{\ap \leq \Y \leq \mathbf{b}\}$}, where  {\small $\mathbf{a}=(a_1,\ldots,a_p)^\top$} and {\small $\mathbf{b}=(b_1,\ldots,b_p)^\top$}. Here, we say that  the distribution of {\small $\Y$} is doubly truncated. Analogously we define {\small $\{\Y \geq \mathbf{a}\}$} and {\small $\{\Y \leq \mathbf{b}\}$}. Thus, we say that the distribution of {\small $\Y$} is truncated from below and truncated from above, respectively. For convenience, we also use the notation {\small $\Y \sim  {\TESN}_{p}(\bmu,\bSigma,\blambda,\tau;[\ap,\bp])$}. In particular, we denote {\small $\W$} to follow a truncated {\small $p$}-variate normal distribution on {\small $[\ap,\bp]$} as {\small $\W \sim  {\TN}_{p}(\bmu,\bSigma;[\ap,\bp])$}.

For the general doubly truncated case, we define the normalizing constant {\small $\LL_p(\ap,\bp;\\\bmu,\bSigma,\blambda,\tau)= P(\ap \leq \Y \leq \mathbf{b}) $} as
{\small $$\mathcal{L}_p(\ap,\bp;\bmu,\bSigma,\blambda,\tau)= \int_{\ap}^{\bp}
ESN_p(\mathbf{y};\bmu,\bSigma,\blambda,\tau)\dr\by.$$}
When all {\small $\blambda$} and {\small $\tau$} are equal to zero, we have a normal integral {\small $\LL_p(\ap,\bp;\bmu,\bSigma,\zero,0) = L_p(\ap,\bp;\bmu,\bSigma) = \int_{\ap}^{\bp}{{{\phi}_p}(\mathbf{y};\bmu,\bSigma)\textrm{d}\by}$}. Note that we use calligraphic style {\small $\LL_p$} when we work with the skewed extended version and Roman style {\small $L_p$} for the symmetric case. 

The following properties of the  truncated multivariate extended skew-normal distributions are useful for  implementation of the EM-algorithm. The proofs are given in \cite{GalarzaTrunSN2019}.

\begin{prop}\label{propdenscond4}
Let {\small $\Y \sim TESN_p(\bmu,\bSigma,\blambda,\tau;[\ap,\bp])$}. For any measurable function {\small $g(\cdot)$}, we have that
{\small\begin{equation}\label{eq:lema1}
\mathbb{E}
\left[
g(\Y)
\frac{\phi_1(\tau + \blambda^{\top}\bSigma^{-1/2}(\Y-\bmu))}
{\Phi_1(\tau + \blambda^{\top}\bSigma^{-1/2}(\Y-\bmu))}
\right] = \eta
\frac{ L_p(\ap,\bp;\bmu-\bmu_b,\bGamma)}{\mathcal{L}_p(\ap,\bp;\bmu,\bSigma,\blambda,\tau)}\mathbb{E}[g(\bW)],
\end{equation}}
with {\small $\eta=\phi_1(\tau;0,1 + {\boldsymbol{\lambda}^\top\boldsymbol{\lambda}})/\xi$}, {\small $\bmu_b = \tautil\bDelta$, $\bGamma = \bSigma - \bDelta\bDelta^{\top}$} and {\small $\bW \sim TN_{p}(\bmu-\bmu_b,\bGamma;[\ap,\bp])$}.
\end{prop}	

\begin{prop} \label{propdenscond5}
Let  {\small $\Y\sim ESN_{p}(\bmu,\bSigma,\blambda, \tau; \ [\ap,\bp])$}, where {\small $\ \Y$} is partitioned as {\small $\Y=(\Y^{\top}_1,\Y^{\top}_2)^{\top}$} of dimensions {\small $p_1$} and {\small $p_2$} ({\small $p_1+p_2=p$}), with corresponding partitions of {\small $\ap$, $\bp$, $\bmu$, $\bSigma$, $\blambda$} and {\small $\bvarphi$}. Then, for any measurable function {\small $g(\cdot)$}, we have that
{\small\begin{equation}\label{eq:lema2}
\mathbb{E}_{\Y_2}
\left[\left.
g(\Y_2)
\frac{\phi_1(\tau + \blambda^{\top}\bSigma^{-1/2}(\Y-\bmu))}{\Phi_1(\tau + \blambda^{\top}\bSigma^{-1/2}(\Y-\bmu))}\right|\Y_1\right] = \frac{\eta_{2.1} L_{2.1}}{\LL_{2.1}}\mathbb{E}[g(\W_2)],
\end{equation}}
where {\small $L_{2.1} = L_{p_2}(\ap_2,\bp_2;\bmu_{2.1}-\bmu_{b2.1},\bGamma_{22.1})$, $\LL_{2.1} = \LL_{p_2}(\ap_2,\bp_2;\bmu_{2.1},\bSigma_{22.1},\blambda_{2.1},\tau_{2.1})$} and {\small $\W_2 \sim TN_{p}(\bmu_{2.1}-\bmu_{b2.1},\bGamma_{22.1},[\ap_2,\bp_2])$} with {\small $\blambda_{2.1} = \bSigma^{1/2}_{22.1}\bvarphi_2$, $\bmu_{2.1}$, $\bSigma_{22.1}$}, and {\small $\tau_{2.1}$} as in proposition \ref{proposition2}, and {\small $\eta_{2.1}$, $\bmu_{b2.1}$} and {\small $\bGamma_{22.1}$} can be computed as expressions {\small $\eta$, $\bmu_b$} and {\small $\bGamma$} in proposition \ref{propdenscond4} but using the new set of parameters {\small $\bmu_{2.1}$, $\bSigma_{22.1}$, $\blambda_{2.1}$} and {\small $\tau_{2.1}$} (instead of {\small $\bmu$, $\bSigma$, $\blambda$} and {\small $\tau$}).
\end{prop}	

Observe that Propositions 4 and 5 depend on formulas for {\small $g(\Y)$}, where {\small $\Y \sim TESN(\bmu,\bSigma, \blambda,\tau; [\ap,\bp])$}. Closed form expressions for these expectations were obtained recently by \cite{GalarzaTrunSN2019}, for which the \texttt{meanvarTMD()} function of the \texttt{R MomTrunc} library can be used.	

\section{Multivariate skew-normal model for censored and missing responses} \label{sec:mod:spec}

Now we present the robust multivariate skew-normal model for censored data. So, we write
{\small\begin{equation}
\Y_1, \ldots , \Y_n \sim SN_{p}(\bmu,\bSigma,\blambda), \label{modeleq}
\end{equation}}
where for each {\small $i \in \{ 1, \ldots , n\}$, $\Y_i=(Y_{i1},\ldots,Y_{ip})^{\top}$} is a {\small $p\times 1$} vector of responses for sample unit {\small $i$, $\bmu=(\mu_1,\ldots,\mu_p)^{\top}$} is the location vector and the dispersion matrix {\small $\bSigma=\bSigma(\balpha)$} depends on an unknown and reduced parameter vector {\small $\balpha$} and skewness parameter {\small $\blambda$}. We assume that  {\small $\Y_1, \ldots, \Y_n$}  are independent and identically distributed. We consider a similar approach to that proposed by \cite{lachos2017finite} to model the censored responses. Thus, the observed data for the {\small $i$}th subject are given by {\small $(\bV_i,\bC_i),$} where each element of {\small $\bV_i=(V_{i1},\ldots,V_{ip})^\top$} represents either 
the vector of uncensored observations {\small $(V_{ik} = V_{0i})$} or the interval censoring level {\small $(V_{ik} \in [V_{1ik}, V_{2ik}])$}, and {\small $\bC_i=(C_{i1},\ldots,C_{ip})^\top$} is the vector of censoring indicators, satisfying
{\small\begin{eqnarray}\label{modelcen}
C_{ik} = \left\{ \begin{array}{ll}
1 & \mbox{if $ V_{1ik} \leq Y_{ik} \leq V_{2ik}$};\\
0 & \mbox{if $ Y_{ik} = V_{0i}$}.\end{array} \right.
\end{eqnarray}}
for all {\small $i \in \{1, \ldots, n\}$} and {\small $k \in \{1, \ldots, p\}$}, i.e., {\small $C_{ik} = 1$} if {\small $Y_{ik}$} is located within a specific interval. In this case, (\ref{modeleq}) and (\ref{modelcen}) define the multivariate skew-normal interval censored model (hereafter, the MSNC model). Missing observations can be handled by considering {\small $V_{1ik} = -\infty$ and $V_{2ik} = +\infty$}.

\subsection{The likelihood function}\label{Likelihood_tMLC}

Let {\small $\yp=(\yp^{\top}_1,\ldots,\yp^{\top}_n)^{\top}$}, where {\small $\y_i=(y_{i1},\ldots,y_{ip})^\top$} is  a realization of {\small $\Y_i\sim \SN_{p}(\bmu,\ \bSigma,\\ \blambda)$}. To obtain the likelihood function of the MSNC model, we first treat the observed and censored components of {\small $\y_i$}, separately, i.e., {\small $\y_i=(\y^{o^\top}_i,\y^{c^\top}_i)^\top$}, where {\small $C_{ik}=0$} for all elements in the {\small $p_i^o$}-dimensional vector {\small $\y^o_i$}, and {\small $C_{ik}=1$} for all elements in the {\small $p_i^c$}-dimensional vector {\small $\y^c_i$}. Accordingly, we write {\small $\bV_i=\mathrm{vec}(\bV^o_i,\bV^c_i)$}, where {\small $\bV^c_i=(\bV^c_{1i},\bV^c_{2i})$} with
{\small\begin{eqnarray}\label{partpars}
\bmu_i&=&(\bmu^{o\top}_i,\bmu^{c\top}_i)^{\top}\mbox{,}\quad
\bSigma =\bSigma(\balpha)=
{\bSigma^{oo}_i \bSigma^{oc}_i\choose \bSigma^{co}_i
\bSigma^{cc}_i}\mbox{,}\nonumber\\ 
\blambda_i&=&(\blambda^{o\top}_i,\blambda^{c\top}_i)^{\top} \quad\mbox{and }\quad \bvarphi_i=(\bvarphi^{o\top}_i,\bvarphi^{c\top}_i)^{\top}.
\end{eqnarray}}
Then, using Proposition \ref{proposition2}, we have that {\small $\Y^o_i\sim \SN_{p_i^o}(\bmu^o_i,\bSigma^{oo}_{i}, c^{oc}_i\bSigma_i^{oo\,1/2}\tilde{\bvarphi}^o_i)$} and {\small $\Y^c_i\mid\Y^o_i=\y^o_i\sim \ESN_{p^c_i}(\bmu^{co}_i,$ $\bSigma^{cc.o}_i, \bSigma^{cc.o\,1/2}_{i}\bvarphi^c_i,$ $\tau^{co}_i)$}, where
{\small\begin{equation}
\bmu^{co}_i = \bmu^c_i+\bSigma^{co}_i\bSigma^{oo-1}_{i}(\y^o_i-\bmu^o_i),\quad
\bSigma^{cc.o}_{i} =\bSigma^{cc}_{i}-\bSigma^{co}_{i}(\bSigma^{oo}_{i})^{-1}\bSigma^{oc}_{i}, \quad
\label{eqn mu.co S.co}
\end{equation}}
{\small\begin{equation}
\tilde{\bvarphi}^o_i=\bvarphi^o_i+\bSigma^{oo\,-1}_i\bSigma^{oc}_i\bvarphi^c_i, \quad c^{oc}_i=(1+\bvarphi^{c\top}_i\bSigma^{cc.o}_i\bvarphi^c_i)^{-1/2}  \quad \mbox{and}\quad
\tau^{co}_i= \tilde{\bvarphi}^{o\top}_i(\by^o_i-\bmu^o_i).
\label{eq:Sig:co1}
\end{equation}}

Let {\small $\bV=\mathrm{vec}(\bV_1,\ldots,\bV_n)$} and {\small $\bC=\mathrm{vec}(\bC_1,\ldots,\bC_n)$} denote the observed data. Therefore, the log-likelihood function of {\small $\btheta=(\bmu^{\top},\balpha^{\top},\blambda^{\top})^{\top}$}, given the observed data {\small $(\bV,\bC)$} is
{\small\begin{equation}
\ell(\btheta\mid\bV,\bC)=\sum_{i=1}^n\ln{L_i},\label{equ8.1}
\end{equation}}
where {\small $L_i$} represents the likelihood function of {\small $\btheta$} for the {\small $i$}th sample, given by
{\small\begin{align*}
{L_i\equiv}L_i(\btheta\mid\bV_i,\bC_i)&=f(\bV_i\mid\bC_i,\btheta)=f(\bV^c_{1i}\leq \yp_i^c\leq \bV^c_{2i}\mid\y^o_i,\btheta)f(\y^o_i\mid\btheta) \nonumber\\
&=\LL_{p^c_i}(\bV^c_{1i},\bV^c_{2i};\bmu^{co}_i,\bSigma^{cc.o}_i,
\bSigma^{cc.o\,1/2}_{i}\bvarphi^c_i, \tau^{co}_i)\\
&\times SN_{p^o_i}
(\y^o_i;\bmu^o_i,\bSigma^{oo}_{i}, c^{oc}_i\bSigma^{oo\,1/2}_i\tilde{\bvarphi}^o_i).
\end{align*}}

\subsection{Parameter estimation via the EM algorithm}\label{EM-tlcm}

We now describe how to carry out ML estimation for the MSNC model. The EM algorithm, originally proposed by \cite{Dempster77}, is a very popular iterative optimization strategy and commonly used to obtain ML estimates for incomplete-data problems. This algorithm has many attractive features, such as numerical stability, simplicity of implementation and quite reasonable memory requirement \citep{McLachlanKrishnan}.

By the essential property of a multivariate SN distribution, we can write
{\small\begin{equation}\label{stochSN}
\Y_i|(T_i=t_i)\sim N_p(\bmu + \bDelta t_i,\bGamma)\,\,{\it and}\,\,\,
T_i\sim\mbox{HN}(0,1),
\end{equation}}
with HN referring to a half normal distribution and with {\small $\bDelta$} and {\small $\bGamma$} as defined in the previous section. The complete-data log-likelihood function of an equivalent set of parameters {\small $\btheta=(\bmu^{\top},\bDelta^{\top},\balpha_{\scriptscriptstyle\Gamma}^{\top})^{\top}$}, where {\small $\balpha_{\scriptscriptstyle\Gamma}=\mbox{vech}(\bGamma)$}, is given by {\small $\ell_c(\btheta)=\sumas \ell_{ic}(\btheta)$,} where the individual complete-data log-likelihood is
{\small\begin{equation*}
\ell_{ic}(\btheta)=-\frac{1}{2}\bigl\{ \ln|\bGamma| + (\yp_i-\bmu - \bDelta t_i)^{\top}\bGamma^{-1}(\yp_i-\bmu - \bDelta t_i)\bigr\} + c,
\end{equation*}}
with {\small $c$} being a constant that does not depend on {\small $\btheta$}. Subsequently, the EM algorithm for the MSNC model can be summarized as follows: \\

\noindent{\bf E-step:} Given the current estimate  {\small $\widehat{\btheta}^{(k)}=(\widehat{\bmu}^{{(k)}},\widehat{\bDelta}^{\scriptscriptstyle(k)},\widehat{\balpha}_{\scriptscriptstyle\Gamma}^{\scriptscriptstyle(k)})$} at the {\small $k$}th step of the algorithm, the E-step provides the conditional expectation of the complete data log-likelihood function
{\small\begin{equation*}
Q(\btheta\mid\widehat{\btheta}^{(k)})=\mathrm{E} \Bigl[ \ell_c(\btheta )\mid\bV,\bC,\widehat{\btheta}^{(k)} \Bigr]
=\sumas{Q_i(\btheta\mid\widehat{\btheta}^{(k)})},\label{eq:Em:Q}
\end{equation*}}
where 
{\small\begin{eqnarray*}
Q_i(\btheta\mid\widehat{\btheta}^{(k)}) &\propto&
-\frac{1}{2}\ln{|\widehat{\bGamma}^{\scriptscriptstyle(k)}|}-\frac{1}{2}\tr\left[\left\{
\widehat{\yp_i^{2}}^{(k)}
+\widehat{\bmu}^{{(k)}} \widehat{\bmu}^{{(k)}\top}
+ \widehat{t_i^{2}}^{(k)}{\widehat{\bDelta}}^{\scriptscriptstyle(k)}{\widehat{\bDelta}}^{{(k)}\top}\right.\right.\\
&-& \left.\left. \widehat{\bmu}^{{(k)}}\widehat{\yp_i}^{{(k)}\top}-\widehat{\yp_i}^{{(k)}}\widehat{\bmu}^{{(k)\top}}
-  \widehat{t\yp}^{(k)}_i\widehat{\bDelta}^{{(k)}\top}-  \widehat{\bDelta}^{{(k)}}\widehat{t\yp}^{(k)\top}_i\right.\right.\\
&+&\left.\left.
\widehat{t_i}^{(k)}\widehat{\bDelta}^{{(k)}}\widehat{\bmu}^{{(k)}\top}+\widehat{t_i}^{(k)}\widehat{\bmu}^{{(k)}}\widehat{\bDelta}^{{(k)\top}}\right\}
\widehat{\bGamma}^{{-1}{(k)}}\right],\nonumber
\end{eqnarray*}}
with {\small $\widehat{\yp_i^{r}}^{(k)}=\mathbb{E}_{T_i\Y_i}[\displaystyle \Y_i^r\,|\,\bV_i,\bC_i,\widehat{\btheta}^{(k)}]$, $\widehat{t_i^{r}}^{(k)}=\mathbb{E}_{T_i\Y_i}[\displaystyle T_i^r\,|\,\bV_i,\bC_i,\widehat{\btheta}^{(k)}]$} (for {\small $r=\{0,1,2\}$}, with {\small $\Y_i^0=1$, $\Y_i^1=\Y_i$} and {\small $\Y_i^2 = \Y_i\Y_i^{\top}$}) and {\small $\widehat{t\yp}^{(k)}_i=\mathbb{E}_{T_i\Y_i}[\displaystyle T_i\Y_i\,|\,\bV_i,\bC_i,\widehat{\btheta}^{(k)}]$}. Then, we can use Propositions 4 and 5 to obtain closed form expressions for these conditional expectations as follows:

\begin{itemize}
\item[1.] If the \CG{$i$th subject}
has only non-censored components, \CG{then}
{\small	\begin{align*}
\widehat{\yp_i^{r}}^{(k)} &= \mathbb{E}_{\Y_i}[\Y_i^r|\bV_i,\C_i,\widehat{\btheta}^{(k)}] = {\yp_i^{r}}, \\
\widehat{t_i^{r}}^{(k)} &= \mathbb{E}_{T_i\Y_i}[T_i^r|\bV_i,\C_i,\widehat{\btheta}^{(k)}] = \mathbb{E}_{T_i}[T_i^r|\Y_i,\widehat{\btheta}^{(k)}], \\
\widehat{{t\yp}_i}^{(k)} &= \mathbb{E}_{T_i\Y_i}[T_i\Y_i|\bV_i,\C_i,\widehat{\btheta}^{(k)}] = \yp_i\mathbb{E}_{T_i}[T_i|\Y_i,\widehat{\btheta}^{(k)}],
\end{align*}}
with {\small $\yp_i^0=1$, $\yp_i^1=\yp_i$} and {\small $\yp_i^2 = \yp_i\yp_i^{\top}$} and {\small $\mathbb{E}_{T_i}[T_i^r|\Y_i,\widehat{\btheta}^{(k)}] = \left.\mathbb{E}_{T_i}[T_i^r|\Y_i]\right|_{\btheta=\widehat{\btheta}^{(k)}}$} for {\small $r=\{1,2\}$}. These last conditional expectations can be obtained directly from the results given in \cite{Cabral2012}.

\item[2.] If the {\small $i$}th subject has only censored components, from Proposition \ref{propdenscond4} we have
{\small\begin{align*}
\widehat{\yp_i^{r}}^{(k)} &= \mathbb{E}_{\Y_i}[\Y_i^r|\bV_i,\C_i,\widehat{\btheta}^{(k)}] =   \widehat{{\bw_i^r}}^{(k)}, \\
\widehat{t_i}^{(k)} &=
M^2(\widehat{\btheta}^{(k)})\widehat\bDelta^{(k)\top}\widehat\bGamma^{-1(k)}(\widehat{{\bw_i}}^{(k)}-\widehat\bmu^{(k)}) +
\widehat\gamma_i^{(k)}M(\widehat{\btheta}^{(k)}),\\
\widehat{t_i^{2}}^{(k)} &=
M^4(\widehat{\btheta}^{(k)})\widehat\bDelta^{(k)\top}\widehat\bGamma^{-1(k)}(\widehat{{\bw_i^2}}^{(k)}- 2\widehat{{\bw_i}}^{(k)}\widehat\bmu^{(k)\top} + \widehat\bmu^{(k)}\widehat\bmu^{(k)\top})\widehat\bGamma^{-1(k)}\widehat\bDelta^{(k)}\\
&+ M^2(\widehat{\btheta}^{(k)})
+ \widehat\gamma_i^{(k)}M^3(\widehat{\btheta}^{(k)})\widehat\bDelta^{(k)\top}\widehat\bGamma^{-1(k)}(\widehat{\w_{0}}_i^{(k)} - \widehat\bmu^{(k)}),\\
\widehat{{t\yp}_i}^{(k)} &= M^2(\widehat{\btheta}^{(k)})(\widehat{{\bw_i}}^{2(k)}-\widehat{{\bw_i}}^{(k)}\widehat\bmu^{(k)\top})\widehat\bGamma^{-1(k)}\widehat\bDelta^{(k)} +
\widehat\gamma_i^{(k)}M(\widehat{\btheta}^{(k)})\widehat{\w_{0}}_i^{(k)},
\end{align*}}
where 
{\small\begin{eqnarray*}
&M^2(\btheta)&=(1 + \bDelta^{\top}\bGamma^{-1}\bDelta)^{-1},\quad \widehat{\bw}_i^{(k)}=\mathbb{E}[\mathbf{W}_i\mid\widehat{\btheta}^{(k)}], \nonumber\\ 
&\widehat{\bw}_i^{2{(k)}}&=\mathbb{E}[\bW_i
\bW_i^\top\mid\widehat{\btheta}^{(k)}] \quad \text{and} \quad \widehat{\w_{0}}_i^{(k)} = \mathbb{E}[\mathbf{W}_{0i}\mid\widehat{\btheta}^{(k)}],\label{w,w2}
\end{eqnarray*}}
with {\small $\mathbf{W}_i\sim
\TSN_{p}(\widehat{\bmu}^{(k)},\widehat{\bSigma}^{(k)},\widehat{\blambda}^{(k)},[\mathbf{v}_{1i},\mathbf{v}_{2i}])$,
$\W_{0i} \sim \TN_{p}(\widehat\bmu^{(k)},\widehat\bGamma^{(k)},[\mathbf{v}_{1i},\mathbf{v}_{2i}])$}
and
{\small$$
\widehat\gamma_i^{(k)} =
\frac{1}{\sqrt{{\frac{\pi}{2}\big(1 + \widehat\blambda^{(k){\top}}\widehat\blambda^{(k)}\big)}}}
\frac{L_p(\mathbf{v}_{1i},\mathbf{v}_{2i},\widehat\bmu^{(k)},\widehat\bGamma^{(k)})}
{\LL_p(\mathbf{v}_{1i},\mathbf{v}_{2i},\widehat\bmu^{(k)},\widehat\bSigma^{(k)},\widehat\blambda^{(k)},0)}.$$}

\item[3.] If the {\small $i$}th subject has both censored and uncensored components and given that {\small $(\Y_i\,|\,\bV_i,\bC_i)$, $(\Y_i\,|\,\bV_i,\bC_i,\Y^o_i)$}, and {\small $(\Y^c_i\,|\,\bV_i,\bC_i,\Y^o_i)$} are equivalent processes, we have from Proposition \ref{propdenscond5} that
{\small\begin{align*}
\widehat{\yp}^{(k)}_i&=\mathrm{E}(\displaystyle \Y_i\,|\,\Y^o_i,\bV_i,\bC_i,\widehat{\btheta}^{(k)})=\mathrm{vec}(\yp^o_i,\widehat{\mathbf{w}}^{c(k)}_i),\,\, \\
\widehat{\yp_i^{2}}^{(k)}&=\mathrm{E}(\displaystyle
\Y_i\Y_i^{\top}\,|\,\Y^o_i,\bV_i,\bC_i,\widehat{\btheta}^{(k)})=\left(\begin{array}{cc}
\yp^o_i\yp^{o\top}_i & \yp^o_i\widehat{\mathbf{w}}^{c(k)\top}_i \\
\widehat{\mathbf{w}}^{c(k)}_i\yp^{o\top}_i  & \widehat{\bw}_i^{2{c(k)}}\\
\end{array}\right ),\\
\widehat{\yp}^{(k)}_{0i}&=\mathrm{vec}(\yp^o_i,\widehat{\mathbf{w}}^{c(k)}_{0i}),\,\, \\
\widehat{t_i}^{(k)} &=
M^2(\widehat{\btheta}^{(k)})\widehat\bDelta^{(k)\top}\widehat\bGamma^{-1(k)}(\widehat{{\yp_i}}^{(k)}-\widehat\bmu^{(k)}) +
\widehat\gamma_{i}^{(k)}M(\widehat{\btheta}^{(k)}),\\
\widehat{t_i^{2}}^{(k)} &=
M^4(\widehat{\btheta}^{(k)})\widehat\bDelta^{(k)\top}\widehat\bGamma^{-1(k)}(\widehat{{\yp_i^2}}^{(k)}- 2\widehat{{\yp_i}}^{(k)}\widehat\bmu^{(k)\top} + \widehat\bmu^{(k)}\widehat\bmu^{(k)\top})\widehat\bGamma^{-1(k)}\widehat\bDelta^{(k)}\\
&+ M^2(\widehat{\btheta}^{(k)}) + \widehat\gamma_{i}^{(k)}M^3(\widehat{\btheta}^{(k)})\widehat\bDelta^{(k)\top}\widehat\bGamma^{-1(k)}(\widehat{\yp_{0}}_i^{(k)} - \widehat\bmu^{(k)}),\\
\widehat{{t\yp}_i}^{(k)} &= M^2(\widehat{\btheta}^{(k)})(\widehat{{\yp_i}}^{2(k)}-\widehat{{\yp_i}}^{(k)}\widehat\bmu^{(k)\top})\widehat\bGamma^{-1(k)}\widehat\bDelta^{(k)} +
\widehat\gamma_{i}^{(k)}M(\widehat{\btheta}^{(k)})\widehat{\yp_{0}}_i^{(k)},
\end{align*}}
where
{\small\begin{equation*}
\widehat{\bw}_i^{c(k)}=\mathbb{E}[\mathbf{W}^c_i\mid\widehat{\btheta}^{(k)}], \quad \widehat{\bw}_i^{2c{(k)}}=\mathbb{E}[\bW_i^c
\bW_i^{c\top}\mid\widehat{\btheta}^{(k)}] \quad \text{and} \quad \widehat{\w_{0}}_i^{c(k)} = \mathbb{E}[\mathbf{W}^c_{0i}\mid\widehat{\btheta}^{(k)}],\label{w,w2}
\end{equation*}}
with {\small $\ \mathbf{W}^c_i \ \sim \ 
\TESN_{p^c_i}\big(\widehat\bmu^{co(k)}_i, \
\widehat\bSigma^{cc.o(k)}_i, \ \widehat\blambda^{co(k)}_i, \ 
\widehat\tau_{i}^{co(k)}, \ [\mathbf{v}_{1i}^c,\mathbf{v}_{2i}^c]\big)$,
$\ \W_{0i}^c \ \sim \ \TN_{p}(\widehat{\mathbf{m}}^{co(k)}_i, \\ \widehat\bGamma^{cc.o(k)}_i,$ $[\mathbf{v}_{1i}^c,\mathbf{v}_{2i}^c])$} and
{\small$$
\widehat\gamma_{i}^{(k)} =
\frac{\eta_i^{co}\,L_p(\mathbf{v}_{1i}^c,\mathbf{v}_{2i}^c;\widehat{\mathbf{m}}^{co(k)}_i,\widehat\bGamma^{cc.o(k)}_i)}
{\LL_p(\mathbf{v}_{1i}^c,\mathbf{v}_{2i}^c;
\widehat\bmu^{co(k)}_i,
\widehat\bSigma^{cc.o(k)}_i,\widehat\blambda^{co(k)}_i,\widehat\tau_{i}^{co(k)}
)},
$$}
where
{\small $\blambda^{co}_i = \bSigma^{cc.o^{1/2}}_{i}\bvarphi^{c}_i$, ${\mathbf{m}}^{co}_i = \bmu^{co}_i - \bmu^{co}_{bi}$,} and {\small $\eta_i^{co}$, $\bmu_{bi}^{co}$} and {\small $\bGamma^{cc.o}_i$} can be computed as expressions {\small $\eta$, $\bmu_b$} and {\small $\bGamma$} in Proposition \ref{propdenscond4} but using the new set of parameters {\small $\bmu_i^{co}$, $\bSigma_i^{cc.o}$, $\blambda_i^{co}$} and {\small $\tau_i^{co}$} (instead of {\small $\bmu$, $\bSigma$, $\blambda$} and {\small $\tau$}).

To compute  {\small $\mathbb{E}[\mathbf{W}_{0i}]$, $\mathbb{E}[\mathbf{W}_i]$} and {\small $\mathbb{E}[\mathbf{W}_i\mathbf{W}_i^{\top}]$} in items 2 and 3, we use the R library \texttt{ MomTrunc}.
\end{itemize}

\noindent{\bf M-step:} Conditionally maximizing {\small $Q(\btheta\mid\widehat{\btheta}^{(k)})=\sumas Q_i(\btheta\mid\widehat{\btheta}^{(k)})$} with respect to each entry of {\small $\btheta$}, we update the estimate {\small $\widehat{\btheta}^{(k)}=(\hat\bmu^{\scriptscriptstyle(k)},\widehat{\bDelta}^{\scriptscriptstyle(k)},\widehat{\balpha}_{\scriptscriptstyle\Gamma}^{\scriptscriptstyle(k)})$} by
{\small\begin{eqnarray}
\widehat{\bmu}^{(k+1)}&=& \frac{1}{n}\sumas \left\{\widehat{\yp_i}^{{(k)}} - \widehat{t_i}^{(k)}\widehat{\bDelta}^{(k)}\right\},\label{eq:beta_tMLC0}\\
\widehat{\bDelta}^{(k+1)}&=&\left\{\sumas \widehat{t_i^{2}}^{(k)}\right\}^{-1}\sumas
\left\{\widehat{t\yp_i}^{(k)}-\widehat{t}^{(k)}_i\widehat{\bmu}^{(k+1)}\right\},\label{eq:delta_tMLC}\\
\widehat{\bGamma}^{(k+1)}&=&\frac{1}{n}\sumas
\left\{
\widehat{\yp_i^{2}}^{(k)}
+\widehat{\bmu}^{{(k)}} \widehat{\bmu}^{{(k)}\top}
+ \widehat{t_i^{2}}^{(k)}{\widehat{\bDelta}}^{\scriptscriptstyle(k)}{\widehat{\bDelta}}^{{(k)}\top}
-\widehat{\bmu}^{{(k)}}\widehat{\yp_i}^{{(k)}\top}-\widehat{\yp_i}^{{(k)}}\widehat{\bmu}^{{(k)\top}}\right. \nonumber\\
&-& \left. \widehat{t\yp}^{(k)}_i\widehat{\bDelta}^{{(k)}\top}-  \widehat{\bDelta}^{{(k)}}\widehat{t\yp}^{(k)\top}_i
+\widehat{t_i}^{(k)}\widehat{\bDelta}^{{(k)}}\widehat{\bmu}^{{(k)}\top}+\widehat{t_i}^{(k)}\widehat{\bmu}^{{(k)}}\widehat{\bDelta}^{{(k)\top}}
\right\}.\label{eq:Gamma_tMLC}
\end{eqnarray}}

The algorithm is iterated until a suitable convergence rule is satisfied. In the later analysis, the algorithm is terminated when the relative distance between two successive evaluations of the log-likelihood defined in (\ref{equ8.1}) is less than a tolerance, i.e., {\small $|\ell(\widehat{\btheta}^{(k+1)}\mid\bV,\bC)/\ell(\widehat{\btheta}^{(k)}\mid\bV,\bC)-1|<\epsilon$}, for example, {\small $\epsilon=10^ {-6}$}. Once converged, we can recover {\small $\widehat\blambda$} and {\small $\widehat\bSigma$} using the expressions
{\small\begin{equation*}\label{recover}
\widehat\bSigma=\widehat\bGamma + \widehat\bDelta\widehat\bDelta^{\top}
\qquad\text{and}\qquad
\widehat\blambda = \frac{{\widehat\bSigma}^{-1/2}\widehat\bDelta}{(1-{\widehat{\bDelta}}^{\top}\widehat\bSigma^{-1}\widehat\bDelta)^{1/2}}.
\end{equation*}}

It is important to stress that,  from {Eqs.}~(\ref{eq:beta_tMLC0})-(\ref{eq:Gamma_tMLC}), the E-step
reduces to the computation of {\small $\widehat{\yp_i^{2}}$, $\widehat{\yp}_i$, $\widehat{t}_i$, $\widehat{t^2_i}$} and {\small $\widehat{t\yp}_i$}, for which we have implementable expressions. As pointed out for an anonymous referee, since missing values as treated as interval censored data, the computation burden relies heavily on the dimension of censored vector for evaluating the expectations of TESN and TN random vectors. In next subsection, we briefly discuss how to circumvent this problem, such that missing values do not represent neither a mathematical or computational burden.

\subsection{\bf Efficient computation of expectations}

In the event that there are missing values, we can partition the censored vector as {\small $\Y_{cens} = (\Y_c^\top,\Y_m^\top)^\top$}, that is, as missing and (truly) censored, in order to avoid unnecessary calculation of integrals for obtaining its expectation. Considering the partition above such that {\small $dim(\Y_{c}) = p^c_c$, $dim(\Y_m) = p^c_m$}, where {\small $p^c_c + p^c_m = p^c$}, it follows that
{\small\begin{equation}\label{EEE}
\mathbb{E}[\Y_{cens}|\Y_{obs}] = \mathbb{E}\left[
\begin{array}{c}
\mathbb{E}[\Y_m|\Y_c,\Y_{obs}]
\\
\Y_c|\Y_{obs}
\end{array}
\right]
\end{equation}}
and {\small $\mathrm{var}[\Y_{cens}|\Y_{obs}]$} is given by
{\small\begin{equation}\label{VVV}
\left[\begin{array}{cc}
\mathbb{E}[\mathrm{var}[\Y_m|\Y_c,\Y_{obs}]] + \mathrm{var}[\mathbb{E}[\Y_m|\Y_c,\Y_{obs}]]&
\mathrm{cov}[\mathbb{E}[\Y_m|\Y_c,\Y_{obs}],\Y_c|\Y_{obs}]\\
\mathrm{cov}[\Y_c|\Y_{obs},\mathbb{E}[\Y_m|\Y_c,\Y_{obs}]] &
\mathrm{var}[\Y_c|\Y_{obs}]
\end{array}
\right].
\end{equation}}

By noting that {\small $\Y_m = (\bV = (-\boldsymbol{\infty},\boldsymbol{\infty}),\C = \mathbf{1})$}, we have that {\small $\Y_m|\Y_c,\Y_{obs}$} is a non-truncated partition following a ESN distribution which moments have closed forms. Then, the computation of the first two moments of {\small $\Y_{cens}|\Y_{obs}$} can be calculated using Eqs. \eqref{EEE} and \eqref{VVV}, these last only depending on the computation of the truncated moments of {\small $\Y_c|\Y_{obs}$}, these are {\small $\mathbb{E}[\Y_c|\Y_{obs}]$} and {\small $\mathrm{var}[\Y_c|\Y_{obs}]$}. As can be seen, we can use the latter equations to treat missing data as censored in a neat manner, where the truncated moments are computed only over the {\small $p^c_c$}-variate partition, avoiding some unnecessary integrals and saving a significant computational effort.

\begin{remark}
In general, TESN distributions are not closed under marginalization but conditioning. For instance, {\small $\Y_m|\Y_{obs}$} does not follow a TESN distribution but its conditional distribution {\small $\Y_m|\Y_c,\Y_{obs}$} does. Furthermore, since {\small $\bV = (-\boldsymbol{\infty},\boldsymbol{\infty})$} for missing observations, we have that {\small $\Y_m|\Y_c,\Y_{obs}$} is a (conditionally) non-truncated partition, following a ESN distribution. For this particular case, {\small $\Y_c|\Y_{obs}$} follow a TESN distribution due to the aforementioned condition.
\end{remark}

\section{The FM-MSNC model}\label{sectionFM-tMC}

Ignoring censoring for the moment, we consider a more general and robust framework for the multivariate response variable {\small $\Y_i$} of the model defined in (\ref{modeleq}), which is assumed to follow a mixture of multivariate skew normal distributions:
{\small\begin{equation}\label{mixerror}
\Y_i\sim \sumasj \pi_j\, SN_{p}(\bmu_j,\bSigma_j,\blambda_j),
\end{equation}}
where {\small $\pi_j$} are weights adding to 1 and {\small $G$} is the number of groups, also called components in mixture models. The mixture model considered in (\ref{mixerror}) can also be by letting {\small $Z_{ij}$} be a latent class variable, such that
{\small$$Z_{ij}=\begin{cases}1&  \mbox{if the
$i$th observation is from the $j$th component,} \\ 0
&\mbox{otherwise}.
\end{cases}$$}
Thus, given {\small $Z_{ij} = 1$}, the response {\small $\Y_i$} follows a multivariate skew-normal distribution
{\small\begin{eqnarray}\label{linearmodelnormal}
\Y_i\sim SN_{p}(\bmu_j,\bSigma_j,\lambda_j),\quad \ii,\quad \jj.
\end{eqnarray}}

Now, suppose {\small $\mathrm{Pr}(Z_{ij} = 1) = \pi_j$}. Then the density of {\small $\Y_i$}, without observing {\small $Z_{ij}$}, is given by
{\small\begin{eqnarray}\label{modelmixturenormal}
f(\yp_i\mid\btheta)=\sum_{j=1}^G \pi_j\,
SN_{p}(\yp_i;\bmu_j,\bSigma_j,\blambda_j),
\end{eqnarray}}
where {\small $\btheta=(\btheta^{\top}_1,\ldots, \btheta^{\top}_G)^{\top},$}
with {\small $\btheta_j=(\pi_j, \bmu^{\top}_j, \bSigma_j,\blambda_j)^{\top}$}.

We treat the observed and censored components of {\small $\ \Y_i $}, separately, i.e. {\small $\y_i=(\y^{o^\top}_i,\y^{c^\top}_i)^\top$}, with respective partitioned parameters as in \eqref{partpars}. Following  \cite{lachos2017finite}, we define the  mixture model for censored data as a mixture of the MSNC models given in (\ref{equ8.1}), {viz.}
{\small\begin{eqnarray}\label{eqG}
f(\bV_i\mid\bC_i,\btheta)=\sum_{j=1}^G \pi_j f_{ij}(\bV_i\mid\bC_i,\btheta),
\end{eqnarray}}
with
{\small\begin{align*}
f_{ij}(\bV_i\mid\bC_i,\btheta)
&=\LL_{p^c_i}(\bV^c_{1i},\bV^c_{2i};\bmu^{co}_i,\bSigma^{cc.o}_i,
\bSigma^{cc.o\,1/2}_{i}\bvarphi^c_i, \tau^{co}_i)\\
&\times SN_{p^o_i}
(\y^o_i;\bmu^o_i,\bSigma^{oo}_{i}, c^{oc}_i\bSigma^{oo\,1/2}_i\tilde{\bvarphi}^o_i),
\end{align*}}
where, for each component $j$,  the arguments are defined as (\ref{eqn mu.co S.co}) and (\ref{eq:Sig:co1}), respectively. The model defined in (\ref{eqG}) will be called the FM-MSNC model.
Thus, the log-likelihood function given the observed data {\small $(\mathbf{V},\mathbf{C})$} is given by {\small$$\ell(\btheta\mid\bV,\bC)= \sumas \ln \{ f(\bV_i\mid\bC_i,\btheta)\}.$$}

\subsection{Maximum likelihood estimation via the EM algorithm}\label{sectionem}

In this section, we present an EM algorithm for the ML estimation of the FM-MSNC model.   To do so, we present the FM-MSNC model in an incomplete-data framework, using the results presented in {Section}~\ref{sec:mod:spec}.  We recall that the likelihood associated with finite mixtures of {skew-normal}  distributions  may be unbounded, as shown by \cite{Cabral2012}. Using a straightforward extension of their  argument, it   can be  shown   that the likelihood may be unbounded in the FM-MSNC  case as well. Despite this,  following \cite{peel2000robust} (p.~41), we  shall henceforth refer to the solution provided by the EM algorithm  as the ML estimate  even in situations where it may not globally maximize the likelihood.

Using the stochastic representation of the {skew-normal} distribution given in (\ref{stochSN}), it follows that the complete data log-likelihood function  is {\small $\ell_c(\btheta)=\sum_{i=1}^n \ell_{ic}(\btheta)$}, where, for each {\small $i \in \{ 1, \ldots, n\}$},
{\small\begin{align}\label{log-likecomple}
\ell_{ic}(\btheta)&=c+\sum_{j=1}^G
z_{ij}\ln{\pi_j}-\frac{1}{2}\sum_{j=1}^G z_{ij}
\ln{(|\bGamma_j|)}\nonumber\\
&-\frac{1}{2}\sum_{j=1}^G
z_{ij}{(\yp_i-\bmu_j-\bDelta_jt_i)^{\top}\bGamma^{-1}_j(\yp_i-\bmu_j-\bDelta_jt_i)},
\end{align}}
with {\small $c$} being a constant which is independent of the parameter vector {\small $\btheta$}.

For each {\small $j \in \{ 1, \ldots, G \}$}, let {\small ${\widehat{\btheta}}^{(k)}_j=({\widehat{\pi}}^{(k)}_j,{\widehat{\bmu}}^{(k)}_j, {\widehat{\bSigma}}^{(k)}_j,{\widehat{\blambda}}^{(k)}_j)^{\top}$}, and let {\small ${\widehat{\btheta}}^{(k)}=({\widehat{\btheta}}^{(k)\top}_1,\ldots,\\ {\widehat{\btheta}}^{(k)\top}_G)^{\top}$} be the estimate of {\small $\btheta$} at the {\small $k$}th iteration. It follows, after  some simple algebra, that the conditional expectation of the complete log-likelihood function has the form
{\small\begin{align*}
Q(\btheta\mid{\widehat{\btheta}}^{(k)})&\propto \sum_{i=1}^n\sum_{j=1}^G
{\cal
Z}_{ij}(\widehat{\btheta}^{(k)})\ln{\pi_j}-\frac{1}{2}\sum_{i=1}^n\sum_{j=1}^G
{\cal Z}_{ij}(\btheta^{(k)})
\ln{(|\widehat{\bGamma_j}^{(k)}|)}\nonumber\\
&-\displaystyle\frac{1}{2} \sum_{i=1}^n\sum_{j=1}^G
\textrm{tr}\left[ \widehat{\bGamma}_j^{-1(k)} \left\{ {\cal
E}_{2ij}(\widehat{\btheta}^{(k)})-\widehat{\bmu}_j^{(k)} {\cal
E}^{\top}_{1ij}(\widehat{\btheta}^{(k)})- {\cal
E}_{1ij}(\widehat{\btheta}^{(k)})\widehat{\bmu}_j^{(k)\top}\right.  \right.\\ 
&- \left. \left. {\cal
E}_{3ij}(\widehat{\btheta}^{(k)})\widehat{\bDelta}_j^{(k)\top} -\widehat{\bDelta}_j^{(k)}{\cal
E}_{3ij}(\widehat{\btheta}^{(k)\top})
+ {\cal
Z}_{ij}(\widehat{\btheta}^{(k)})
\widehat{\bmu}_j^{(k)}\widehat{\bmu}_j^{(k)\top} \right.  \right.\\ 
&+ \left. \left.{\cal
E}_{4ij}(\widehat{\btheta}^{(k)})\widehat{\bDelta}_j^{(k)}\widehat{\bDelta}_j^{(k)\top} +   {\cal
E}_{5ij}(\widehat{\btheta}^{(k)})\widehat{\bDelta}_j^{(k)}\widehat{\bmu}_j^{(k)\top} +{\cal
E}_{5ij}(\widehat{\btheta}^{(k)})\widehat{\bmu}_j^{(k)}\widehat{\bDelta}_j^{(k)\top}\right\}
\right],
\end{align*}} where
{\small \begin{align*}
{\cal E}_{1ij}(\widehat{\btheta}^{(k)}) &=  \mathrm{E}(Z_{ij} \Y_i\mid\bV_i,\bC_i,\widehat{\btheta}^{(k)}),\quad
{\cal E}_{2ij}(\widehat{\btheta}^{(k)}) =  \mathrm{E}(Z_{ij} \Y_i\Y_i^{\top}\mid\bV_i,\bC_i,\widehat{\btheta}^{(k)}), \\
{\cal E}_{3ij}(\widehat{\btheta}^{(k)}) &=  \mathrm{E}(Z_{ij}T_i\Y_i\mid\bV_i,\bC_i,\widehat{\btheta}^{(k)}), \quad
{\cal E}_{4ij}(\widehat{\btheta}^{(k)}) =  \mathrm{E}(Z_{ij}T^2_i
\mid\bV_i,\bC_i,\widehat{\btheta}^{(k)}), \\
{\cal E}_{5ij}(\widehat{\btheta}^{(k)}) &=  \mathrm{E}(Z_{ij}T_i
\mid\bV_i,\bC_i,\widehat{\btheta}^{(k)})\quad \textrm{and}\quad
{\cal
Z}_{ij}(\widehat{\btheta}^{(k)})=\mathrm{E}(Z_{ij}\mid\bV_i,\bC_i,\widehat{\btheta}^{(k)}).
\end{align*}}
By using known properties of conditional expectation, we obtain
{\small \begin{align}{\cal Z}_{ij}(\widehat{\btheta}^{(k)})& =\displaystyle\frac{{\widehat{\pi}}^{(k)}_j
f_{ij}(\bV_i\mid\bC_i,\widehat{\btheta}^{(k)}_j)}{\displaystyle\sum_{j=1}^G
{\widehat{\pi}}^{(k)}_j
f_{ij}(\bV_i\mid \bC_i,\widehat{\btheta}^{(k)}_j)},\label{zij}\\
{\cal E}_{1ij}(\widehat{\btheta}^{(k)}) & ={\cal
Z}_{ij}(\widehat{\btheta}^{(k)})\mathrm{E}(\Y_i\mid\bV_i,\bC_i,\widehat{\btheta}^{(k)},
Z_{ij}=1)\nonumber \\
{\cal E}_{2ij}(\widehat{\btheta}^{(k)}) & ={\cal
Z}_{ij}(\widehat{\btheta}^{(k)})\mathrm{E}(\Y_i\Y^{\top}_i\mid\bV_i,\bC_i,\widehat{\btheta}^{(k)},
Z_{ij}=1),\nonumber \\
{\cal E}_{3ij}(\widehat{\btheta}^{(k)}) & ={\cal
Z}_{ij}(\widehat{\btheta}^{(k)})\mathrm{E}(T_i\Y_i\mid\bV_i,\bC_i,\widehat{\btheta}^{(k)},
Z_{ij}=1),\\
{\cal E}_{4ij}(\widehat{\btheta}^{(k)}) & ={\cal
Z}_{ij}(\widehat{\btheta}^{(k)})\mathrm{E}(T^2_i\mid\bV_i,\bC_i,\widehat{\btheta}^{(k)},
Z_{ij}=1)\nonumber
\end{align}}
\noindent and
{\small \begin{align}
{\cal E}_{5ij}(\widehat{\btheta}^{(k)}) &={\cal
Z}_{ij}(\widehat{\btheta}^{(k)})\mathrm{E}(T_{i}\mid\bV_i,\bC_i,\widehat{\btheta}^{(k)},
Z_{ij}=1).   \nonumber
\end{align}}

The conditional expectations
{\small$\mathrm{E}(\Y_i\mid\bV_i,\bC_i, \ \widehat{\btheta}^{(k)}, Z_{ij}=1)$, $\ \mathrm{E}(\Y_i\Y^{\top}_i\mid\bV_i,\bC_i, \ \widehat{\btheta}^{(k)}, \\ Z_{ij}=1)$, $\ \mathrm{E}(T_i\Y_i\mid\bV_i,\bC_i,\widehat{\btheta}^{(k)}, Z_{ij}=1)$, $\ \mathrm{E}(T^2_i\mid\bV_i,\bC_i,\widehat{\btheta}^{(k)}, Z_{ij}=1)$} and {\small$\ \mathrm{E}(T_i\mid\bV_i,\bC_i,\widehat{\btheta}^{(k)}, Z_{ij}=1)$} can be directly obtained from expressions $\widehat{\yp}_i$, $\widehat{\yp_i^{2}}$, {\small$\widehat{t\yp}_i$, $\widehat{t^2}_i$} and {\small$\widehat{t}_i$}, respectively, given in {Subsection}~\ref{EM-tlcm}. Thus, we have closed form expressions  for all the quantities involved in the E-step of the algorithm. Next, we describe the EM algorithm for maximum likelihood estimation of the parameters in  the FM-MSNC model.\\

\noindent \textbf{E-step}: Given {\small$\btheta={\widehat{\btheta}}^{(k)}$}, compute {\small${\cal E}_{sij}(\widehat{\btheta}^{(k)})$} for all {\small$s\in \{1,2,3,4,5\}$} and
{\small${\cal Z}_{ij}(\widehat{\btheta}^{(k)})$} for all {\small$\ii$, $\jj$}.

\noindent \textbf{M-step}: Update {\small${\widehat{\btheta}}^{(k+1)}$} by maximizing {\small$Q(\btheta\mid{\widehat{\btheta}}^{(k)})$} over {\small$\btheta$}, which leads to the following closed  form expressions:
{\small \begin{align*}
{\widehat{\pi}}^{(k+1)}_j&=\frac{1}{n}\sumas {\cal Z}_{ij}(\widehat{\btheta}^{(k)}),\\
{\widehat{\bmu}}^{(k+1)}_j&=\left\{ \sumas {\cal Z}_{ij}(\widehat{\btheta}^{(k)})\right\}^{-1}\sumas\{
{\cal E}_{1ij}(\widehat{\btheta}^{(k)})-{\cal E}_{5ij}(\widehat{\btheta}^{(k)})\widehat{\bDelta}^{(k)}_j\}
\\
{\widehat{\bDelta}}^{(k+1)}_j&=\left\{ \sumas {\cal E}_{4ij}(\widehat{\btheta}^{(k)})\right\}^{-1}\sumas\{
{\cal E}_{3ij}(\widehat{\btheta}^{(k)})-{\cal E}_{5ij}(\widehat{\btheta}^{(k)})\widehat{\bmu}^{(k+1)}_j\}
\\
{\widehat{\bGamma}}^{(k+1)}_j&= \left\{ \sumas {\cal
Z}_{ij}(\widehat{\btheta}^{(k)})\right\}^{-1}
\sum_{i=1}^n
\left\{  {\cal
E}_{2ij}(\widehat{\btheta}^{(k)})-\widehat{\bmu}_j^{(k)} {\cal
E}^{\top}_{1ij}(\widehat{\btheta}^{(k)})- {\cal
E}_{1ij}(\widehat{\btheta}^{(k)})\widehat{\bmu}_j^{(k)\top} \right. \\
&- \left.{\cal
E}_{3ij}(\widehat{\btheta}^{(k)})\widehat{\bDelta}_j^{(k)\top} -\widehat{\bDelta}_j^{(k)}{\cal
E}_{3ij}(\widehat{\btheta}^{(k)\top}) + {\cal
Z}_{ij}(\widehat{\btheta}^{(k)})
\widehat{\bmu}_j^{(k)}\widehat{\bmu}_j^{(k)\top} \right.\\ 
& \left. + {\cal
E}_{4ij}(\widehat{\btheta}^{(k)})\widehat{\bDelta}_j^{(k)}\widehat{\bDelta}_j^{(k)\top} + {\cal
E}_{5ij}(\widehat{\btheta}^{(k)})\widehat{\bDelta}_j^{(k)}\widehat{\bmu}_j^{(k)\top} + {\cal
E}_{5ij}(\widehat{\btheta}^{(k)})\widehat{\bmu}_j^{(k)}\widehat{\bDelta}_j^{(k)\top}\right\} ,
\end{align*}}
for all {\small$j \in \{1,\ldots, G\}$}.\\

It is well known that mixture models can provide a multimodal log-likelihood function. In this sense, the method of maximum likelihood estimation through the EM algorithm may not give global solutions if the starting values are far from the real parameter values. Thus, the choice of starting values for the EM algorithm in the mixture context plays a big role in parameter estimation. In our examples and simulation studies, we consider the following procedure for the FM-MSNC model:
\begin{itemize}
\item [(i)]
Partition the data (censoring levels replacing  the  censored observations)  into {\small$G$} groups using the K-means clustering algorithm \citep{Cabral2012}.
\item [(ii)] Compute the proportion  of data points belonging to the same cluster {\small$j$}, say  {\small$\pi_{j}^{(0)}$, $j \in \{1,\ldots,G\}$}. This gives the initial value for {\small$\pi_{j}$}.
\item [(iii)] For each group {\small$j$}, compute the initial values {\small$\bmu_{j}^{(0)}$, $\bSigma_{j}^{(0)}$, $\blambda_{j}^{(0)}$} using the \verb|R| package  \verb|mixsmsn| \citep{prates2013mixsmsn}.
\end{itemize}

\subsection{Model selection}\label{Mselect}

Because there is no universal criterion for mixture model selection, we chose three criteria to compare the models considered in this work, namely, the Akaike information criterion (AIC) \citep{Akaike74},  Bayesian information criterion (BIC) \citep{Schwarz78} and efficient determination criterion (EDC) \citep{Bai1989}. Like the AIC and BIC, EDC has the form {\small$ -2\ell(\widehat{\btheta})+\rho c_n,$} where {\small$\ell(\btheta)$} is the actual log-likelihood, {\small$\rho$} is the number of free parameters that has to be estimated in the model and the penalty term {\small${c_n}$} is a convenient sequence of positive numbers. Here, we use {\small$c_n = 0.2 \sqrt{n}$}, a proposal that was considered in \cite{Basso2010} and \cite{Cabral2012}. Note that {\small $c_n$} constant is given by {\small$c_n = 2$} for AIC and {\small$c_n = \log n$} for BIC, with {\small$n$} being the sample size.

\subsection{Provision of standard errors}  \label{sec SE}

In this section, we describe how to obtain the standard errors of the ML estimates for the FM-MSNC model. We follow the information-based method exploited by \cite{basford1997standard} to compute the asymptotic covariance of the ML estimates. The empirical information matrix, according to \cite{meilijson1989fast}'s formula, is defined as
{\small\begin{eqnarray}\label{eq:IM}
\bI_e(\btheta|\yp) = \sum_{i = 1}^{n} s(\yp_i|\btheta)s^\top (\yp_i|\btheta) - \frac{1}{n} S(\yp_i|\btheta)S^\top (\yp_i|\btheta),
\end{eqnarray}}
where {\small$S(\yp_i|\btheta) = \sum_{i = 1}^{N} s(\yp_i|\btheta)$} and {\small$s(\yp_i|\btheta)$} is the empirical score function for the {\small$i$}th subject. It is noted from the result of \cite{louis1982finding} that the individual score can be determined as
{\small\begin{eqnarray}
s(\yp_i|\btheta) = \E\left( \left. \frac{\partial\ell_i(\btheta|\yp_c)}{\partial \btheta}\right| \bV_i, \bC_i,\btheta\right).
\end{eqnarray}}
Using the ML estimates {\small$\widehat{\btheta}$} in {\small$s(\yp_i|\btheta)$}, leads to {\small$S(\yp_i|\widehat{\btheta}) = 0$}, so from (\ref{eq:IM}) we have that
{\small \begin{eqnarray}\label{eq:oim}
\bI_e(\widehat{\btheta}|\yp) = \sum_{i = 1}^{n} \widehat{\se}_i \widehat{\se}^\top_i,
\end{eqnarray}}
where {\small$\widehat{\se}_i$} is an individual score vector given by {\small $\widehat{\se}_i = (\widehat{s}_{i,\bmu_{1}}, \ldots, \widehat{s}_{i,\bmu_{G}}, \widehat{s}_{i,{\bsigma^2}_{1}}, \ldots, \widehat{s}_{i,{\bsigma^2}_{G}},\\ \widehat{s}_{i,\blambda_{1}}, \ldots,  \widehat{s}_{i,\blambda_{G}}, \widehat{s}_{i,\pi_1}, \ldots, \widehat{s}_{i,\pi_{G-1}})^\top$}, where {\small$\bsigma_j^2$} is a vector with {\small$p(p + 1)/2$} distinct elements of {\small$\bSigma_j$}.

First we reparameterize {\small$\bSigma_j = \bF_j^2$} for ease of computation and theoretical derivation, where {\small$\bF_j$} is the square root of {\small$\bSigma_j$} containing {\small$p(p + 1)/2$} distinct elements.

Now we have that {\small$\widehat{\se}_i = (\widehat{s}_{i,\bmu_{1}}, \ldots, \widehat{s}_{i,\bmu_{G}}, \widehat{s}_{i,{\balpha}_{1}}, \ldots, \widehat{s}_{i,{\balpha}_{G}}, \widehat{s}_{i,\blambda_{1}}, \ldots,  \widehat{s}_{i,\blambda_{G}}, \widehat{s}_{i,\pi_1},\  \ldots \ , \\\widehat{s}_{i,\pi_{G-1}})^\top$}. So, the expressions for the elements of {\small$\widehat{\se}_i$} are given by:
{\small\begin{eqnarray*}
\widehat{s}_{i,\pi_j} &=& \frac{{\cal
Z}_{ij}(\widehat{\btheta})}{\widehat{\pi}_j} - \frac{{\cal
Z}_{ij}(\widehat{\btheta})}{\widehat{\pi}_G},\\
\widehat{s}_{i,\bmu_{j}} &=& (\widehat{s}_{i,\mu_{j1}}, \ldots, \widehat{s}_{i,\mu_{jp}}) = \widehat{\bGamma}_j^{-1}\left( {\cal E}_{1ij}(\widehat{\btheta}) - {\cal
Z}_{ij}(\widehat{\btheta})\widehat{\bmu}_j - {\cal E}_{5ij}(\widehat{\btheta})\widehat{\bDelta}_j\right), \\
\widehat{s}_{i,\balpha_{j}} &=& (\widehat{s}_{i,\alpha_{j,11}}, \ldots, \widehat{s}_{i,\alpha_{j,pp}}) = -\frac{{\cal
Z}_{ij}(\widehat{\btheta})}{2}\mathrm{tr}\left(\widehat{\bGamma}_j^{-1}A_{j}(\widehat{\btheta})\right)
-\frac{1}{2}\left\lbrace {\cal E}_{1ij}(\widehat{\btheta})^\top\widehat{\bGamma}_j^{-1}A_{j}(\widehat{\btheta})\widehat{\bGamma}_j^{-1}\widehat{\bmu}_j \right. \\ 
&+& \left.
\widehat{\bmu}_j^\top\widehat{\bGamma}_j^{-1}A_{j}(\widehat{\btheta})\widehat{\bGamma}_j^{-1}{\cal E}_{1ij}(\widehat{\btheta}) - {\cal E}_{3ij}(\widehat{\btheta})^\top\left(\widehat{\bGamma}_j^{-1}\dot{\bF}_j(r)\widehat{\bdelta}_j - \widehat{\bGamma}_j^{-1}A_{j}(\widehat{\btheta})\widehat{\bGamma}_j^{-1}\widehat{\bDelta}_j \right) \right.
\\ &-&\left. \mathrm{tr}\left({\cal E}_{2ij}(\widehat{\btheta})\widehat{\bGamma}_j^{-1}A_{j}(\widehat{\btheta})\widehat{\bGamma}_j^{-1}\right) - \left(\widehat{\bdelta}_j^\top\dot{\bF}_j(r)\widehat{\bGamma}_j^{-1} - \widehat{\bDelta}_j^\top\widehat{\bGamma}_j^{-1}A_{j}(\widehat{\btheta})\widehat{\bGamma}_j^{-1}\widehat{\bDelta}_j \right){\cal E}_{3ij}(\widehat{\btheta}) \right. \\ &-&\left.  
\widehat{\bmu}_j^\top\widehat{\bGamma}_j^{-1}A_{j}(\widehat{\btheta})\widehat{\bGamma}_j^{-1}\widehat{\bmu}_j
+{\cal E}_{5ij}(\widehat{\btheta})\widehat{\bmu}_j^\top\left(\widehat{\bGamma}_j^{-1}\dot{\bF}_j(r)\widehat{\bdelta}_j - \widehat{\bGamma}_j^{-1}A_{j}(\widehat{\btheta})\widehat{\bGamma}_j^{-1}\widehat{\bDelta}_j\right) \right. \\ 
&+& \left. {\cal E}_{4ij}(\widehat{\btheta})\left(\widehat{\bdelta}_j^\top\dot{\bF}_j(r)\widehat{\bGamma}_j^{-1}\widehat{\bDelta}_j - \widehat{\bDelta}_j^\top\widehat{\bGamma}_j^{-1}A_{j}(\widehat{\btheta})\widehat{\bGamma}_j^{-1}\widehat{\bDelta}_j + \widehat{\bDelta}_j^\top\widehat{\bGamma}_j^{-1}\dot{\bF}_j(r)\widehat{\bdelta}_j  \right) \right. \\
&+& \left. \left(\widehat{\bdelta}_j^\top\dot{\bF}_j(r)\widehat{\bGamma}_j^{-1} - \widehat{\bDelta}_j^\top\widehat{\bGamma}_j^{-1}A_{j}(\widehat{\btheta})\widehat{\bGamma}_j^{-1}\right)\widehat{\bmu}_j{\cal E}_{5ij}(\widehat{\btheta})
\right\rbrace,\\
\end{eqnarray*}}
{\small\begin{eqnarray*}
\widehat{s}_{i,\blambda_{j}} &=& (\widehat{s}_{i,\lambda_{j1}}, \ldots, \widehat{s}_{i,\lambda_{jp}}) = \frac{{\cal
Z}_{ij}(\widehat{\btheta})}{2}\mathrm{tr}\left(\widehat{\bGamma}_j^{-1}B_j(\widehat{\btheta})\right)
-\frac{1}{2}\left\lbrace \mathrm{tr} \left( {\cal E}_{2ij}(\widehat{\btheta})\widehat{\bGamma}_j^{-1}B_{j}(\widehat{\btheta})\widehat{\bGamma}_j^{-1}\right) \right. \\ 
&-& \left.{\cal E}_{1ij}(\widehat{\btheta})^\top\widehat{\bGamma}_j^{-1}B_{j}(\widehat{\btheta})\widehat{\bGamma}_j^{-1}\widehat{\bmu}_j
+ \widehat{\bmu}_j^\top\widehat{\bGamma}_j^{-1}B_{j}(\widehat{\btheta})\widehat{\bGamma}_j^{-1}\widehat{\bmu}_j -
\widehat{\bmu}_j^\top\widehat{\bGamma}_j^{-1}B_{j}(\widehat{\btheta})\widehat{\bGamma}_j^{-1}{\cal E}_{1ij}(\widehat{\btheta}) \right. \\
&-& \left. 
{\cal E}_{3ij}(\widehat{\btheta})^\top\left( \widehat{\bGamma}_j^{-1}B_{j}(\widehat{\btheta})\widehat{\bGamma}_j^{-1}\widehat{\bDelta}_j + \widehat{\bGamma}_j^{-1}b_{j}(\widehat{\btheta})\right) \right. \\
&-& \left. \left(\widehat{\bDelta}_j^\top \widehat{\bGamma}_j^{-1}B_{j}(\widehat{\btheta})\widehat{\bGamma}_j^{-1} + b_{j}(\widehat{\btheta})^\top\widehat{\bGamma}_j^{-1}\right) {\cal E}_{3ij}(\widehat{\btheta})\right. \\
&+& \left. {\cal E}_{5ij}(\widehat{\btheta})\widehat{\bmu}_j^\top\left( \widehat{\bGamma}_j^{-1}B_{j}(\widehat{\btheta})\widehat{\bGamma}_j^{-1}\widehat{\bDelta}_j + \widehat{\bGamma}_j^{-1}b_{j}(\widehat{\btheta})\right) \right. \\
&+& \left. \left(\widehat{\bDelta}_j^\top \widehat{\bGamma}_j^{-1}B_{j}(\widehat{\btheta})\widehat{\bGamma}_j^{-1} + b_{j}(\widehat{\btheta})^\top\widehat{\bGamma}_j^{-1}\right)\widehat{\bmu}_j{\cal E}_{5ij}(\widehat{\btheta}) \right. \\
&+& \left.{\cal E}_{4ij}(\widehat{\btheta})\left(b_{j}(\widehat{\btheta})^\top\widehat{\bGamma}_j^{-1}\widehat{\bDelta}_j + \widehat{\bDelta}_j^\top\widehat{\bGamma}_j^{-1}B_{j}(\widehat{\btheta})\widehat{\bGamma}_j^{-1}\widehat{\bDelta}_j +  \widehat{\bDelta}_j^\top\widehat{\bGamma}_j^{-1}b_{j}(\widehat{\btheta})\right)  \right\rbrace,
\end{eqnarray*}}
where
{\small\begin{eqnarray*}
A_j(\widehat{\btheta}) &=& \left(\dot{\bF}_j(r)(\mathbf{I} - \widehat{\bdelta}_j\widehat{\bdelta}_j^\top)\widehat{\bF}_j +  \widehat{\bF}_j(\mathbf{I} - \widehat{\bdelta}_j\widehat{\bdelta}_j^\top)\dot{\bF}_j(r)\right),\\
B_j(\widehat{\btheta}) &=& \widehat{{\bF}}_j\left( \frac{\dot{\bR}_{j}(r)(1 + \widehat{\blambda}_{j}^\top\widehat{\blambda}_{j}) - 2\lambda_{jr}\widehat{\blambda}_{j}\widehat{\blambda}_{j}^\top}{(1 + \widehat{\blambda}_{j}^\top\widehat{\blambda}_{j})^2}\right)\widehat{{\bF}}_j,\\
b_j(\widehat{\btheta}) &=& \widehat{{\bF}}_j\left( \frac{\dot{\blambda}_{j}(r)(1 + \widehat{\blambda}_{j}^\top\widehat{\blambda}_{j}) - \lambda_{jr}\widehat{\blambda}_{j}}{(1 + \widehat{\blambda}_{j}^\top\widehat{\blambda}_{j})^{3/2}}\right),
\end{eqnarray*}}
{\small$\left.\dot{\bF}_j(r) = \frac{\partial\bF_j}{\partial\sigma^2_{jr}}\right|_{\sigma^2=\widehat{\sigma^2}}$, $\left. \dot{\bR}_{j}(r) = \frac{\partial\blambda_j\blambda_j^\top}{\partial\lambda_{jr}}\right|_{\lambda=\widehat{\lambda}}$}, and {\small$\left. \dot{\blambda}_{j}(r) = \frac{\partial\blambda_j}{\partial\lambda_{jr}}\right|_{\lambda=\widehat{\lambda}}$}, with {\small$r = 1, 2, \ldots, p$}.

\section{Simulation studies}\label{sec-simStudy}

In order to study the performance of our proposed method, we present five simulation studies. The first and second study investigates whether we can estimate the true parameter values and their respective standard errors accurately by using the proposed EM algorithm and approximated empirical information matrix, respectively involving censoring and missing data. The third one investigates the number of mixture components by comparing the {\small FM-MSNC} with two groups and {\small FM-MNC} with various groups. The fourth study investigates  the  ability  of  the {\small FM-MSNC}  model  to  cluster  observations. Finally, the last one shows the asymptotic behavior of the EM estimates for the proposed model. The computations were done using the \verb|R| package \verb|CensMFM|.

\subsection{Performance of the ML Estimates over censoring data}\label{simu.1}

This simulation study is designed to verify if we can estimate the true parameter values  of the {\small FM-MSNC} model accurately when we have censoring data by using the proposed EM algorithm. We simulated several datasets considering mixtures with
two components from model (\ref{eqG}) with two left-censoring proportion settings ({\small$5\%$} and  {\small$30\%$}), taken in each mixture
component, and different samples sizes {\small$n$ $\in (500, 1000, 2000)$}. For each combination, we generated {\small $500$} Monte
Carlo (MC) samples. Summary statistics of the estimates across the {\small $500$} MC samples were computed, such as the mean estimate (MC mean), the empirical standard error (MC Sd), and the mean of the approximate standard errors of the estimates, obtained through the method described in Section \ref{sec SE} (IM SE). 

\begin{figure}[!h]
\centering
\center{
\subfigure[n = 500]{\includegraphics[width = 0.3\textwidth]{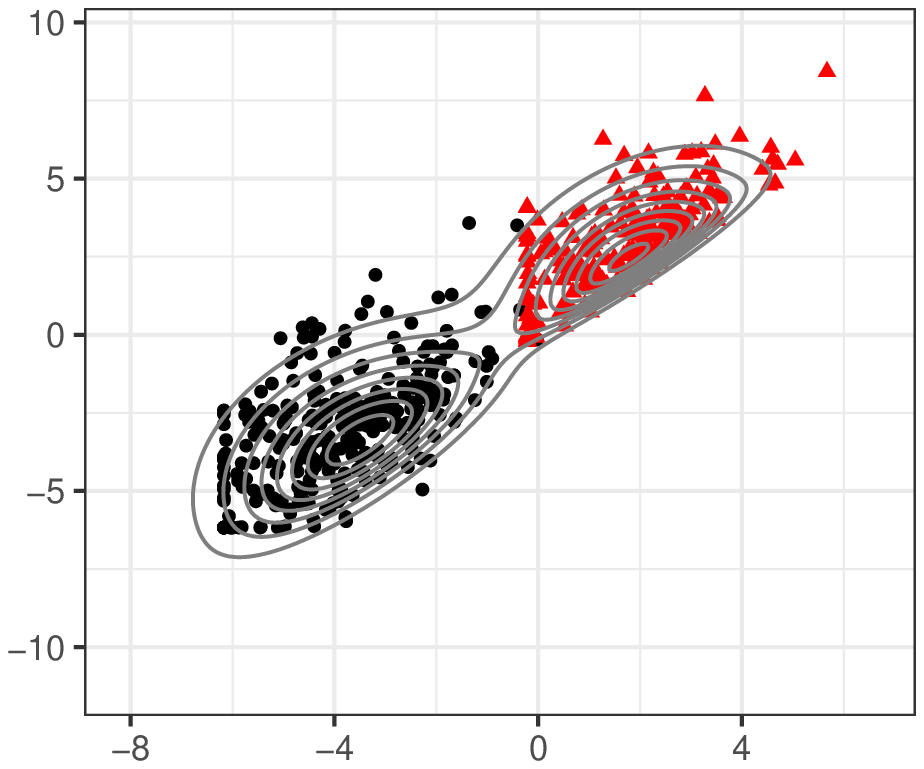}\label{fig:betadensity2}}
\subfigure[n = 1000]{\includegraphics[width = 0.3\textwidth]{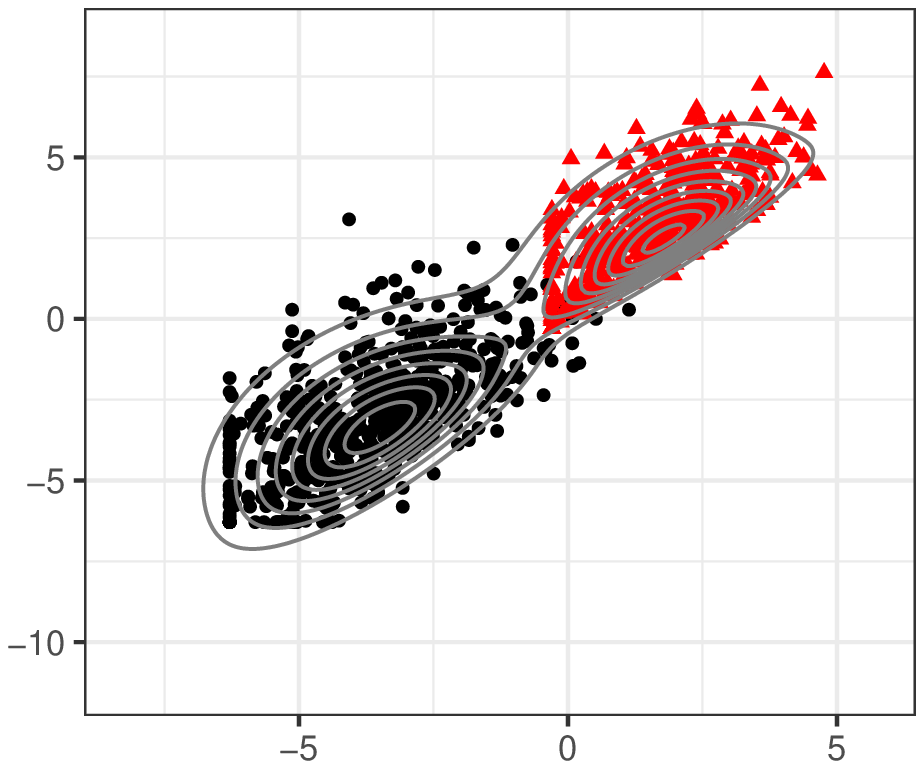}\label{fig:betadensity2}}
\subfigure[n = 2000]{\includegraphics[width = 0.3\textwidth]{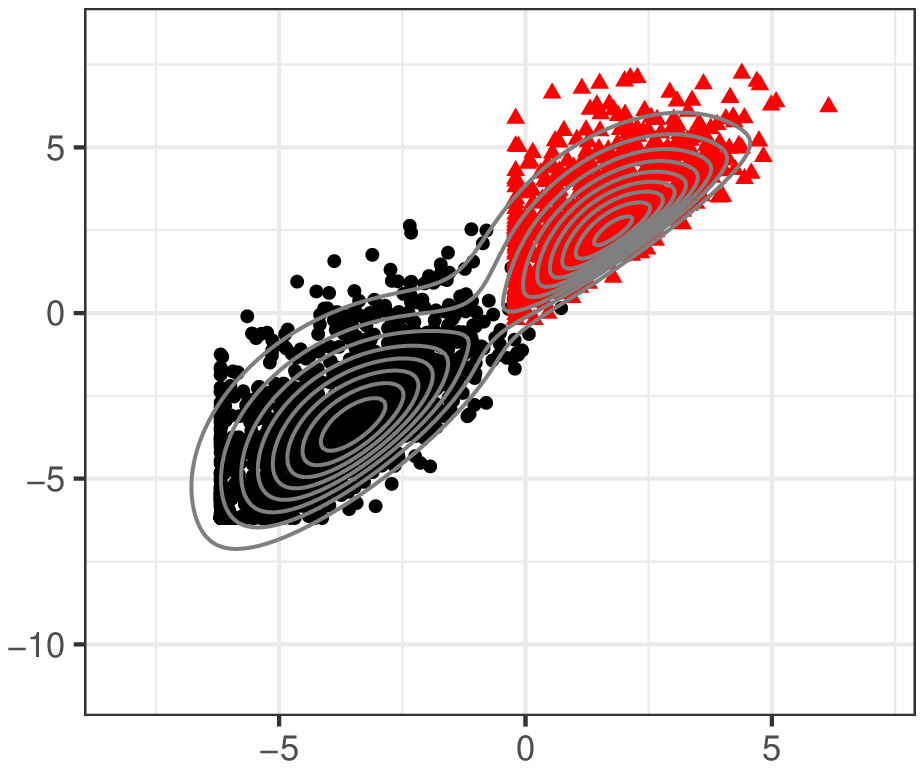}\label{fig:betadensity2}}\\
\subfigure[n = 500]{\includegraphics[width = 0.3\textwidth]{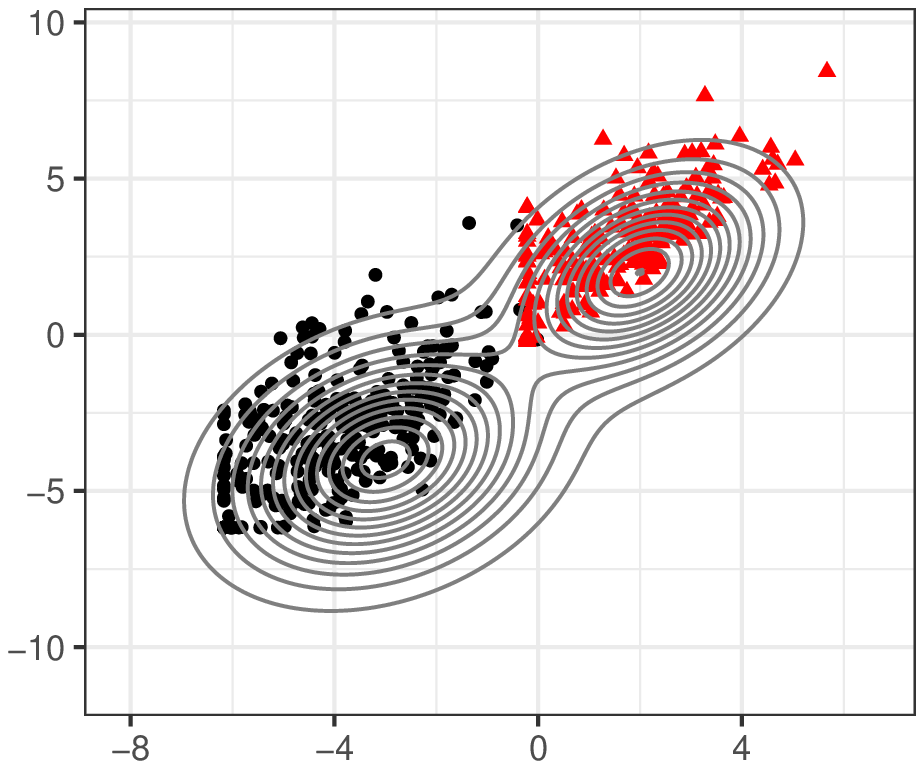}\label{fig:betadensity2}}
\subfigure[n = 1000]{\includegraphics[width = 0.3\textwidth]{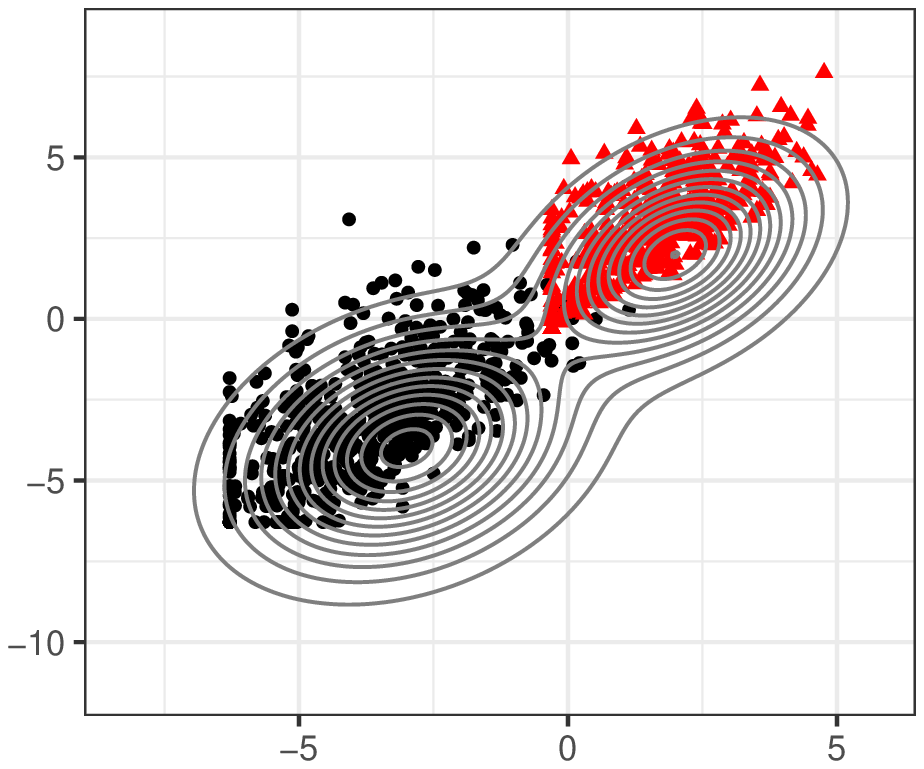}\label{fig:betadensity2}}
\subfigure[n = 2000]{\includegraphics[width = 0.3\textwidth]{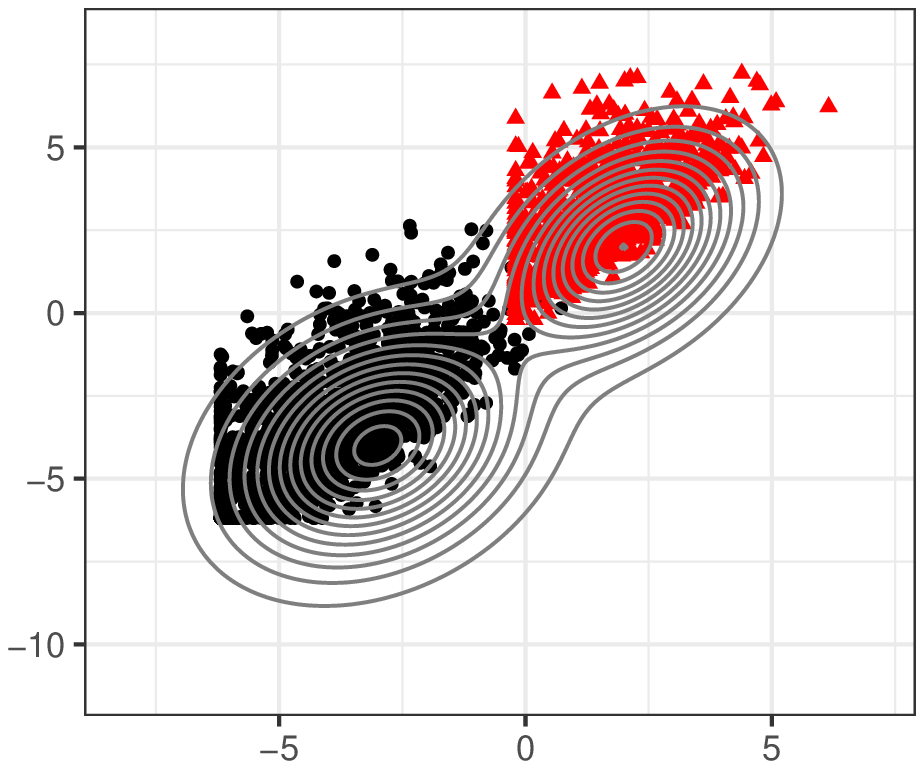}\label{fig:betadensity2}}}
\caption{\small{Simulated data: Performance of the ML Estimates over censoring data. Scatter plot for some simulated data from FM-MSNC model with the respective density contours, skew-normal (top panel) and normal (bottom panel) with $5\%$ censoring level.}}
\label{fig:data_simu1}
\end{figure}

We consider small and different variances with the following parameter setup:
{\small
\begin{eqnarray*}
0.65\,\, SN_{2}\left( \begin{bmatrix}
-3 \\
-4 \\
\end{bmatrix},
\begin{bmatrix}
3 & 1 \\
1 & 4.5 \\
\end{bmatrix}, \begin{bmatrix}
-2 \\
2 \\
\end{bmatrix}
\right) +
0.35\,\, SN_{2} \left( \begin{bmatrix}
2 \\
2 \\
\end{bmatrix},
\begin{bmatrix}
2 & 1 \\
1 & 3.5 \\
\end{bmatrix}, \begin{bmatrix}
-3 \\
4 \\
\end{bmatrix}
\right).
\end{eqnarray*}}

Figure \ref{fig:data_simu1} shows the simulated data from the {\small FM-MSNC}  model with their respective density contours for the skew-normal (top panel) and normal (bottom panel) distributions and the allocations in each group for samples sizes {\small$500$, $1000$}  and {\small$2000$} with left-censoring proportion of {\small$5\%$}. The black points represent the first component and the red triangles represent the second component of the mixture. One can note that the contour lines of the skew-normal distribution are more appropriate to  represent the shape assumed by the generated data. 

The results are presented in Table \ref{tab:results_simu1}. This table shows that, regardless the sample size, the Monte Carlo mean of the parameter estimates deviates further the true values as the censoring level increases, i.e., the parameter estimates are affected by the censoring level. In particular, the estimates of {\small $\bmu_1$} and {\small $\bmu_2$} appear to be less affected by increasing the censoring level than the other parameters. Furthermore, the estimates of the standard errors, i.e., MC Sd and IM SE, provide relatively close results, which may indicate that the asymptotic approach proposed for the standard errors of the ML estimates is reliable.

\begin{center}
\tiny
\begin{longtable}{ccccccccccc}
\caption{Simulated data: Performance of the ML Estimates over censoring data. Parameter estimates based on 500 simulated samples. Monte Carlo (MC) mean, MC Sd are the respective
mean estimates and standard deviations. IM SE is the average value of the approximate standard error obtained through the
information-based method.}
\label{tab:results_simu1}\\
\toprule
\multirow{2}{*}{Censoring} & \multirow{2}{*}{$j$} & \multirow{2}{*}{Measure} & \multicolumn{8}{c}{Parameter} \\ \cmidrule{4-11}
& & & $\mu_{j1}$ & $ \mu_{j2}$ & $\alpha_{j,11}$ & $\alpha_{j,12}$ & $\alpha_{j,22}$ & $\lambda_{j1}$ & $\lambda_{j2}$ & $\pi$ \\ \midrule
\endfirsthead
\multicolumn{11}{c}%
{\tablename\ \thetable\ -- \textit{Continued from previous page}} \\
\midrule
\multirow{2}{*}{Censored} & \multirow{2}{*}{j} & \multirow{2}{*}{Measure} & \multicolumn{8}{c}{Parameter} \\ \cmidrule{4-11}
& & & $\mu_{j1}$ & $ \mu_{j2}$ & $\alpha_{j,11}$ & $\alpha_{j,12}$ & $\alpha_{j,22}$ & $\lambda_{j1}$ & $\lambda_{j2}$ & $\pi$ \\ \midrule
\endhead
\hline \multicolumn{11}{r}{\textit{Continued on next page}} \\
\endfoot
\endlastfoot
\multicolumn{11}{c}{$n = 500$} \\
\midrule

\multirow{8}{*}{$5\%$} & \multirow{4}{*}{1} & True & ($-3$) & ($-4$) & ($1.7121$) & ($0.2620$) & ($2.1051$) & ($-2$) & ($2$) &($0.65$)\\
& & MC mean & -3.1797 & -4.1235 & 1.6262 & 0.2923 & 2.2499 & -1.7236 & 2.1533 & 0.6607\\
& & MC Sd &  0.3101 & 0.3705 & 0.1510 & 0.1138 & 0.1906 & 0.5262 & 0.5739 & 0.0274  \\
& & IM SE & 0.2844 & 0.3406 & 0.1361 & 0.0924 & 0.1944 & 0.5397 & 0.6202 & 0.0236 \\ \addlinespace[0.1cm]
& \multirow{4}{*}{2} & True & ($2$) & ($2$) & ($1.3798$) & ($0.3101$) & ($1.8449$) & ($-3$) & ($4$)\\
& & MC mean & 2.0196 & 2.0954 & 1.3152 & 0.3177 & 1.7883 & -2.7361 & 3.7058 \\
& & MC Sd &0.2381 & 0.2808 & 0.1431 & 0.0928 & 0.1595 & 1.0896 & 1.3303\\
& & IM SE & 0.2677 & 0.2922 & 0.1391 & 0.0981 & 0.1816 & 1.2411 & 1.5434 \\
\midrule
\multirow{8}{*}{$30\%$} & \multirow{4}{*}{1} & True & ($-3$) & ($-4$) & ($1.7121$) & ($0.2620$) & ($2.1051$) & ($-2$) & ($2$) &($0.65$)\\
& & MC mean & -3.3139 & -4.1708 & 1.5445 & 0.3483 & 2.2909 & -1.3744 & 2.0529 & 0.7030 \\
& & MC Sd &0.3360 & 0.4545 & 0.1580 & 0.1413 & 0.2700 & 0.5082 & 0.7388 & 0.0238 \\
& & IM SE & 0.4719 & 0.5176 & 0.2080 & 0.1396 & 0.2762 & 0.7304 & 0.8737 & 0.0238\\ \addlinespace[0.1cm]
& \multirow{4}{*}{2} & True & ($2$) & ($2$) & ($1.3798$) & ($0.3101$) & ($1.8449$) & ($-3$) & ($4$)\\
& & MC mean & 2.2651 & 2.5165 & 1.3373 & 0.2533 & 1.6079 & -2.3831 & 3.1945  \\
& & MC Sd & 0.5131 & 0.5107 & 0.2661 & 0.1682 & 0.2207 & 1.5120 & 1.4642\\
& & IM SE & 0.3213 & 0.3134 & 0.1929 & 0.1202 & 0.1726 & 1.4296 & 1.4575\\ \midrule
\multicolumn{11}{c}{$n = 1000$} \\
\midrule
\multirow{8}{*}{$5\%$} & \multirow{4}{*}{1} & True & ($-3$) & ($-4$) & ($1.7121$) & ($0.2620$) & ($2.1051$) & ($-2$) & ($2$) &($0.65$)\\
& & MC mean & -3.1812 & -4.1450 & 1.6219 & 0.2804 & 2.2407 & -1.6988 & 2.1169 & 0.6594\\
& & MC Sd & 0.1876 & 0.2248 & 0.0911 & 0.0615 & 0.1354 & 0.3611 & 0.4228 & 0.0165\\
& & IM SE & 0.1988 & 0.2351 & 0.0962 & 0.0607 & 0.1369 & 0.3660 & 0.4177 & 0.0167\\ \addlinespace[0.1cm]
& \multirow{4}{*}{2} & True & ($2$) & ($2$) & ($1.3798$) & ($0.3101$) & ($1.8449$) & ($-3$) & ($4$)\\
& & MC mean & 2.0488 & 2.1190 & 1.3265 & 0.3071 & 1.7823 & -2.6601 & 3.5305 \\
& & MC Sd & 0.1698 & 0.2069 & 0.0912 & 0.0683 & 0.1091 & 0.6830 & 0.8175 \\
& & IM SE & 0.1938 & 0.2113 & 0.0994 & 0.0676 & 0.1266 & 0.7723 & 0.9274\\
\midrule
\multirow{8}{*}{$30\%$} & \multirow{4}{*}{1} & True &($-3$) & ($-4$) & ($1.7121$) & ($0.2620$) & ($2.1051$) & ($-2$) & ($2$) &($0.65$)\\
& & MC mean & -3.3168 & -4.2133 & 1.5402 & 0.3276 & 2.2758 & -1.3758 & 2.0334 & 0.7021 \\
& & MC Sd & 0.2029 & 0.2736 & 0.0978 & 0.0776 & 0.1758 & 0.3513 & 0.5272 & 0.0162\\
& & IM SE & 0.3071 & 0.3289 & 0.1374 & 0.0878 & 0.1840 & 0.4791 & 0.5681 & 0.0160 \\\addlinespace[0.1cm]
& \multirow{4}{*}{2} & True & ($2$) & ($2$) & ($1.3798$) & ($0.3101$) & ($1.8449$) & ($-3$) & ($4$)\\
& & MC mean & 2.3043 & 2.5243 & 1.3347 & 0.2208 & 1.5994 & -2.3618 & 3.0762 \\
& & MC Sd & 0.3216 & 0.2956 & 0.1899 & 0.1000 & 0.1513 & 1.0722 & 0.8913 \\
& & IM SE & 0.2144 & 0.2025 & 0.1295 & 0.0743 & 0.1172 & 0.8982 & 0.8670\\ 
\midrule
\multicolumn{11}{c}{$n = 2000$} \\
\midrule
\multirow{8}{*}{$5\%$} & \multirow{4}{*}{1} & True & ($-3$) & ($-4$) & ($1.7121$) & ($0.2620$) & ($2.1051$) & ($-2$) & ($2$) &($0.65$)\\
& & MC mean & -3.2080 & -4.1723 & 1.6098 & 0.2779 & 2.2581 & -1.6429 & 2.1188 & 0.6600 \\
& & MC Sd & 0.1577 & 0.1728 & 0.0678 & 0.0525 & 0.1010 & 0.2864 & 0.3225 & 0.0118 \\
& & IM SE & 0.1411 & 0.1681 & 0.0672 & 0.0429 & 0.0978 & 0.2515 & 0.2926 & 0.0118 \\ \addlinespace[0.1cm]
& \multirow{4}{*}{2} & True & ($2$) & ($2$) & ($1.3798$) & ($0.3101$) & ($1.8449$) & ($-3$) & ($4$)\\
& & MC mean & 2.0480 & 2.1143 & 1.3204 & 0.3059 & 1.7857 & -2.5958 & 3.4666 \\
& & MC Sd & 0.1283 & 0.1995 & 0.0642 & 0.0553 & 0.0845 & 0.5176 & 0.6986 \\
& & IM SE & 0.2882 & 0.3822 & 0.0739 & 0.0501 & 0.0888 & 0.6434 & 0.8032 \\
\midrule
\multirow{8}{*}{$30\%$} & \multirow{4}{*}{1} & True & ($-3$) & ($-4$) & ($1.7121$) & ($0.2620$) & ($2.1051$) & ($-2$) & ($2$) &($0.65$)\\
& & MC mean & -3.3559 & -4.2548 & 1.4775 & 0.3046 & 2.1378 & -1.3237 & 2.0660 & 0.7036 \\
& & MC Sd & 0.1785 & 0.2527 & 0.1310 & 0.0859 & 0.3274 & 0.2934 & 0.4413 & 0.0109\\
& & IM SE & 0.2138 & 0.2299 & 0.0921 & 0.0641 & 0.1320 & 0.3289 & 0.4023 & 0.0113\\\addlinespace[0.1cm]
& \multirow{4}{*}{2} & True & ($2$) & ($2$) & ($1.3798$) & ($0.3101$) & ($1.8449$) & ($-3$) & ($4$)\\
& & MC mean & 2.2991 & 2.5248 & 1.3155 & 0.2141 & 1.6032 & -2.2326 & 2.9887 \\
& & MC Sd & 0.2436 & 0.2913 & 0.1381 & 0.0824 & 0.1245 & 0.7816 & 0.7760 \\
& & IM SE & 0.1601 & 0.1569 & 0.0922 & 0.0542 & 0.0847 & 0.5911 & 0.5862\\
\bottomrule
\end{longtable}
\end{center}

\subsection{\textbf{Performance of the ML Estimates over missing data}}

In order to evaluate the performance of {\small FM-MSNC} model for dealing with partially incomplete data, a simulation study was conducted. Various ways of using models for imputation are described in \cite{little2019statistical},  among them, one of the most relevant is the missing completely at random (MCAR). We simulated several datasets considering mixtures with two components from model (\ref{eqG}) with two missing data proportion settings ({\small $5\%$} and {\small $20\%$}), taken in each mixture component, and different samples sizes {\small $n$ $\in (500, 700, 900)$}. For each combination, we generated {\small $500$} Monte Carlo (MC) samples. Summary statistics of the estimates across the {\small $500$} MC samples were computed, such as the mean estimate (MC mean), the empirical standard error (MC Sd), and the mean of the approximate standard errors of the estimates, obtained through the method described in Section \ref{sec SE} (IM SE). We consider small and different variances with the following parameter as in the simulation about asymptotic properties in \ref{subsec:Asy_pro}.

Table \ref{tab:results_simu2} shows the results for this simulation. The results obtained are similar to those of simulation \ref{simu.1} and the same conclusions can be drawn. Additionally, we note that {\small $\lambda$} estimates appear to be more strongly affected as we increase the proportion of missing data in the sample.

\begin{center}
\tiny
\begin{longtable}{ccccccccccc}
\caption{Simulated data: Performance of the ML Estimates over missing data. Parameter estimates based on 500 simulated samples. Monte Carlo (MC) mean, MC Sd are the respective
mean estimates and standard deviations. IM SE is the average value of the approximate standard error obtained through the
information-based method.}
\label{tab:results_simu2}\\
\toprule
\multirow{2}{*}{Missing} & \multirow{2}{*}{$j$} & \multirow{2}{*}{Measure} & \multicolumn{8}{c}{Parameter} \\ \cmidrule{4-11}
& & & $\mu_{j1}$ & $ \mu_{j2}$ & $\alpha_{j,11}$ & $\alpha_{j,12}$ & $\alpha_{j,22}$ & $\lambda_{j1}$ & $\lambda_{j2}$ & $\pi$ \\ \midrule
\endfirsthead
\multicolumn{11}{c}%
{\tablename\ \thetable\ -- \textit{Continued from previous page}} \\
\midrule
\multirow{2}{*}{Missing} & \multirow{2}{*}{j} & \multirow{2}{*}{Measure} & \multicolumn{8}{c}{Parameter} \\ \cmidrule{4-11}
& & & $\mu_{j1}$ & $ \mu_{j2}$ & $\alpha_{j,11}$ & $\alpha_{j,12}$ & $\alpha_{j,22}$ & $\lambda_{j1}$ & $\lambda_{j2}$ & $\pi$ \\ \midrule
\endhead
\hline \multicolumn{11}{r}{\textit{Continued on next page}} \\
\endfoot
\endlastfoot
\multicolumn{11}{c}{$n = 500$} \\
\midrule
\multirow{8}{*}{$5\%$} & \multirow{4}{*}{1} & True & ($-5$) & ($-4$) & ($1.7121$) & ($0.2620$) & ($2.1051$) & ($-2$) & ($3$) &($0.65$)\\
& & MC mean & -5.2150 & -3.3411 & 1.6844 & 0.4594 & 1.9442 & -1.0396 & 1.4896 & 0.6495\\
& & MC Sd &  0.5773 & 0.8843 & 0.1075 & 0.2005 & 0.2071 & 1.0020 & 1.5677 & 0.0219 \\
& & IM SE & 0.8618 & 0.9443 & 0.2666 & 0.1876 & 0.2603 & 0.9352 & 1.0664 & 0.0220  \\ \addlinespace[0.1cm]
& \multirow{4}{*}{2} & True & ($2$) & ($3$) & ($1.3798$) & ($0.3101$) & ($1.8449$) & ($-2$) & ($3$)\\
& & MC mean & 1.7699 & 3.4694 & 1.3709 & 0.4857 & 1.7308 & -0.9892 & 1.5876  \\
& & MC Sd & 0.5614 & 0.7703 & 0.1231 & 0.1904 & 0.1922 & 1.1875 & 1.6737\\
& & IM SE & 0.8408 & 0.9545 & 0.2759 & 0.2176 & 0.3247 & 1.2244 & 1.3864 \\
\midrule
\multirow{8}{*}{$20\%$} & \multirow{4}{*}{1} & True & ($-5$) & ($-4$) & ($1.7121$) & ($0.2620$) & ($2.1051$) & ($-2$) & ($3$) &($0.65$)\\
& & MC mean & -5.2797 & -3.2378 & 1.6751 & 0.4903 & 1.9116 & -0.8182 & 1.1958 & 0.6495 \\
& & MC Sd & 0.6079 & 0.9007 & 0.1120 & 0.1908 & 0.2215 & 0.9494 & 1.5060 & 0.0228\\
& & IM SE & 1.0618 & 1.1557 & 0.3267 & 0.2297 & 0.3216 & 1.1195 & 1.2620 & 0.0227\\ \addlinespace[0.1cm]
& \multirow{4}{*}{2} & True & ($2$) & ($3$) & ($1.3798$) & ($0.3101$) & ($1.8449$) & ($-2$) & ($3$)\\
& & MC mean &  1.6992 & 3.5252 & 1.3680 & 0.5125 & 1.7031 & -0.7323 & 1.3092\\
& & MC Sd & 0.5987 & 0.7648 & 0.1359 & 0.1924 & 0.2047 & 1.1693 & 1.5322\\
& & IM SE & 1.2860 & 1.4228 & 0.3530 & 0.2721 & 0.4106 & 1.6672 & 1.7740\\ \midrule
\multicolumn{11}{c}{$n = 700$} \\
\midrule
\multirow{8}{*}{$5\%$} & \multirow{4}{*}{1} & True & ($-5$) & ($-4$) & ($1.7121$) & ($0.2620$) & ($2.1051$) & ($-2$) & ($3$) &($0.65$)\\
& & MC mean & -5.1275 & -3.2608 & 1.6902 & 0.4589 & 1.9299 & -1.0984 & 1.4147 & 0.6493\\
& & MC Sd &  0.5118 & 0.9054 & 0.0931 & 0.2000 & 0.1867 & 0.9515 & 1.5788 & 0.0188\\
& & IM SE & 0.6699 & 0.7212 & 0.2189 & 0.1539 & 0.2158 & 0.7571 & 0.8440 & 0.0186\\ \addlinespace[0.1cm]
& \multirow{4}{*}{2} & True & ($2$) & ($3$) & ($1.3798$) & ($0.3101$) & ($1.8449$) & ($-2$) & ($3$)\\
& & MC mean & 1.8256 & 3.4568 & 1.3733 & 0.4711 & 1.7356 & -1.1133 & 1.6816 \\
& & MC Sd & 0.4959 & 0.7484 & 0.1053 & 0.1725 & 0.1741 & 1.0682 & 1.5869 \\
& & IM SE & 0.6973 & 0.8268 & 0.2228 & 0.1766 & 0.2621 & 1.0000 & 1.1773\\
\midrule
\multirow{8}{*}{$20\%$} & \multirow{4}{*}{1} & True &($-5$) & ($-4$) & ($1.7121$) & ($0.2620$) & ($2.1051$) & ($-2$) & ($3$) &($0.65$)\\
& & MC mean & -5.1948 & -3.2079 & 1.6799 & 0.4817 & 1.8994 & -0.9175 & 1.1938 & 0.6488 \\
& & MC Sd & 0.5467 & 0.8905 & 0.0937 & 0.1907 & 0.1882 & 0.9160 & 1.4792 & 0.0192\\
& & IM SE & 0.8447 & 0.8977 & 0.2686 & 0.1866 & 0.2680 & 0.9194 & 1.0006 & 0.0192 \\\addlinespace[0.1cm]
& \multirow{4}{*}{2} & True & ($2$) & ($3$) & ($1.3798$) & ($0.3101$) & ($1.8449$) & ($-2$) & ($3$)\\
& & MC mean & 1.7787 & 3.5349 & 1.3666 & 0.4987 & 1.7052 & -0.8822 & 1.3254 \\
& & MC Sd & 0.5295 & 0.7466 & 0.1172 & 0.1741 & 0.1832 & 1.0146 & 1.4570\\
& & IM SE & 0.9450 & 1.0692 & 0.2765 & 0.2118 & 0.3275 & 1.2557 & 1.3766\\ 
\midrule
\multicolumn{11}{c}{$n = 900$} \\
\midrule
\multirow{8}{*}{$5\%$} & \multirow{4}{*}{1} & True & ($-5$) & ($-4$) & ($1.7121$) & ($0.2620$) & ($2.1051$) & ($-2$) & ($3$) &($0.65$)\\
& & MC mean & -5.1377 & -3.4011 & 1.6847 & 0.4243 & 1.9552 & -1.1829 & 1.6348 & 0.6488\\
& & MC Sd & 0.4678 & 0.8574 & 0.0794 & 0.1869 & 0.1790 & 0.8858 & 1.4904 & 0.0169\\
& & IM SE & 0.4769 & 0.4857 & 0.1699 & 0.1183 & 0.1781 & 0.5865 & 0.6471 & 0.0163 \\ \addlinespace[0.1cm]
& \multirow{4}{*}{2} & True & ($2$) & ($3$) & ($1.3798$) & ($0.3101$) & ($1.8449$) & ($-2$) & ($3$)\\
& & MC mean & 1.8644 & 3.3518 & 1.3744 & 0.4342 & 1.7513 & -1.2648 & 1.9203 \\
& & MC Sd & 0.4578 & 0.6741 & 0.0891 & 0.1616 & 0.1540 & 0.9946 & 1.4206 \\
& & IM SE & 0.5043 & 0.5860 & 0.1656 & 0.1297 & 0.1999 & 0.7796 & 0.9222 \\
\midrule
\multirow{8}{*}{$20\%$} & \multirow{4}{*}{1} & True &($-5$) & ($-4$) & ($1.7121$) & ($0.2620$) & ($2.1051$) & ($-2$) & ($3$) &($0.65$)\\
& & MC mean & -5.2002 & -3.3434 & 1.6760 & 0.4495 & 1.9244 & -0.9997 & 1.4014 & 0.6485 \\
& & MC Sd & 0.5123 & 0.8387 & 0.0839 & 0.1816 & 0.1866 & 0.8815 & 1.4234 & 0.0175\\
& & IM SE & 0.6459 & 0.6892 & 0.2138 & 0.1502 & 0.2198 & 0.7351 & 0.8213 & 0.0168\\\addlinespace[0.1cm]
& \multirow{4}{*}{2} & True & ($2$) & ($3$) & ($1.3798$) & ($0.3101$) & ($1.8449$) & ($-2$) & ($3$)\\
& & MC mean & 1.7946 & 3.4393 & 1.3699 & 0.4709 & 1.7147 & -0.9902 & 1.5475 \\
& & MC Sd & 0.5079 & 0.6797 & 0.0986 & 0.1640 & 0.1650 & 1.0069 & 1.3840 \\
& & IM SE & 0.7262 & 0.7971 & 0.2178 & 0.1622 & 0.2598 & 1.0105 & 1.0992 \\
\bottomrule
\end{longtable}
\end{center}

To exemplify the predictive accuracies on the imputation of missing values, we compare the {\small FM-MSNC} with the traditional randomization-based mean imputation ({\small MI}) predictor \cite{little2019statistical}, known as a common heuristic by filling in a single value for each missing value with the observed sample mean of the associated attribute. As a measure of precision, we use the mean absolute error (MAE) and the mean absolute relative error (MARE). They are defined as
{\small 
\begin{eqnarray}
MAE = \frac{1}{m}\sum_{i = 1}^{n}\sum_{j = 1}^{g}|y_{ij} - \hat{y}_{ij}| \quad \mathrm{and} \quad MARE = \frac{1}{m}\sum_{i = 1}^{n}\sum_{j = 1}^{g}\left| \frac{y_{ij} - \hat{y}_{ij}}{y_{ij}}\right|,
\end{eqnarray}}
where {\small $m$} is the number of missing entries, {\small $y_{ij}$} is the actual value and {\small $\hat{y}_{ij}$} is the respective predictive value.
The MAE and MARE measures, for both {\small FM-MSCN} and {\small MI} method, are listed in Table \ref{tab:MAE_MARE}. We can see that the {\small FM-MSCN} predictor exhibits considerable promising accuracy in the prediction of missing values when compared with those of MI imputations for all cases.

\begin{table}[!htb]
\small
\begin{center}
\caption{Simulated data: Performance of the ML Estimates over missing data. Average prediction accuracies for the both imputation methods FM-MSNC and mean imputation (MI) with varying sample size, $n \in (500, 700, 900)$, and proportions of missing values $5\%$, $10\%$ and $20\%$.}
\label{tab:MAE_MARE}
\vspace{-0.3cm}
\begin{tabular}{cccccccc}
\toprule
\multirow{2}{*}{Imputation method} & \multirow{2}{*}{Missing rate($\%$)}   & \multicolumn{3}{c}{MAE} & \multicolumn{3}{c}{MARE}\\
\cmidrule(lr){3-5} \cmidrule(lr){6-8}
& & 500 & 700 & 900 & 500 & 700 & 900\\
\midrule
\multirow{3}{*}{FM-MSNC} & $5\%$ &1.9009 & 1.8556 & 1.8444 & 0.8642 & 0.8747 & 0.8252\\
& $10\%$ &  2.0681 & 2.0243 & 2.0025 & 0.9883 & 1.0274 & 0.9678 \\
& $20\%$ & 2.4074 & 2.3693 & 2.3605 & 1.2264 & 1.2386 & 1.2345 \\ \addlinespace[0.1cm]
\multirow{3}{*}{MI} & $5\%$ & 3.2550 & 3.0297 & 2.7708 & 1.6897 & 1.8542 & 1.5337 \\
& $10\%$ & 3.2594 & 3.0498 & 2.7717 & 1.7402 & 1.8841 & 1.5501 \\
& $20\%$ & 3.2567 & 3.0563 & 2.7733 & 1.9881 & 1.8875 & 1.6624 \\
\bottomrule
\end{tabular}
\end{center}
\end{table}

\subsection{Number of mixture components}

In this section, we compare the ability of some classic model selection criteria discussed in Subsection \ref{Mselect} to select the appropriate model. One may argue that an arbitrary multivariate density can always be approximated by a finite mixture of normal multivariate distributions, see \cite[Chapter~1]{Peel2000}, for example. Thus, an interesting comparison can be made if we consider a sample from a two-component {\small FM-MSNC\textbf{(2)}} and use some model choice criteria to compare this model with the {\small FM-MNC} and several components under different censoring levels. Here we consider {\small $100$} samples of size {\small $500$} from a two-component {\small FM-MSNC\textbf{(2)}} model with left censoring levels at {\small $5\%$, $10\%$} or {\small $20\%$}, and parameter values set at 
{\small 
\begin{eqnarray*}
0.65\,\, SN_{2}\left( \begin{bmatrix}
2 \\
2 \\
\end{bmatrix},
\begin{bmatrix}
1.5 & 0 \\
0 & 1.5 \\
\end{bmatrix}, \begin{bmatrix}
-5 \\
10 \\
\end{bmatrix}
\right) +
0.35\,\, SN_{2} \left( \begin{bmatrix}
-2 \\
-1 \\
\end{bmatrix},
\begin{bmatrix}
1.5 & 0 \\
0 & 1.5 \\
\end{bmatrix}, \begin{bmatrix}
-5 \\
10 \\
\end{bmatrix}
\right).
\end{eqnarray*}}

The results are presented in Table \ref{tab:cluster}, under different censoring levels, where it can be seen that all criteria favor the true model, that is, the {\small FM-MSNC\textbf{(2)}} model instead the {\small FM-MNC} model with two, three and four components, as expected. This is evidence that these measures are capable of detecting departures from normality. It is important to emphasize that the {\small FM-MNC} models with three and four components have {\small $17$} and {\small $23$} parameters respectively, while the {\small FM-MSNC\textbf{(2)}} model has {\small $11$} parameters.  

\begin{table}[!htb]
\small
\begin{center}
\caption{Simulated data: Number of mixture components. Percentage when the FM-MSNC model with two components is preferred over the other adjusted FM-MNC models.}
\label{tab:cluster}
\begin{tabular}{cc|ccc|ccc|ccc}
\toprule
& \multirow{1}{*}{Censoring} & \multicolumn{3}{c|}{$5\%$} & \multicolumn{3}{c|}{$10\%$} & \multicolumn{3}{c}{$20\%$}\\
\cmidrule{3-5}
\cmidrule{6-8}
\cmidrule{9-11}
& Group & 2 & 3 & 4 & 2 & 3 & 4 & 2 & 3 & 4\\
\midrule
\multirow{3}{*}{Criteria} & AIC & 100 & 100 & 100  & 95 & 100 & 100& 100 & 100 & 100\\
& BIC & 100 & 100 & 100 & 93 & 100 & 100 & 100 & 100 & 100\\
& EDC & 100 & 100 & 100  & 93 & 100 & 100 & 100 & 100 & 100\\
\bottomrule
\end{tabular}
\end{center}
\end{table}

\begin{figure}[!h]
\centering
\includegraphics[width = 0.62\textwidth]{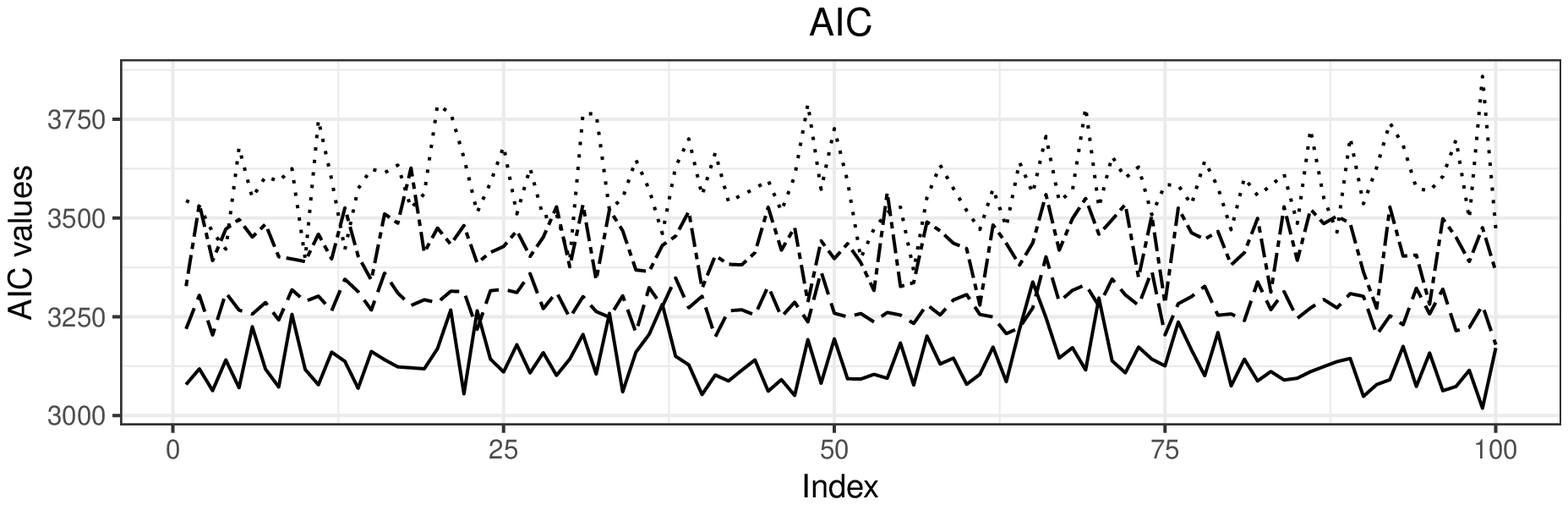}
\vskip8pt
\includegraphics[width = 0.62\textwidth]{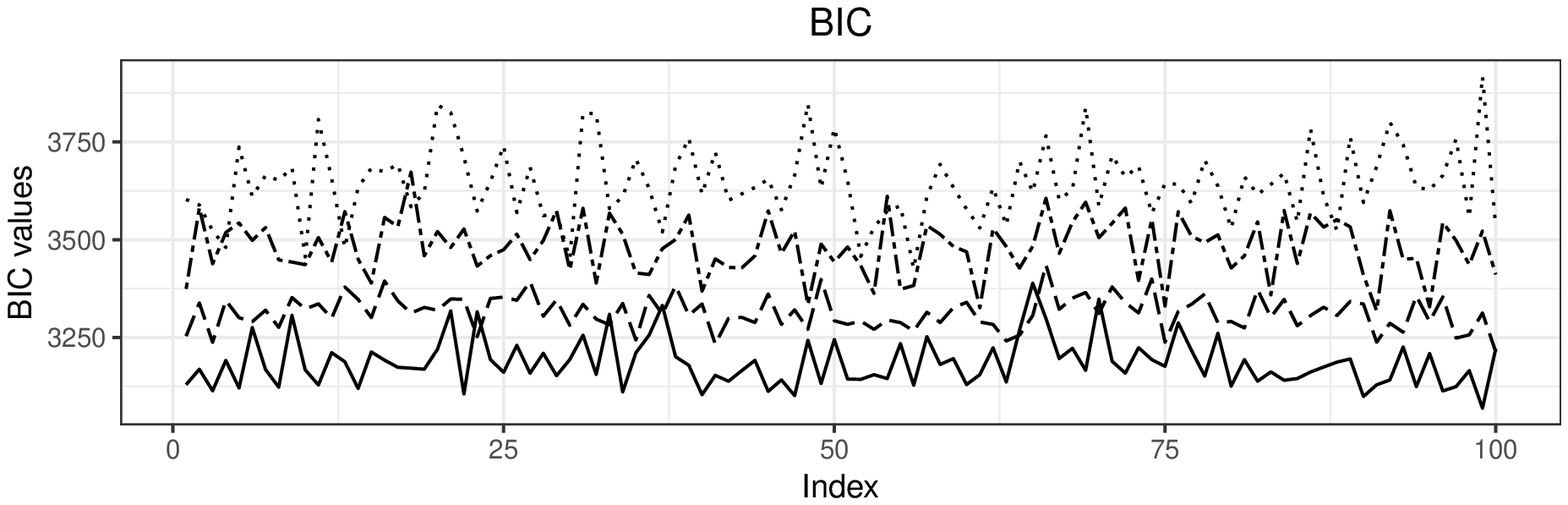}
\vskip8pt
\includegraphics[width = 0.62\textwidth]{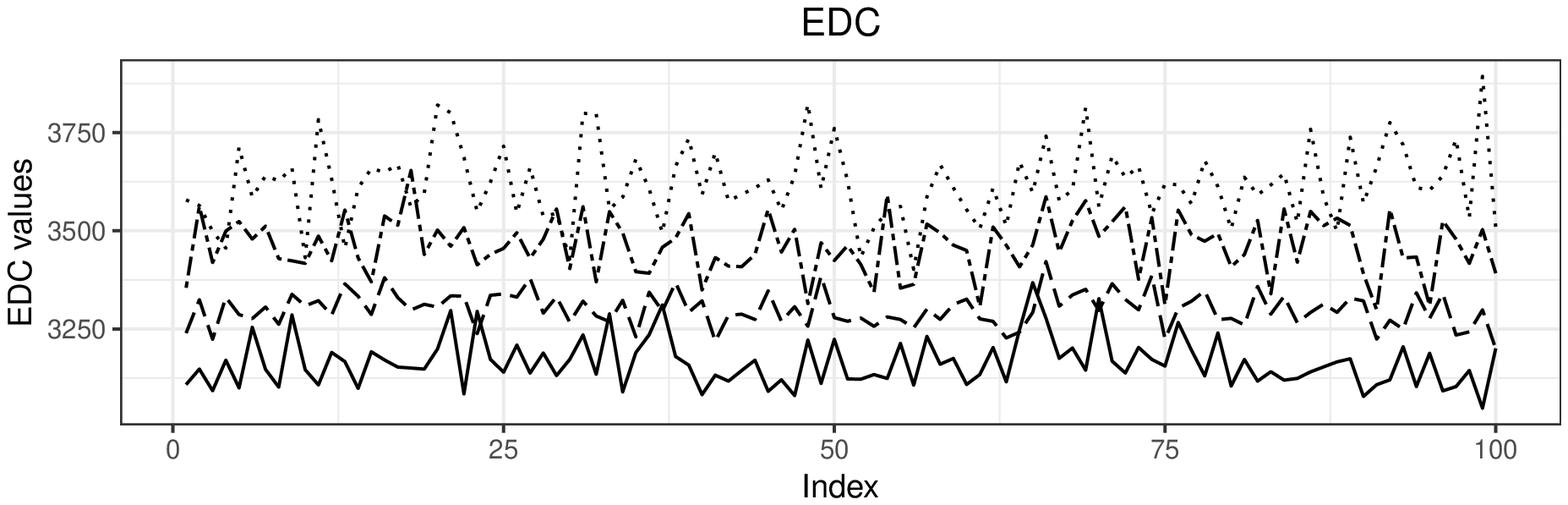}
\caption{\small{Simulated data: Number of mixture components. AIC, BIC and EDC values for 100 samples and left-censoring level $10\%$. Solid line: FM-MSNC, long dashed: FM-MNC(2), dot dashed: FM-MNC(3) and dotted: FM-MNC(4).}}
\label{fig:criteira_N}
\end{figure}

As pointed out for an anonymus referee, we can see in Table  \ref{tab:cluster} that the case with {\small $10\%$} censoring and two components is the only case where the correct model was not preferred by model selection criteria for a small number of instances. According to Figure \ref{fig:criteira_N}, the preferred model in these (atypical) cases was the {\small FM-MNC(2)}, however the differences in the criteria values related to a {\small FM-MSNC(2)} are close to zero. We believe that the amount of data sets generated in the simulation ({\small $100$} data sets) may not be sufficient and a more intensive simulation study would be required. However, due to the computational burden of the simulation, it would be too time consuming to move over to bigger simulations, say, {\small $1000$} data sets.

\begin{figure}[!h]
\centering
\center{
\subfigure[$0\%$]{\includegraphics[width = 0.24\textwidth]{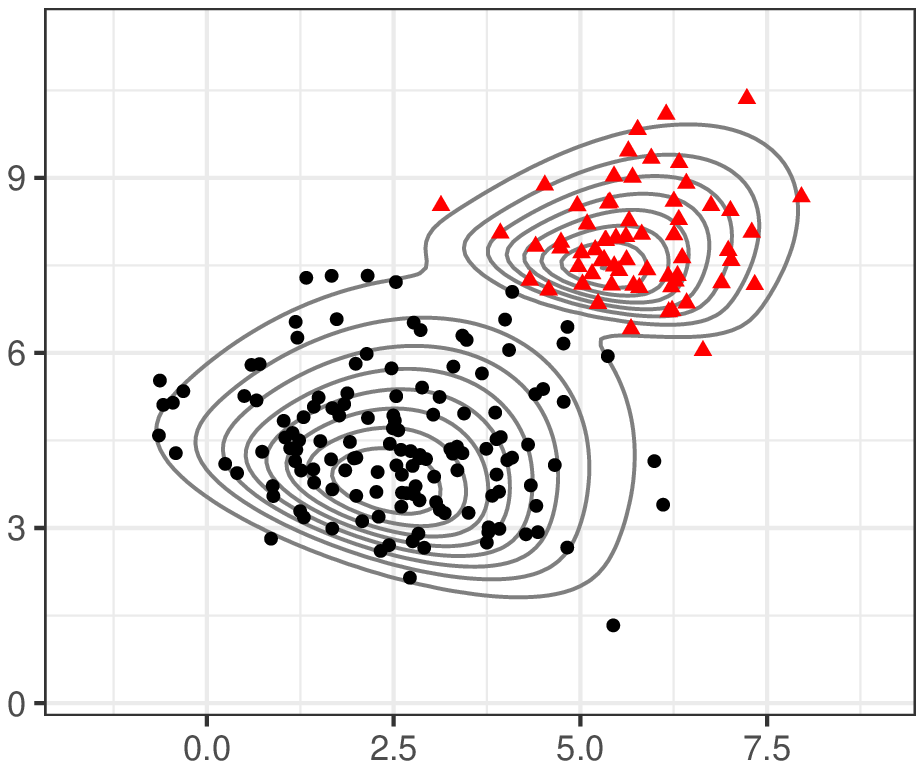}\label{fig:betadensity2}}
\subfigure[$5\%$]{\includegraphics[width = 0.24\textwidth]{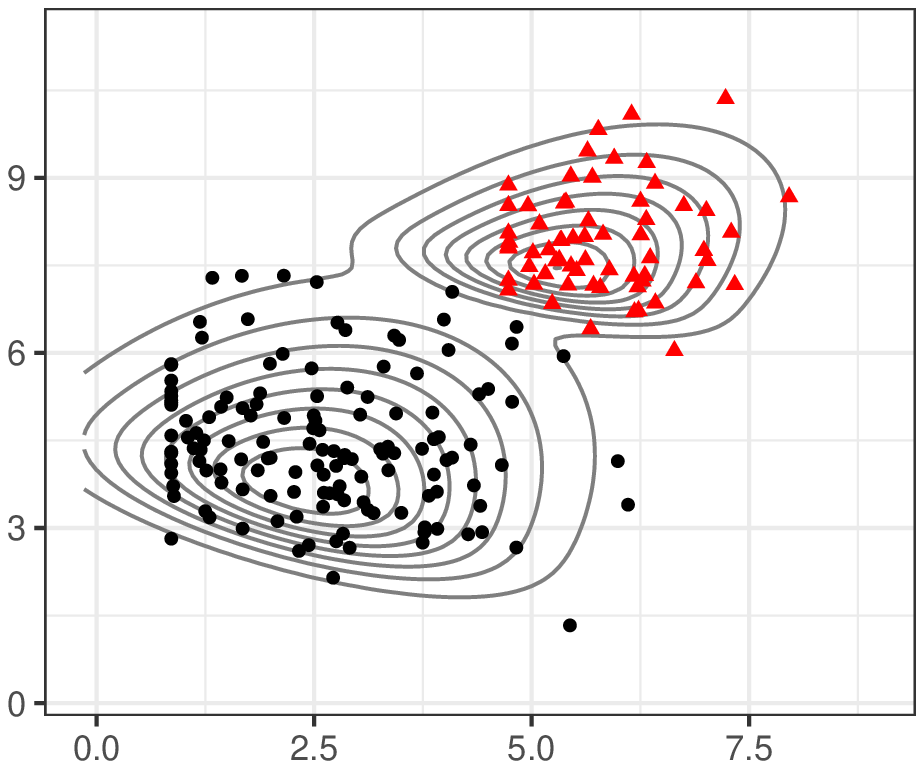}\label{fig:betadensity2}}
\subfigure[$10\%$]{\includegraphics[width = 0.24\textwidth]{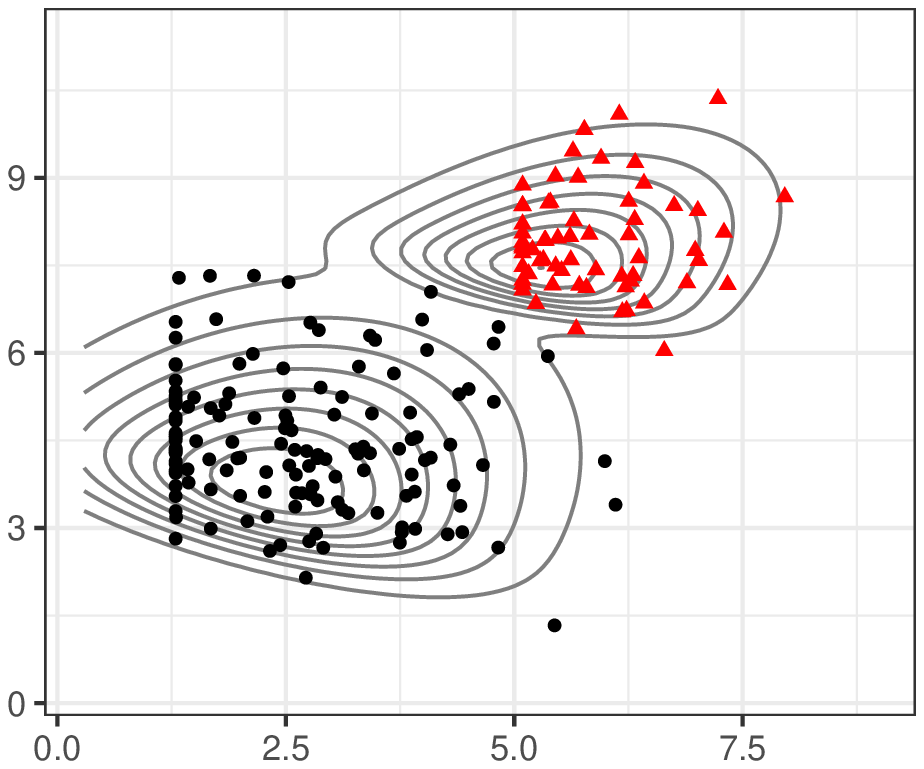}\label{fig:betadensity2}}
\subfigure[$20\%$]{\includegraphics[width = 0.24\textwidth]{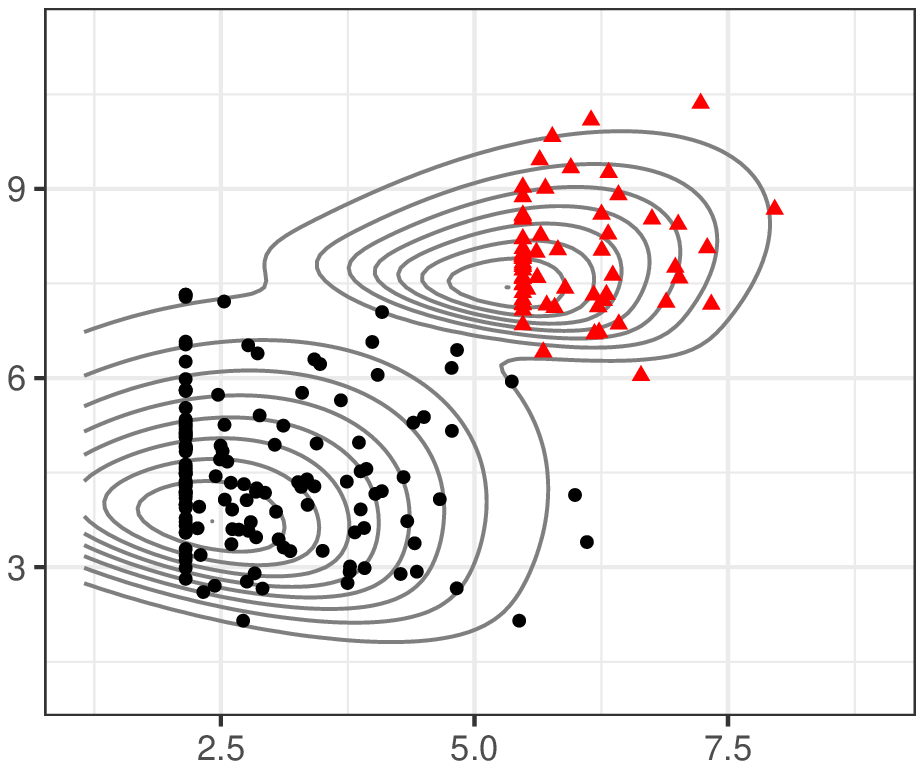}\label{fig:betadensity2}}\\
\subfigure[$0\%$]{\includegraphics[width = 0.24\textwidth]{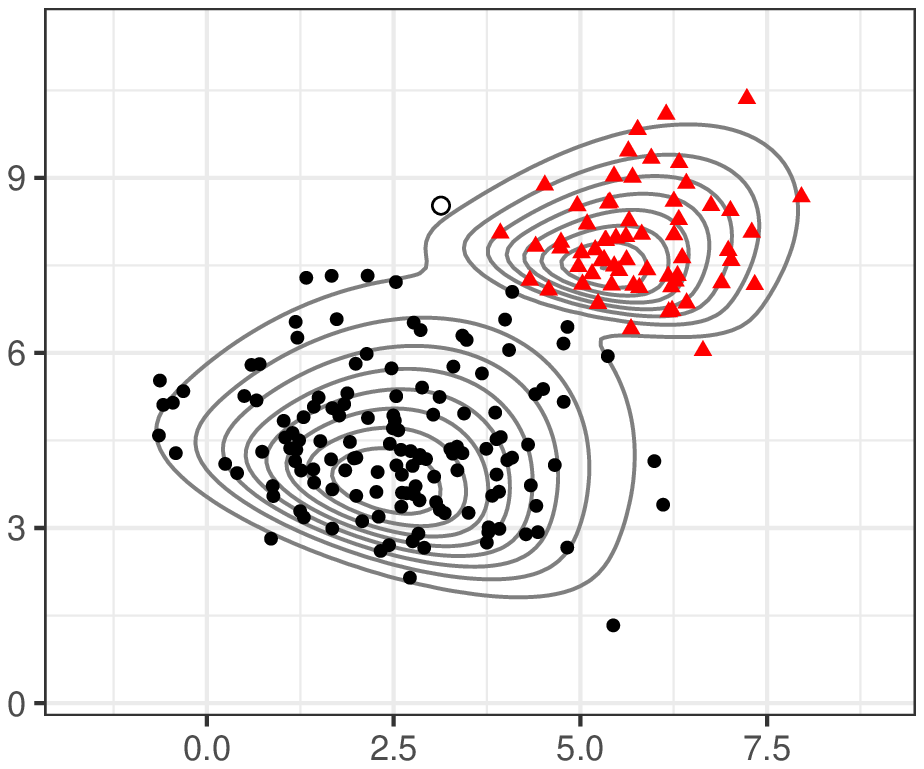}\label{fig:betadensity2}}
\subfigure[$5\%$]{\includegraphics[width = 0.24\textwidth]{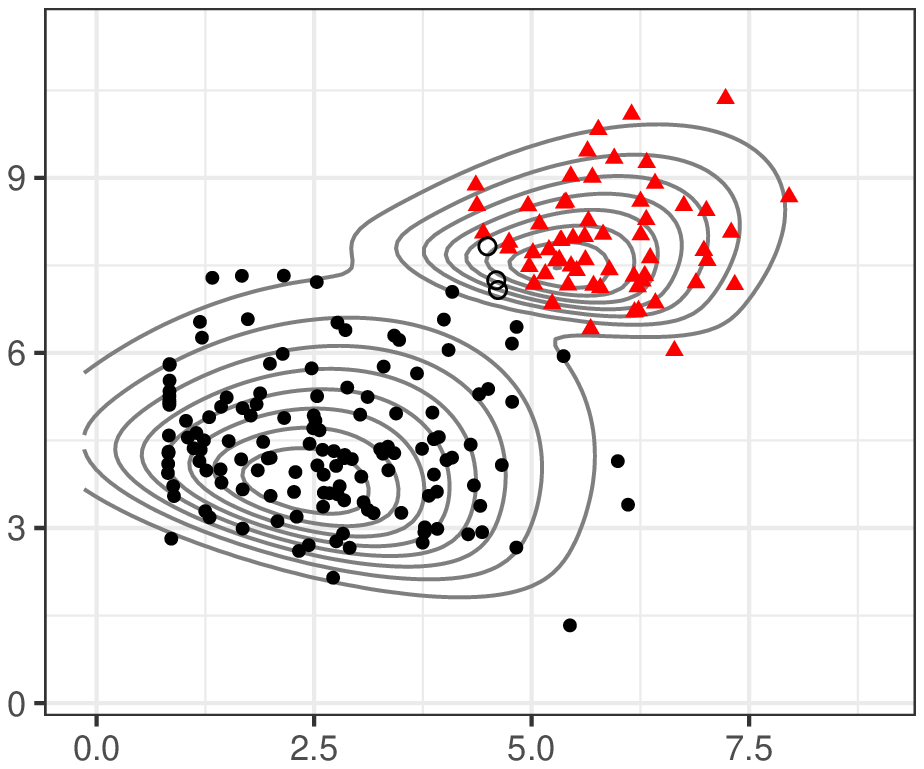}\label{fig:betadensity2}}
\subfigure[$10\%$]{\includegraphics[width = 0.24\textwidth]{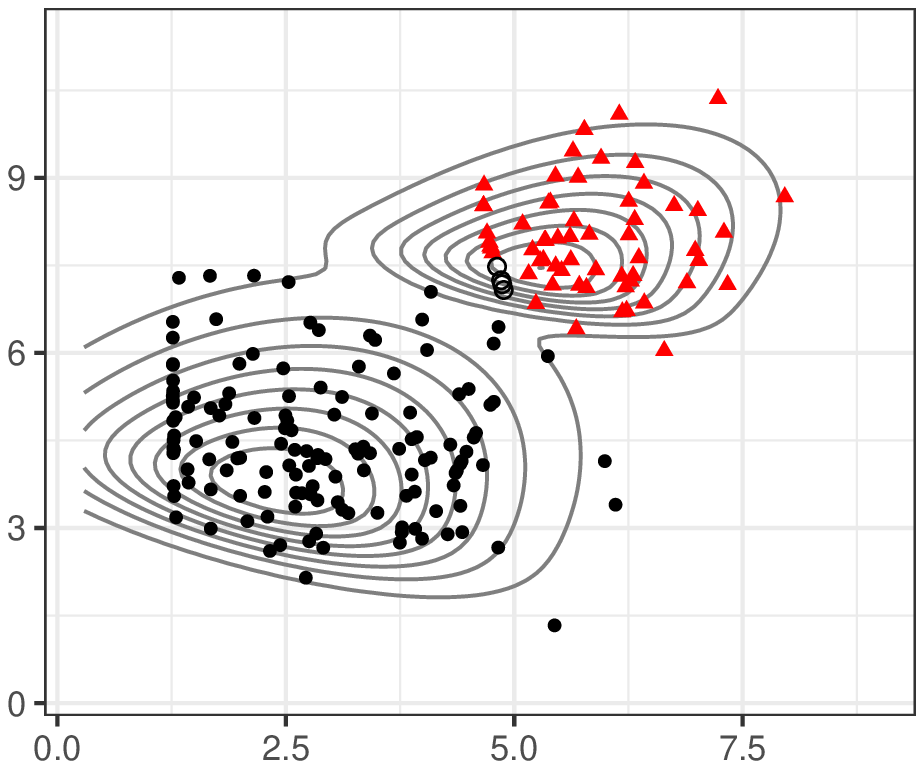}\label{fig:betadensity2}}
\subfigure[$20\%$]{\includegraphics[width = 0.24\textwidth]{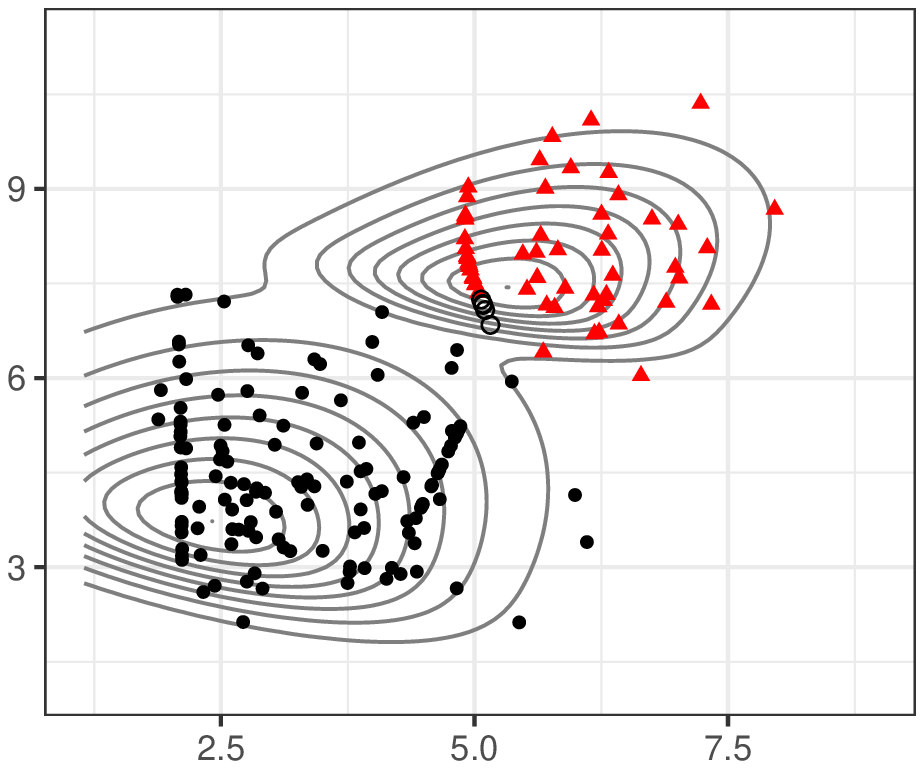}\label{fig:betadensity2}}\\
\subfigure[$0\%$]{\includegraphics[width = 0.24\textwidth]{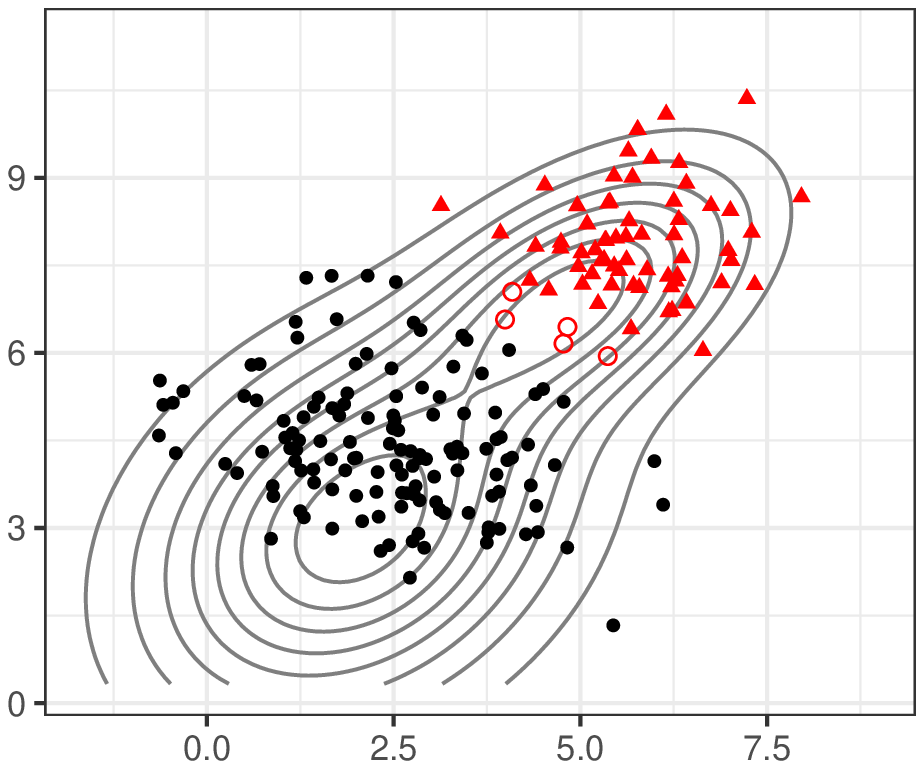}\label{fig:betadensity2}}
\subfigure[$5\%$]{\includegraphics[width = 0.24\textwidth]{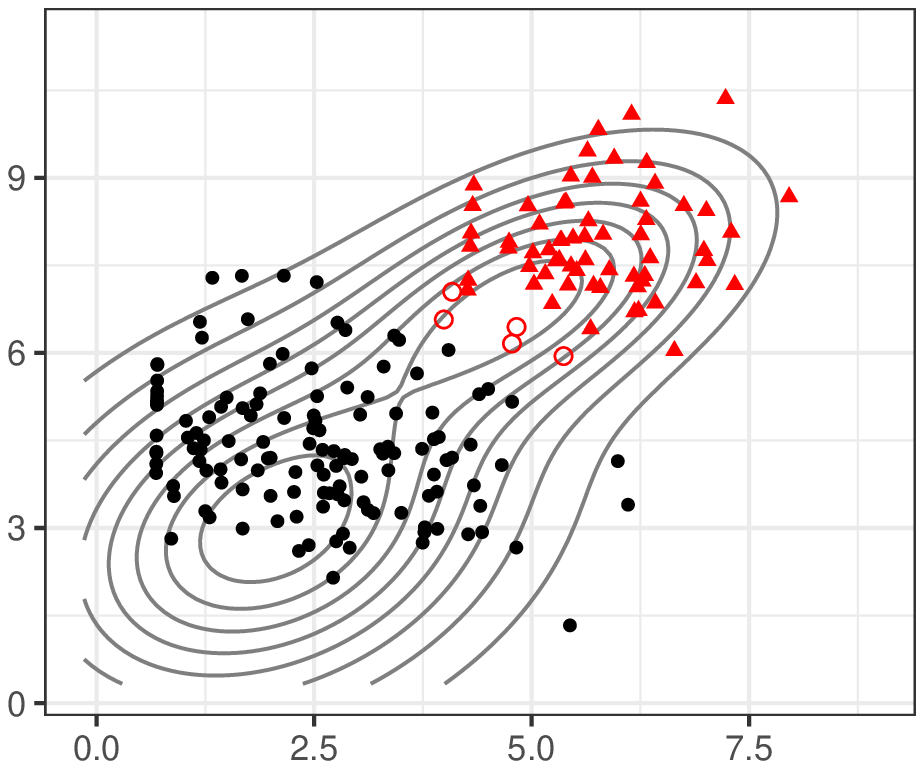}\label{fig:betadensity2}}
\subfigure[$10\%$]{\includegraphics[width = 0.24\textwidth]{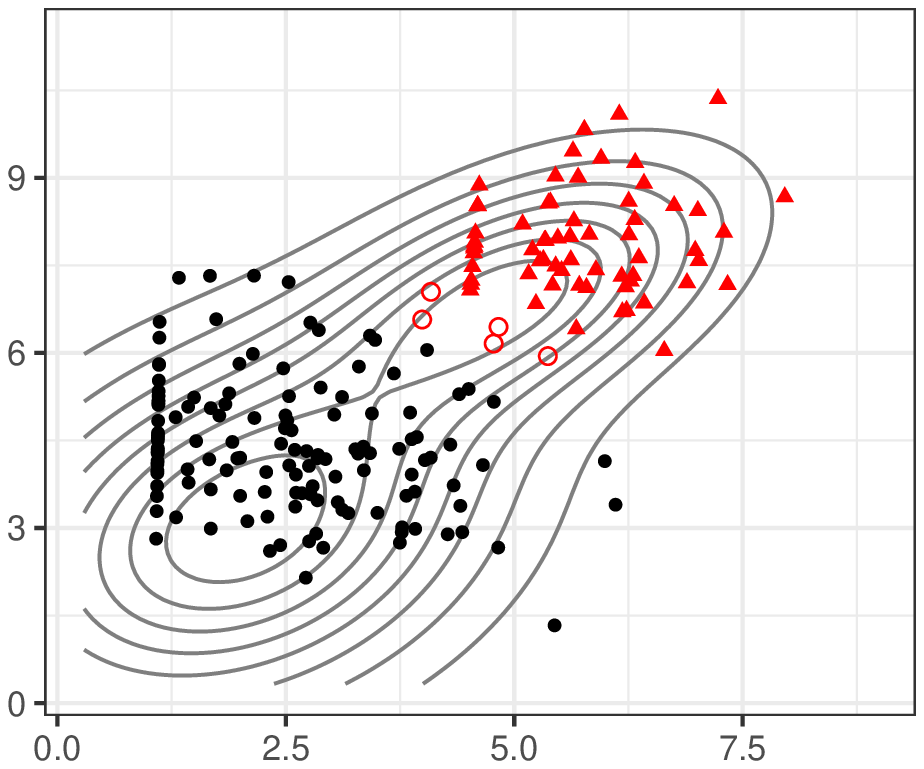}\label{fig:betadensity2}}
\subfigure[$20\%$]{\includegraphics[width = 0.24\textwidth]{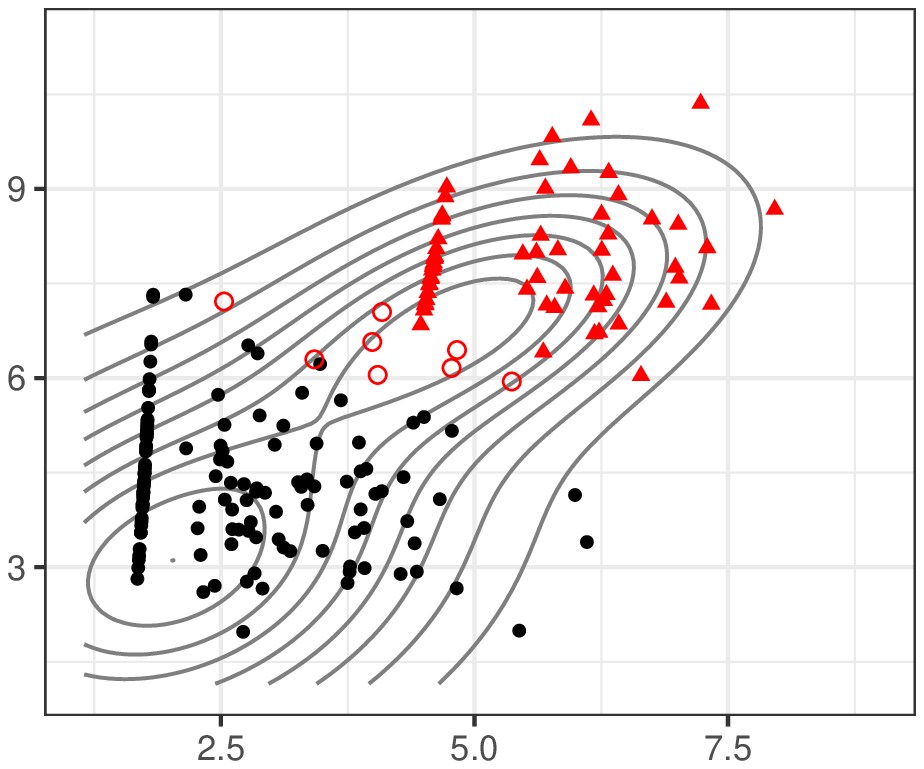}\label{fig:betadensity2}}}
\caption{\small{Simulated data: Clustering. Scatter plots for some simulated data from FM-MSNC model with $n = 200$ and the respective density contours (first line). Clustering scatter plots from fitted skew-normal (second line) and normal (last line) with multiples censoring level.}}
\label{fig:criteria}
\end{figure}

\subsection{Clustering}

Mixture models in general can be used for two main purposes:  1. estimation, and 2. model-based clustering \cite{mclachlan2000finite}. In this section, we investigate the ability of the {\small FM-MSNC} model to cluster observations, that is, to allocate them into groups of observations that are similar in some sense. We know that each data point belongs to {\small$g$} heterogeneous populations, but we don't know how to discriminate between them. Fitting the data with mixture models allows clustering the data in terms of the estimated posterior probability that a single point belongs to a given group. For this purpose, we follow the method proposed by \cite{zeller2016robust}, to assess the quality of the clustering of each mixture model using an index measure called correct classification rate (CCR), which is based on the posterior assigned to each subject. For the investigation of the clustering ability of the {\small FM-MSNC} model, we simulated {\small$500$} MC samples considering mixtures with two components from model (\ref{modelmixturenormal}), with sample size {\small$n \in (100, 200, 300)$}, without censoring and left-censoring proportion settings {\small$(5\%, 10\%, 20\%)$} taken in each mixture component, and parameter values set at
{\small
\begin{eqnarray*}
0.7\,\, SN_{2}\left( \begin{bmatrix}
2 \\
3 \\
\end{bmatrix},
\begin{bmatrix}
3 & 1 \\
1 & 4 \\
\end{bmatrix}, \begin{bmatrix}
2 \\
4 \\
\end{bmatrix}
\right) +
0.3\,\, SN_{2} \left( \begin{bmatrix}
5 \\
7 \\
\end{bmatrix},
\begin{bmatrix}
2 & 1 \\
1 & 2 \\
\end{bmatrix}, \begin{bmatrix}
3 \\
5 \\
\end{bmatrix}
\right).
\end{eqnarray*}}

To fit the data we used the models {\small FM-MSNC} and {\small FM-MNC} amd for each model we obtain the estimate of the posterior probability that an observation {\small$\yp_{i}$} belongs to the {\small$j$}th component of the mixture, {\small$\widehat{{Z}}_{ij}$}. So, if {\small $\max_j \widehat{{Z_{ij}}}$} occurs in component {\small$j$}, then {\small$\yp_{i}$} is classified into group {\small$j$}. For the {\small$m$}th sample of the MC, we computed the correct classification rate, denoted by CCRm, then obtained the average of the correct classification rate (ACCR) of CCRm.
Table \ref{tab:criteria} shows the ACCR values. From this table it is possible to observe that the model produces a high
correct classification rate in both fitted models. We see that the rate decreases when the censoring proportion increases, this decrease is stronger for {\small $n = 100$}.  
Looking at the {\small$n$} samples, keeping the censoring proportion fixed, the rate increased when the sample size increased.

\begin{table}[!htb]
\small
\begin{center}
\caption{Simulated data: Clustering. ACCR for fitted models FM-MSNC and FM-MNC for the simulated.}
\label{tab:criteria}
\begin{tabular}{cccccccccc}
\toprule
\multirow{2}{*}{n} & \multicolumn{4}{c}{FM-MSNC} & \multicolumn{5}{c}{FM-MNC}\\
\cmidrule{2-5} \cmidrule{7-10}
& $0\%$ & $5\%$ & $10\%$ & $20\%$ & & $0\%$ & $5\%$ & $10\%$ & $20\%$\\
100 & 0.9685 & 0.9619 & 0.9558 & 0.945 & & 0.8801 & 0.8809 & 0.8772 & 0.8735 \\
200 & 0.9729 & 0.9661 & 0.9599 & 0.9508 & & 0.9206 & 0.9191 & 0.916 & 0.9016\\
300 & 0.9733 & 0.9661 & 0.9608 & 0.9545 &  & 0.9248 & 0.9229 & 0.9233 & 0.9155 \\
\bottomrule
\end{tabular}
\end{center}
\end{table}

Figure \ref{fig:criteria} shows the allocations in each group for sample size {\small$n = 200$} and left-censoring proportion of {\small$0\%$,$5\%$,$10\%$} and {\small$20\%$}, where the groups are represented by black and red points. The first line of graphics (a - d) contains the scatter plot of the generated real data, where the black circles represent an observation erroneously classified as belonging to the black group. The second line of graphics (e - h) contains the scatter plot of the fitted {\small FM-MSNC} model, where the black circles represent an observation erroneously classified as belonging to the black group. 
The last line of graphics (i - l) contains the scatter plot of the fitted {\small FM-MNC} model, where the red circles represent an observation erroneously classified as belonging to the red group.


\subsection{Asymptotic properties}\label{subsec:Asy_pro}
In this simulation study, we analyze the absolute bias and the mean square error (mse) of the estimates obtained from the {\small FM-MSNC} model through the proposed EM algorithm. The idea of this simulation is to provide empirical evidence about the consistency of the ML estimates. These measures are defined by
{\small \begin{eqnarray}
bias(\theta_i) = \frac{1}{M}\sum_{m = 1}^{M} |\widehat{\theta}^{(m)}_i - \theta_i|\quad \mathrm{and} \quad mse(\theta_i) = \frac{1}{M}\sum_{m = 1}^{M} (\widehat{\theta}^{(m)}_i - \theta_i)^2,
\end{eqnarray}}
where {\small$M$} is the number of MC samples, and {\small$\widehat{\theta}^{(m)}_i$} is the estimated ML of the parameter {\small$\theta_i$} for the {\small$m$}th sample. Four different sample sizes {\small$(n = 300, 600, 900, 1200)$} are considered. For each sample size, we generated {\small$500$} Monte Carlo samples with {\small$5\%, 10\%, 15\%$} censoring proportion. Using the EM algorithm, the absolute bias and mean squared error for each parameter over the {\small$500$} datasets were computed. The parameter setup is as follows
{\small
\begin{eqnarray*}
0.65\,\, SN_{2}\left( \begin{bmatrix}
-5 \\
-4 \\
\end{bmatrix},
\begin{bmatrix}
3 & 1 \\
1 & 4.5 \\
\end{bmatrix}, \begin{bmatrix}
-2 \\
3 \\
\end{bmatrix}
\right) +
0.35\,\, SN_{2} \left( \begin{bmatrix}
2 \\
3 \\
\end{bmatrix},
\begin{bmatrix}
2 & 1 \\
1 & 3.5 \\
\end{bmatrix}, \begin{bmatrix}
-2 \\
3 \\
\end{bmatrix}
\right).
\end{eqnarray*}}

\begin{figure}[!htpb]
\centering
\center{\subfigure{\includegraphics[width = 3cm]{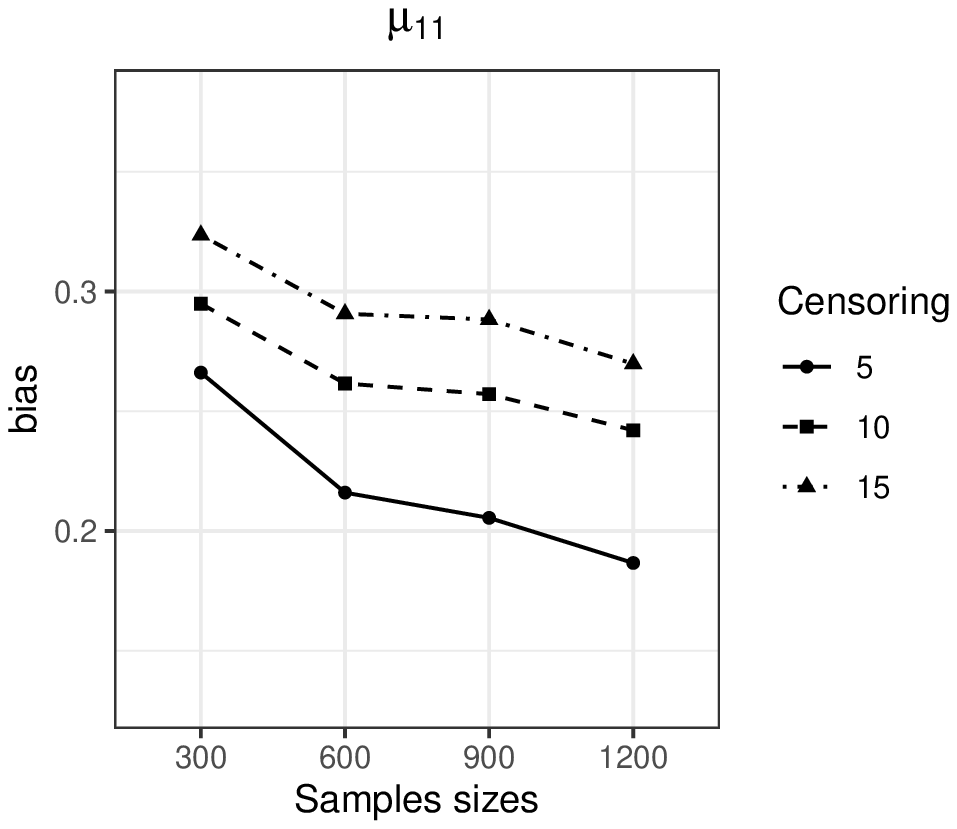}}
\subfigure{\includegraphics[width = 3cm]{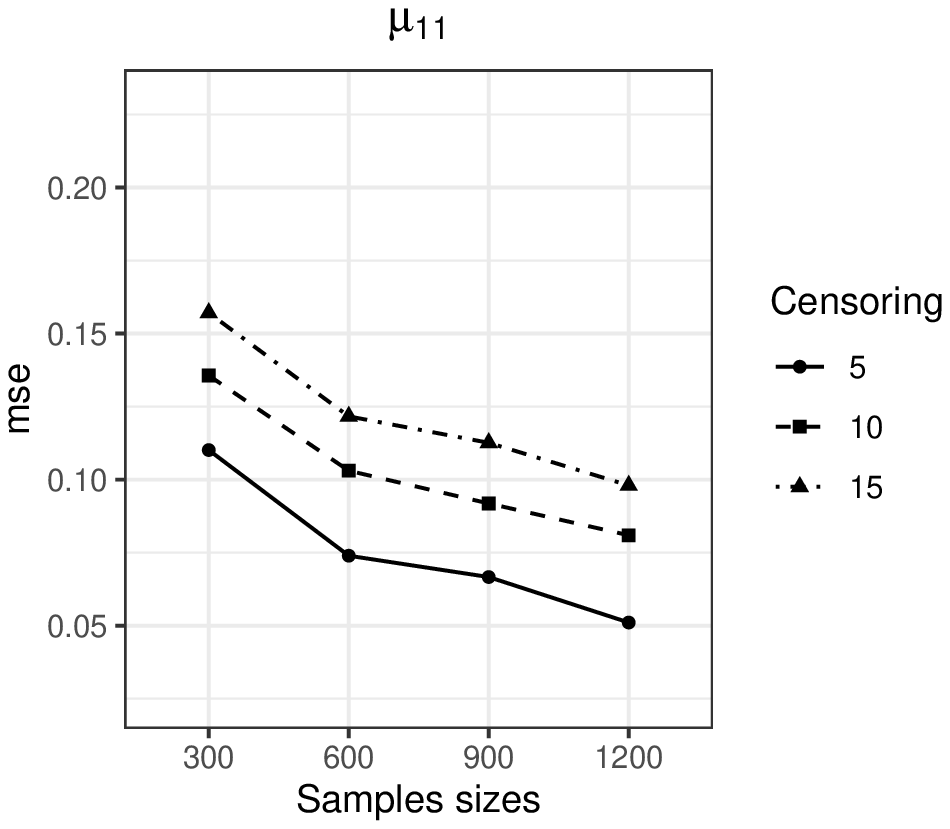}\label{fig:betadensity2}}
\subfigure{\includegraphics[width = 3cm]{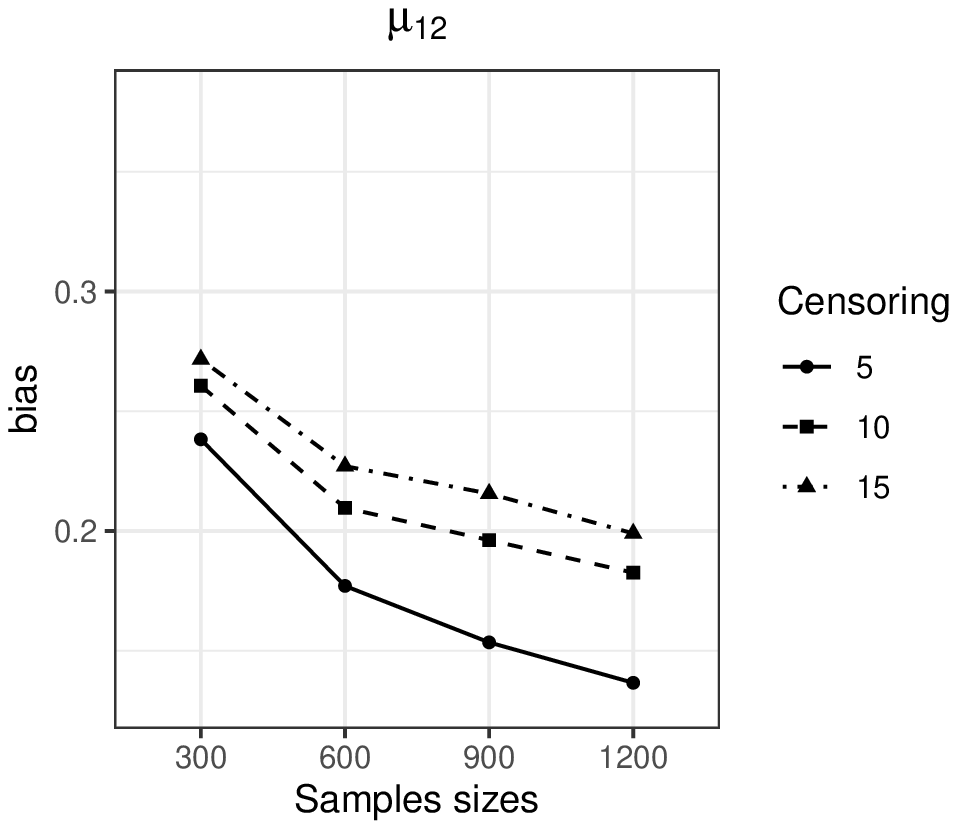}\label{fig:betadensity1}}
\subfigure{\includegraphics[width = 3cm]{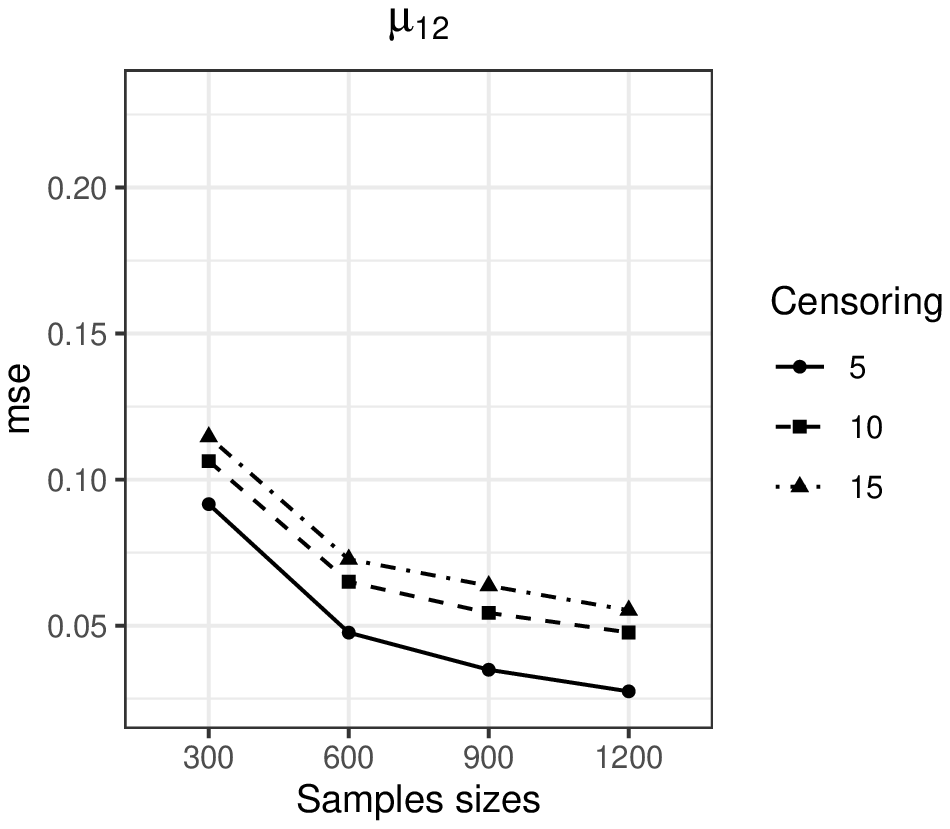}\label{fig:betadensity2}}
\subfigure{\includegraphics[width = 3cm]{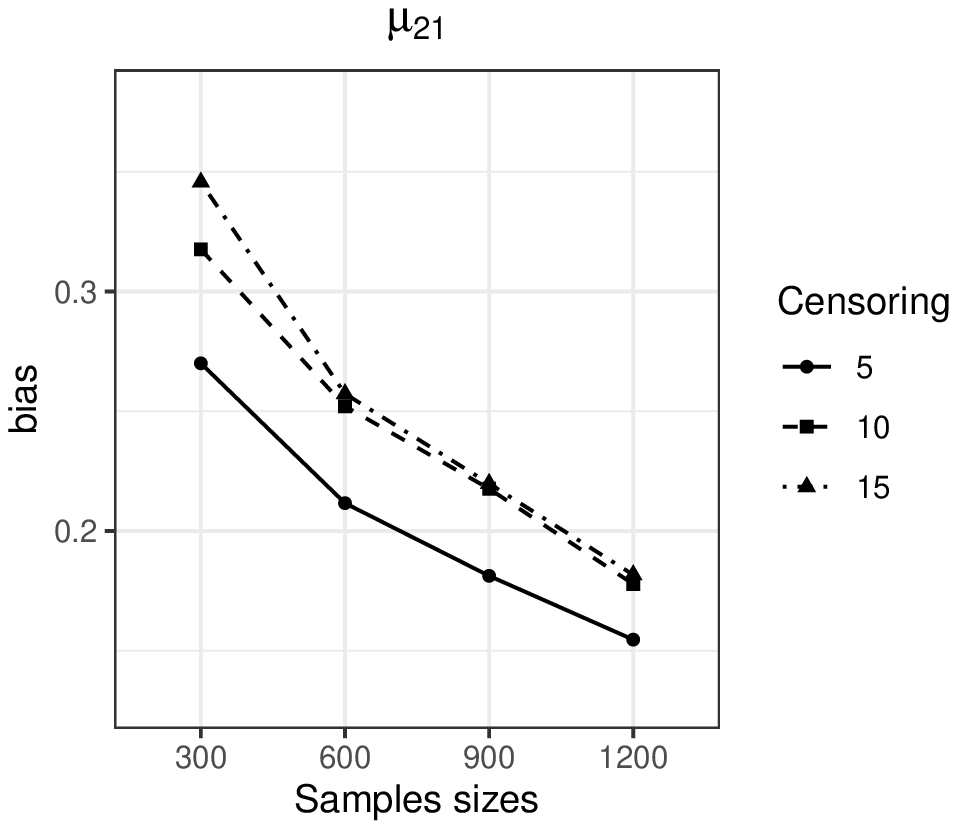}\label{fig:betadensity1}}
\subfigure{\includegraphics[width = 3cm]{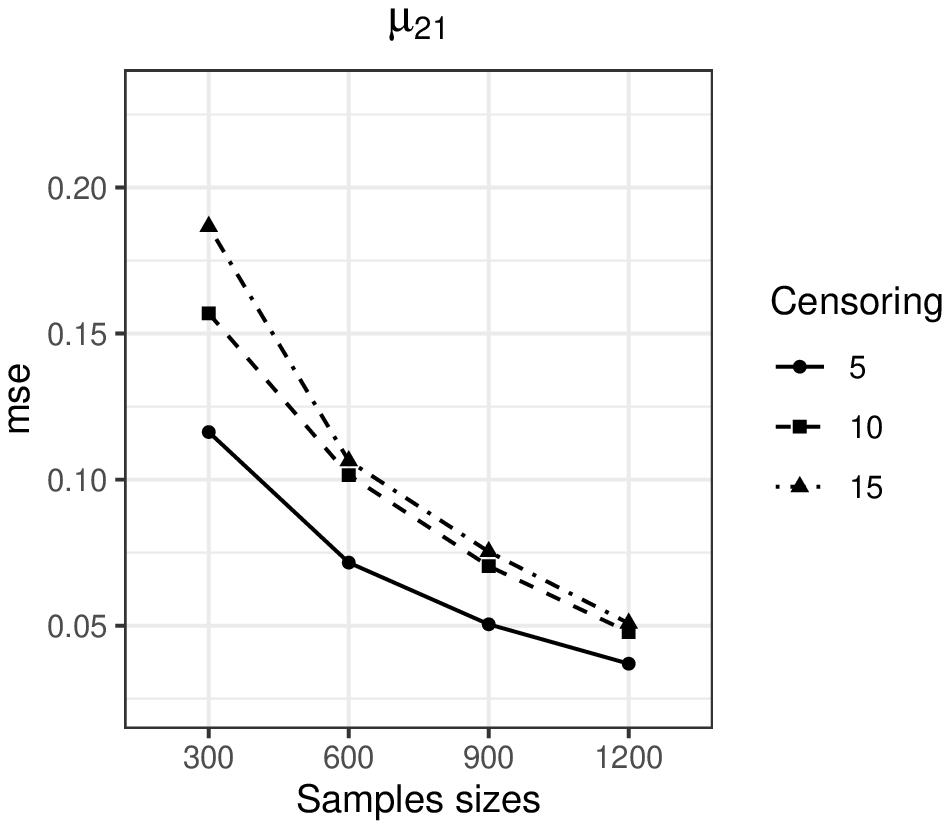}\label{fig:betadensity2}}
\subfigure{\includegraphics[width = 3cm]{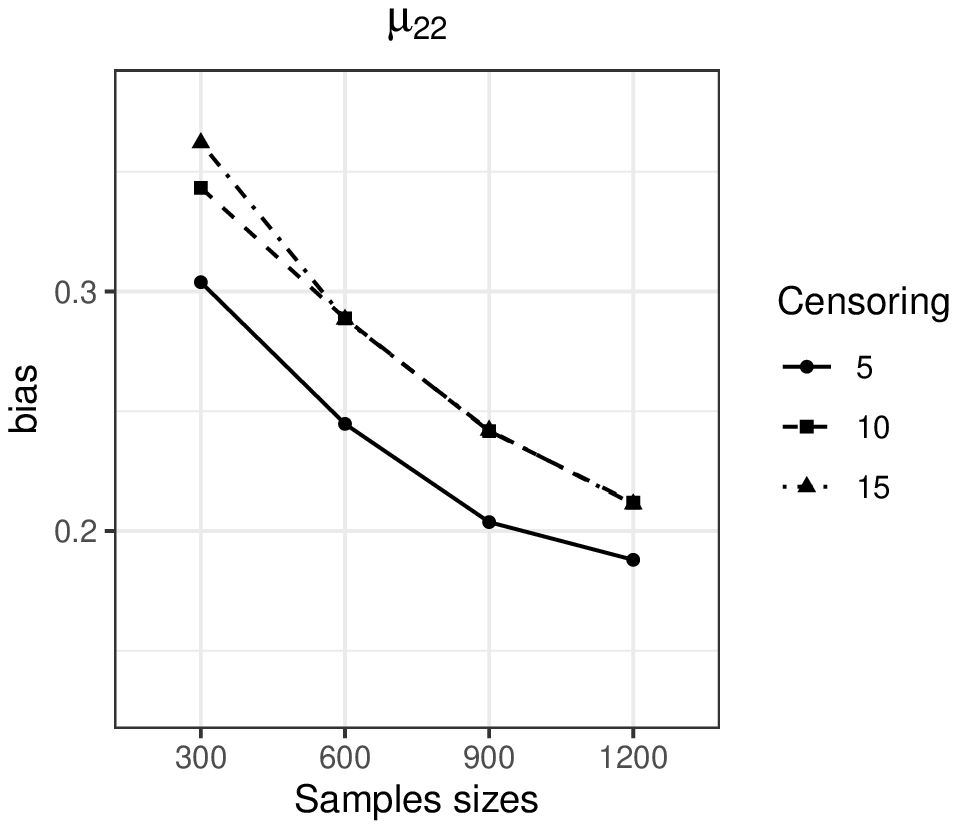}\label{fig:betadensity2}}
\subfigure{\includegraphics[width = 3cm]{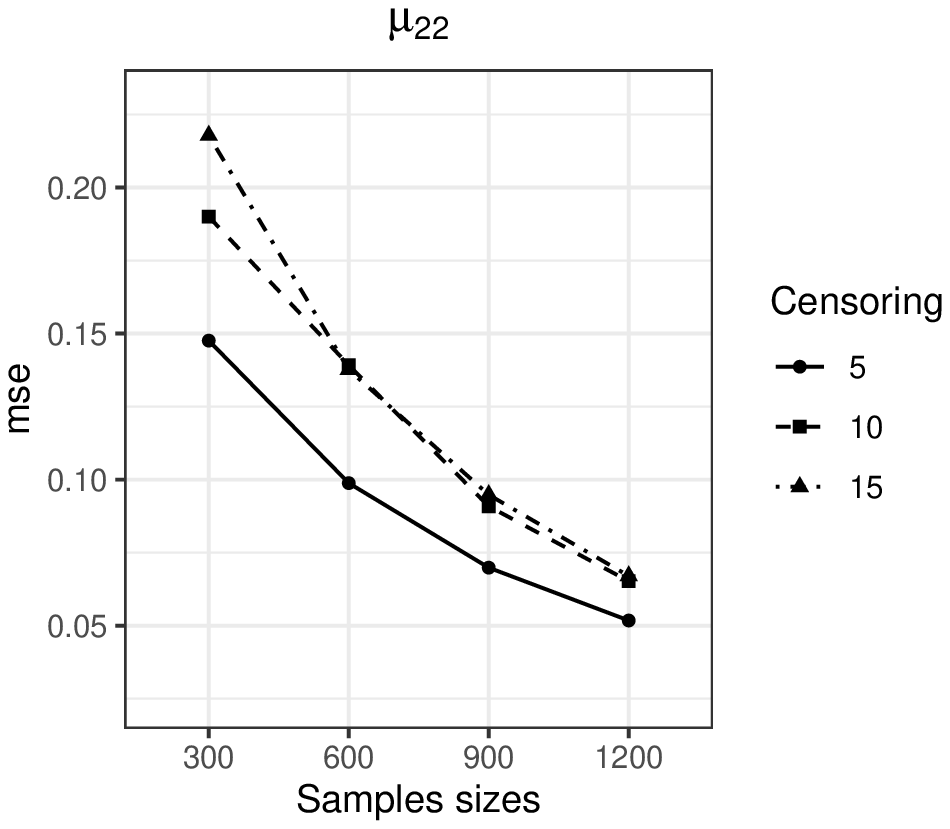}\label{fig:betadensity2}}}
\caption{\small{Simulated data: Asymptotic properties. bias and mse of $\bmu_1$ and $\bmu_2$ estimate in the FM-MSNC model with
different  censoring levels:  $5\%$ (solid line), $10\%$ (dashed line), $15\%$ (dot-dashed line).}}
\label{fig:assinto1}
\end{figure}

\begin{figure}[!htpb]
\centering
\center{\subfigure{\includegraphics[width = 3cm]{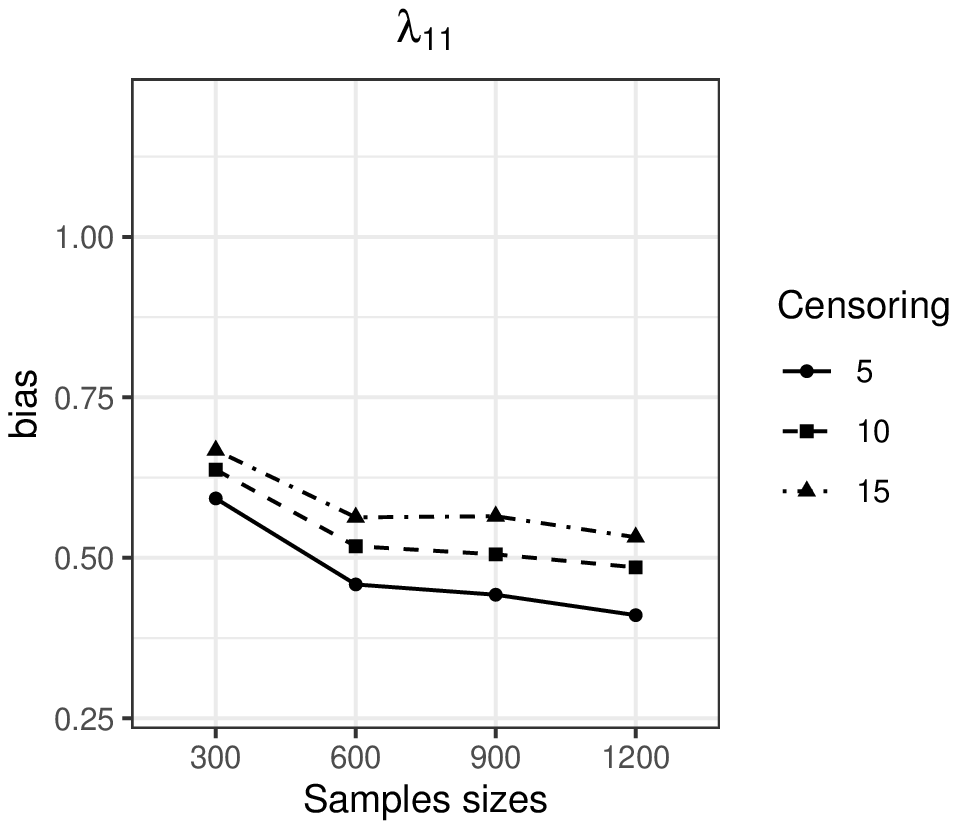}}
\subfigure{\includegraphics[width = 3cm]{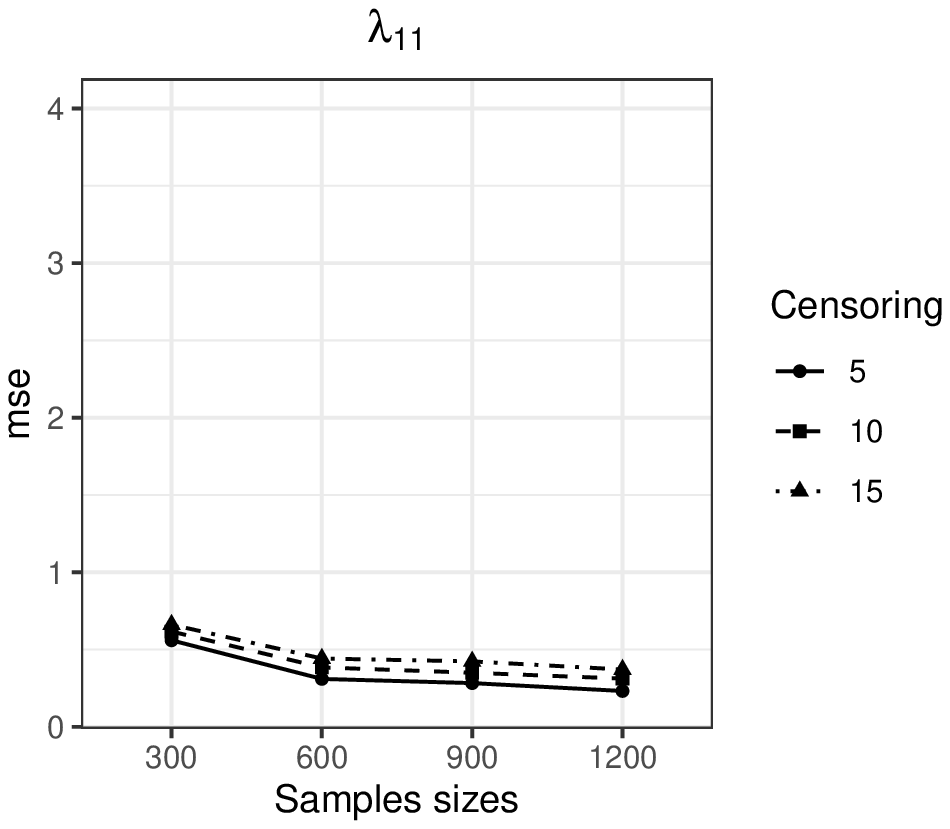}\label{fig:betadensity2}}
\subfigure{\includegraphics[width = 3cm]{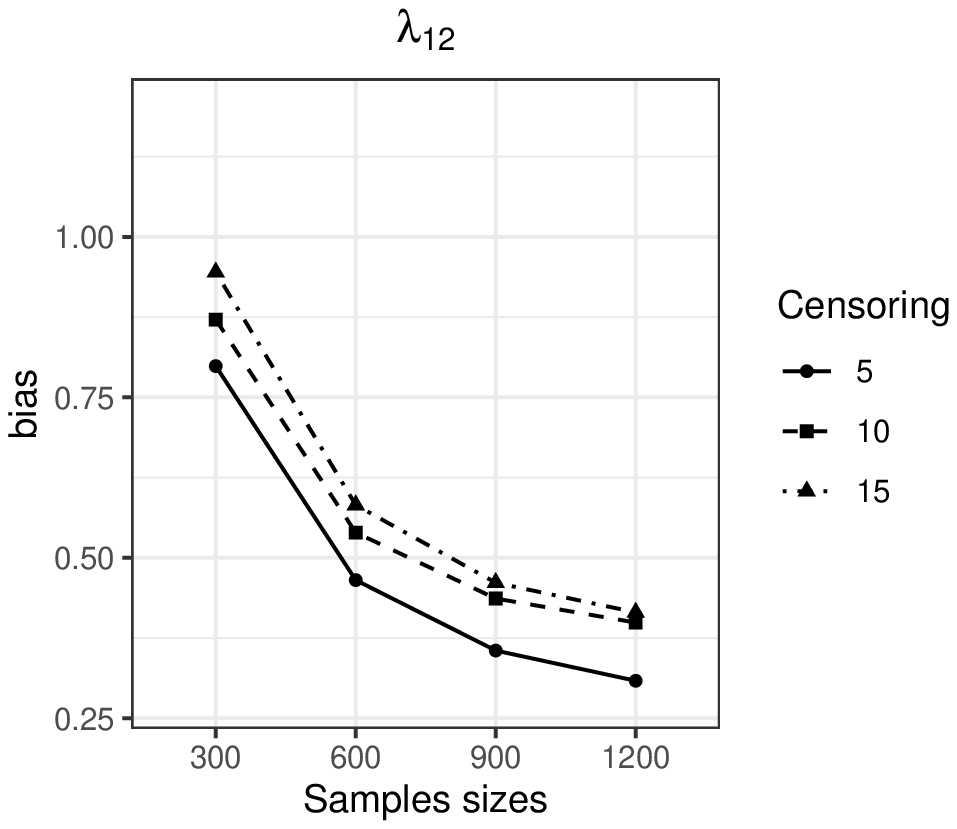}\label{fig:betadensity1}}
\subfigure{\includegraphics[width = 3cm]{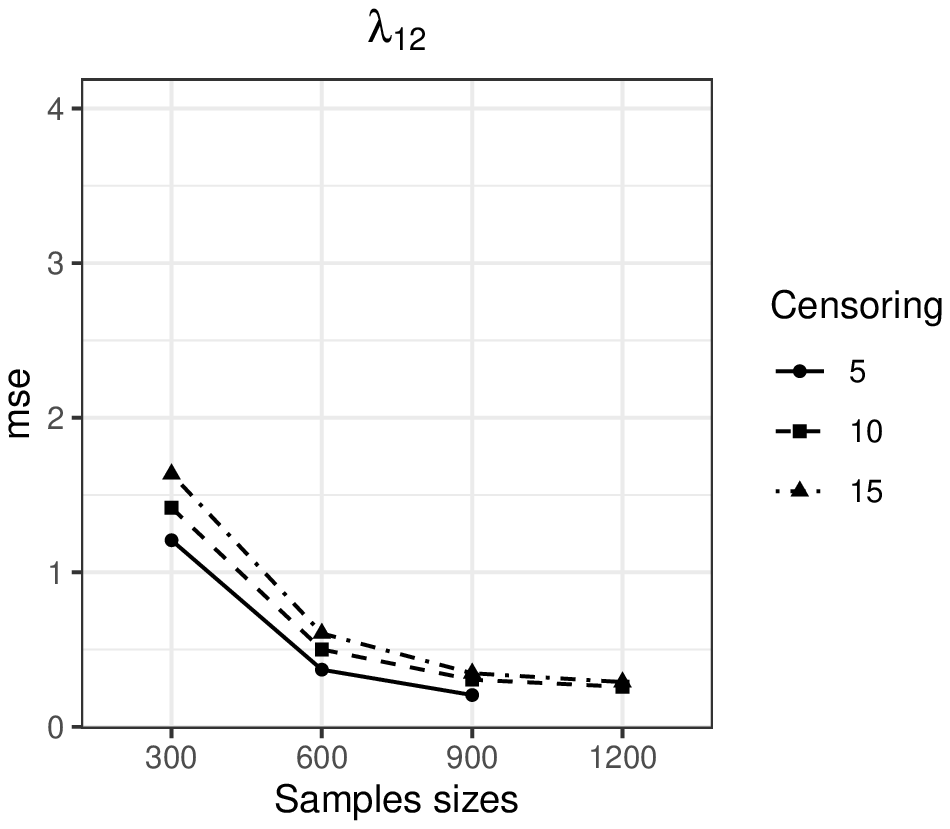}\label{fig:betadensity2}}
\subfigure{\includegraphics[width = 3cm]{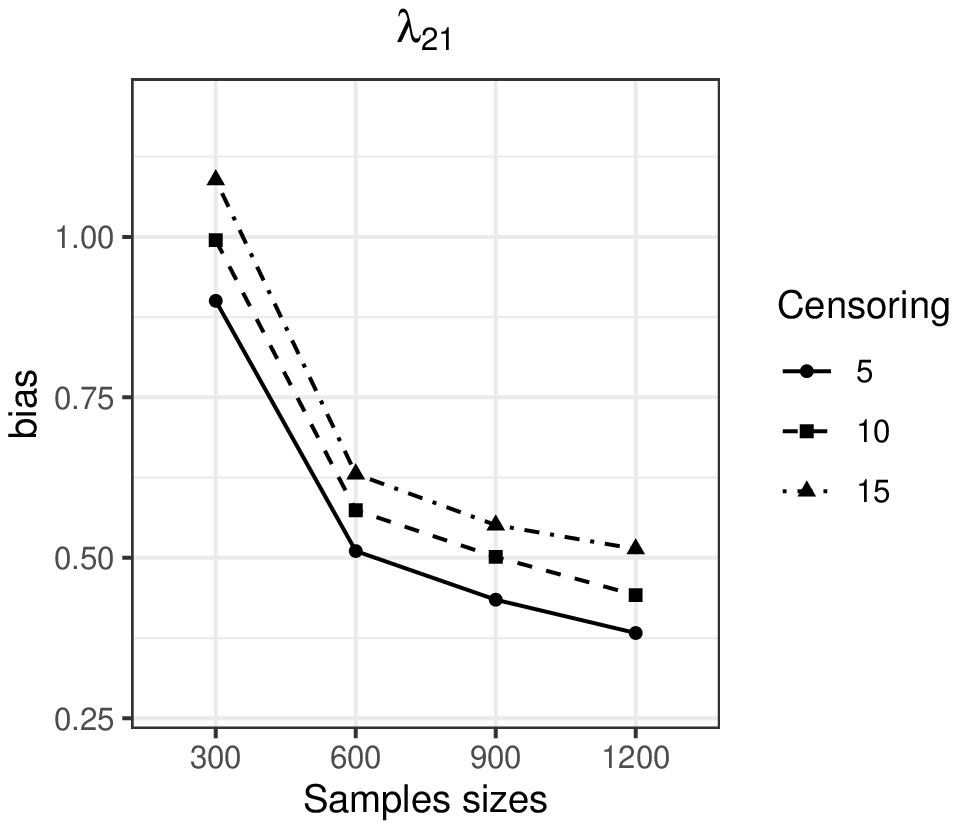}\label{fig:betadensity1}}
\subfigure{\includegraphics[width = 3cm]{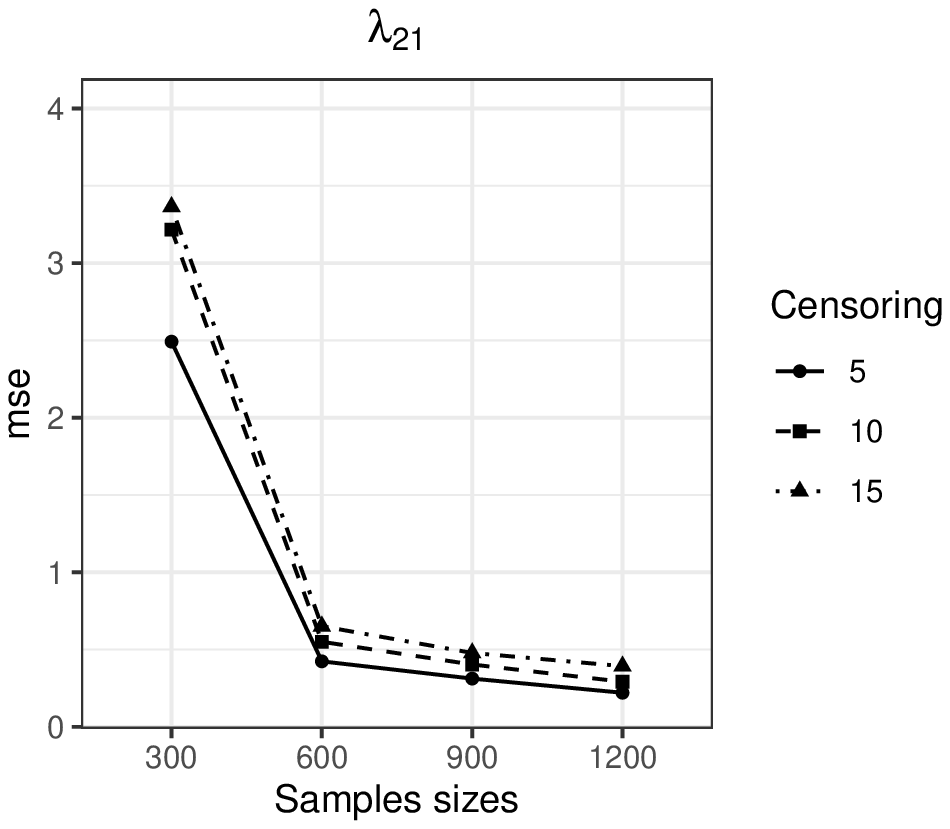}\label{fig:betadensity2}}
\subfigure{\includegraphics[width = 3cm]{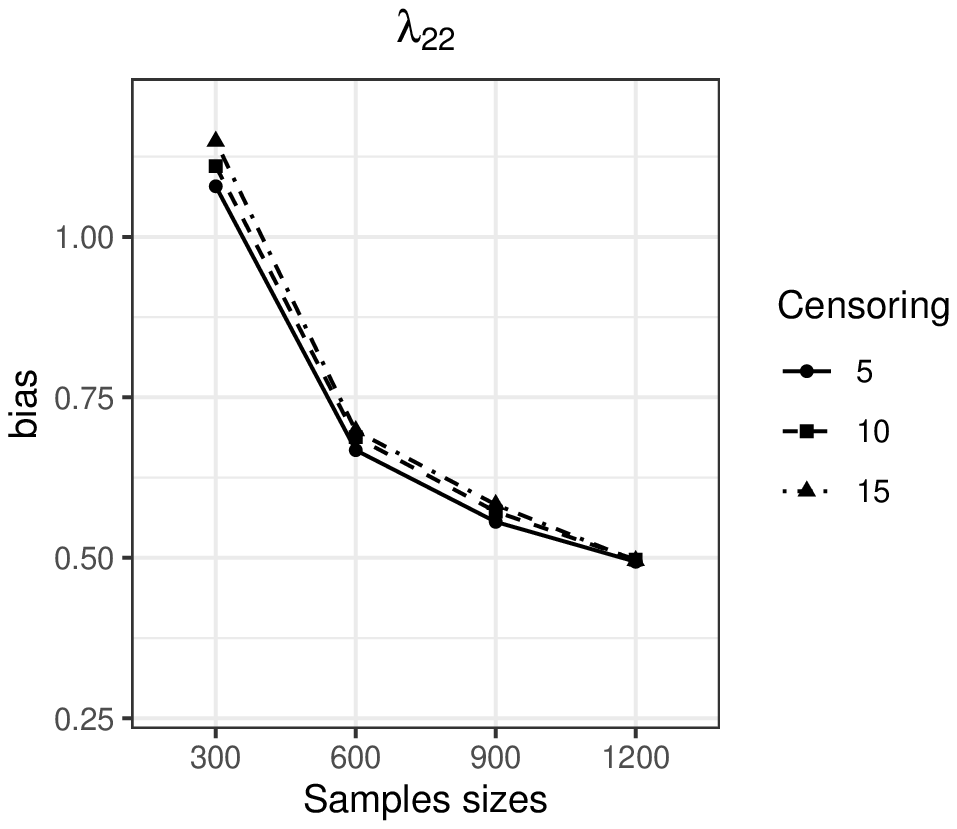}\label{fig:betadensity2}}
\subfigure{\includegraphics[width = 3cm]{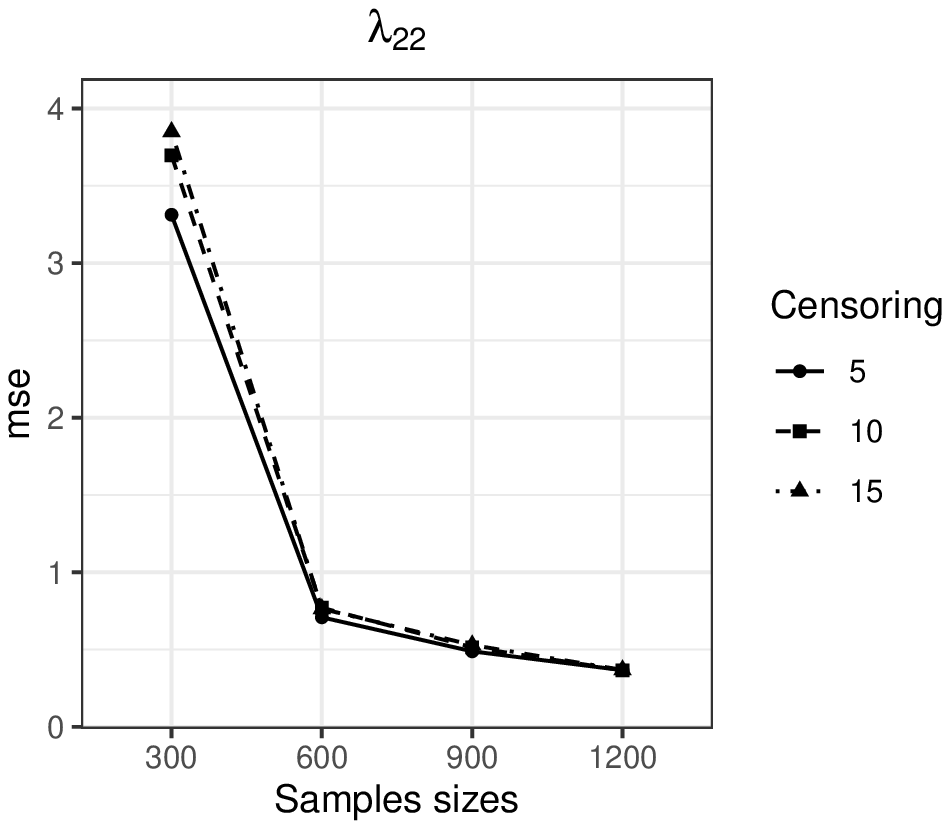}\label{fig:betadensity2}}
}
\caption{\small{Simulated data: Asymptotic properties. bias and mse of $\blambda_1$ and $\blambda_2$ estimate in the FM-MSNC model with
different censoring levels:  $5\%$ (solid line), $10\%$ (dashed line), $15\%$ (dot-dashed line).}}
\label{fig:assinto3}
\end{figure}

\begin{figure}[!t]
\centering
\center{\subfigure{\includegraphics[width = 3cm]{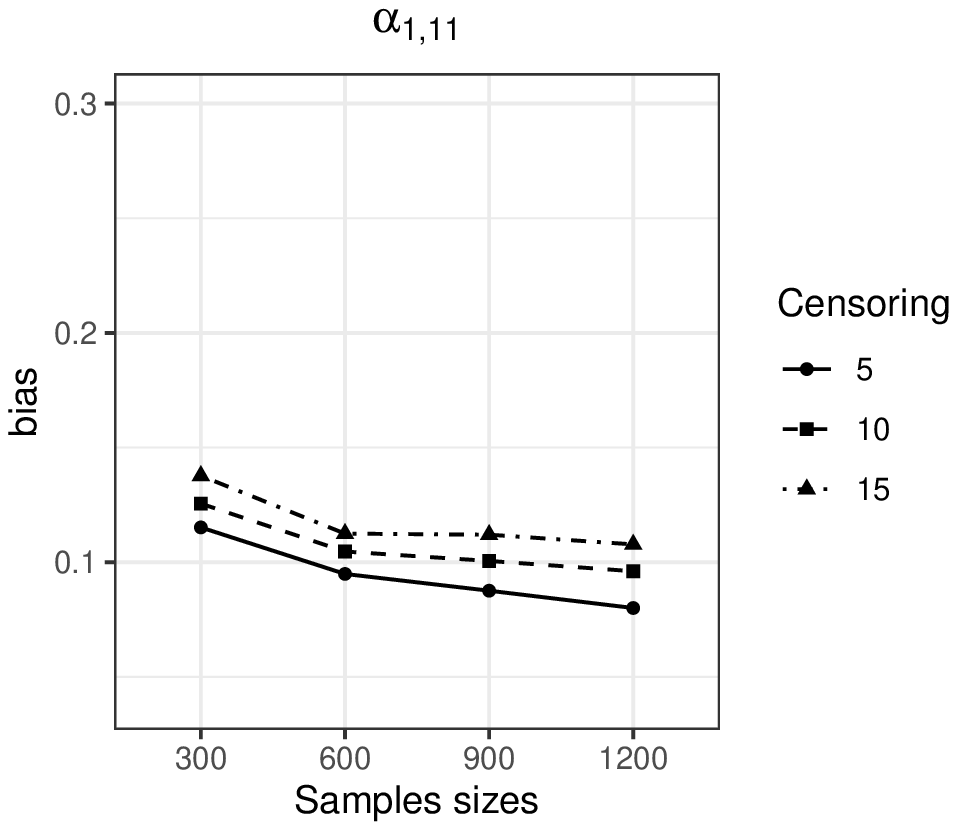}}
\subfigure{\includegraphics[width = 3cm]{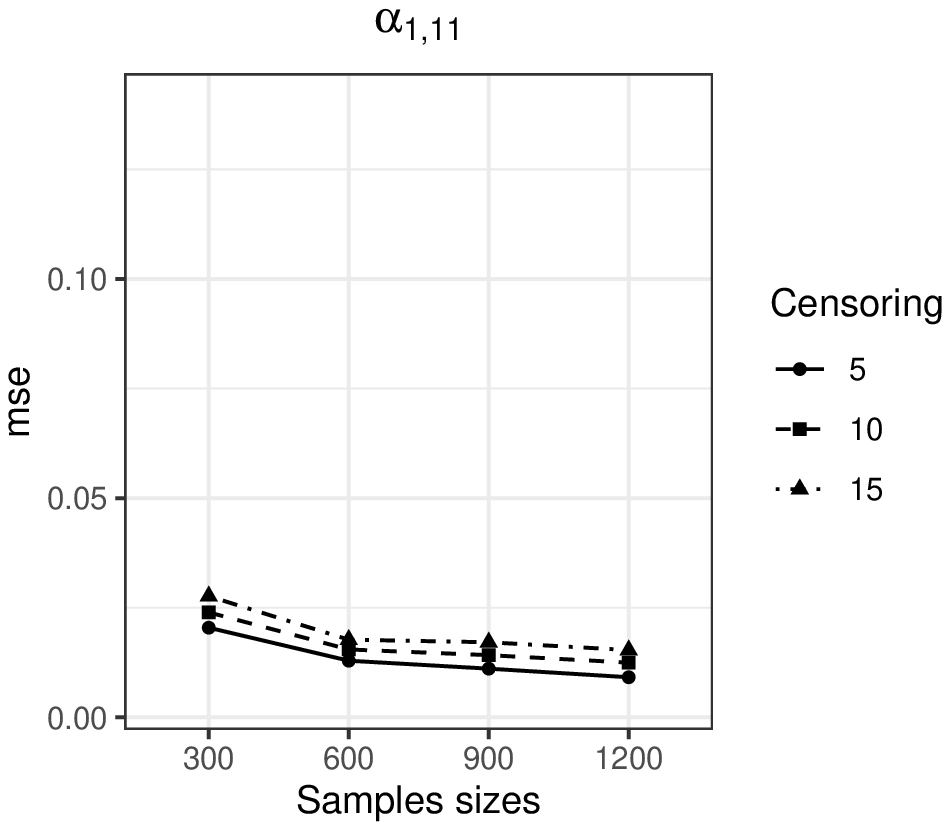}\label{fig:betadensity2}}
\subfigure{\includegraphics[width = 3cm]{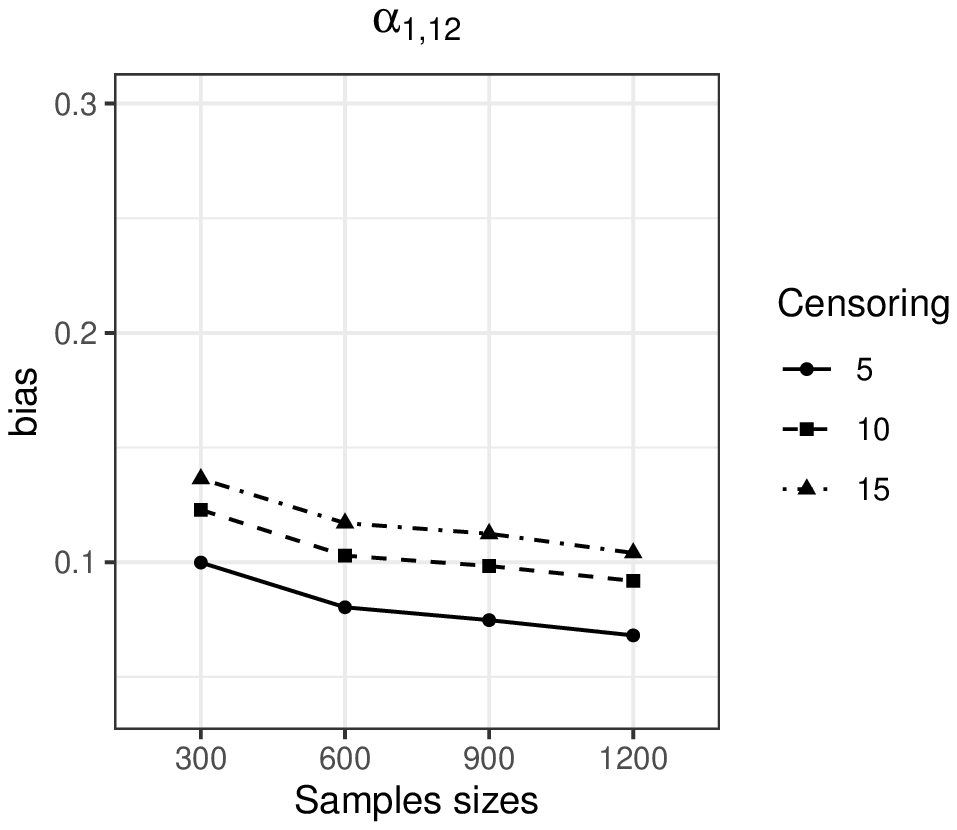}\label{fig:betadensity1}}
\subfigure{\includegraphics[width = 3cm]{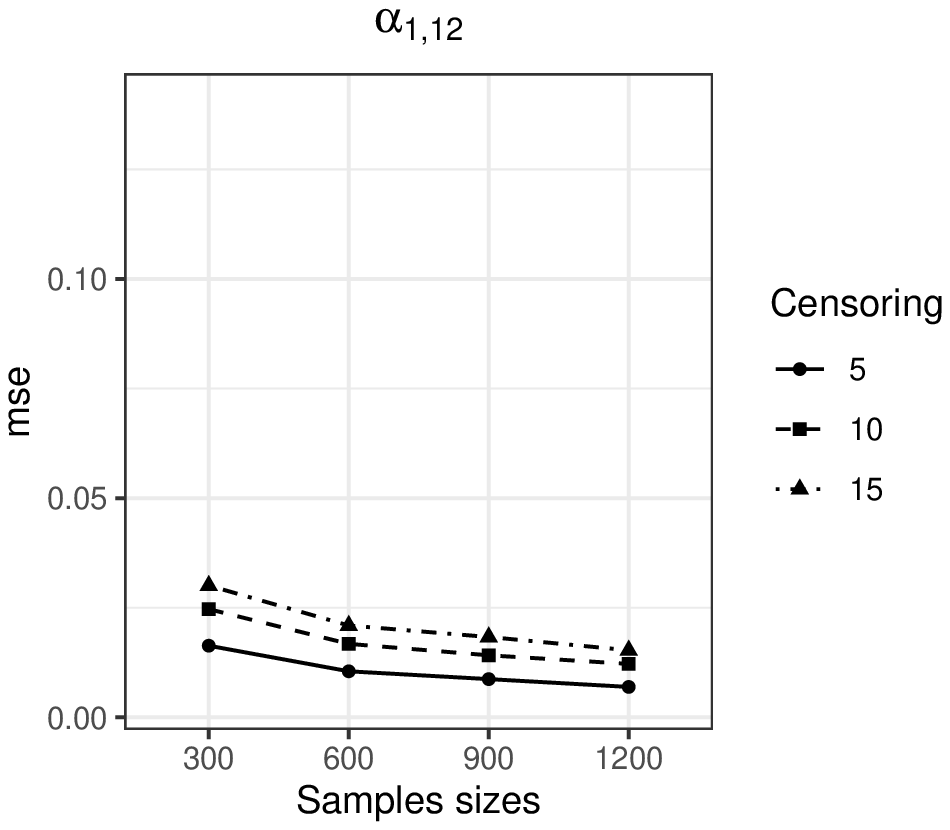}\label{fig:betadensity2}}
\subfigure{\includegraphics[width = 3cm]{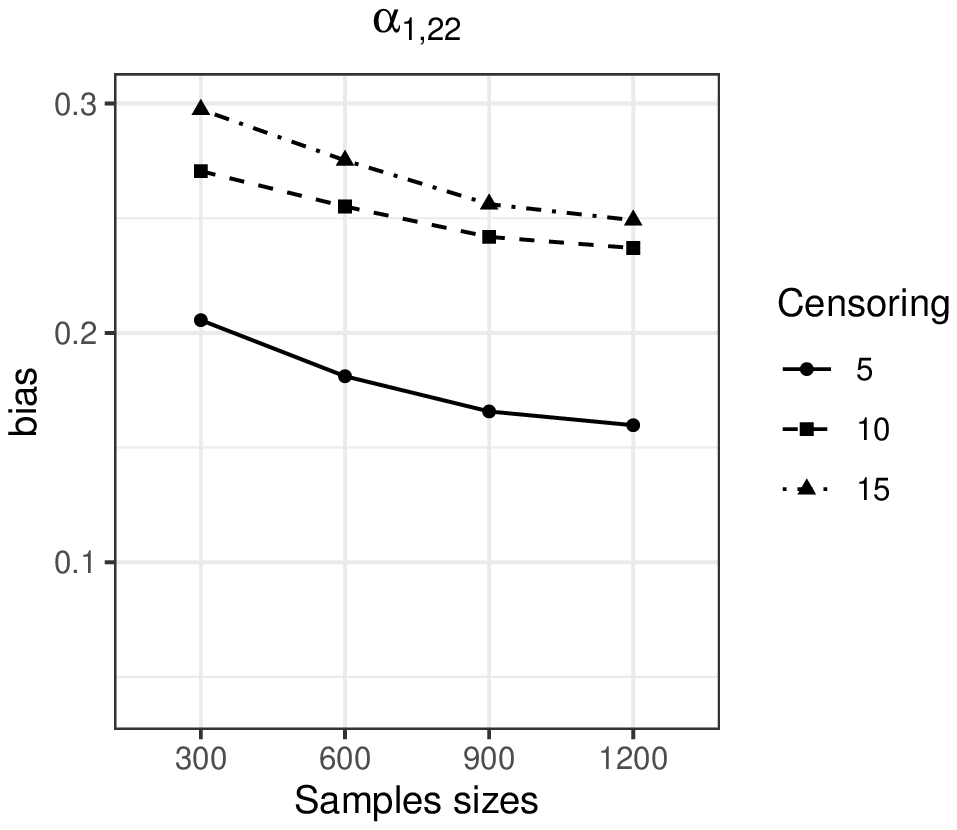}\label{fig:betadensity1}}
\subfigure{\includegraphics[width = 3cm]{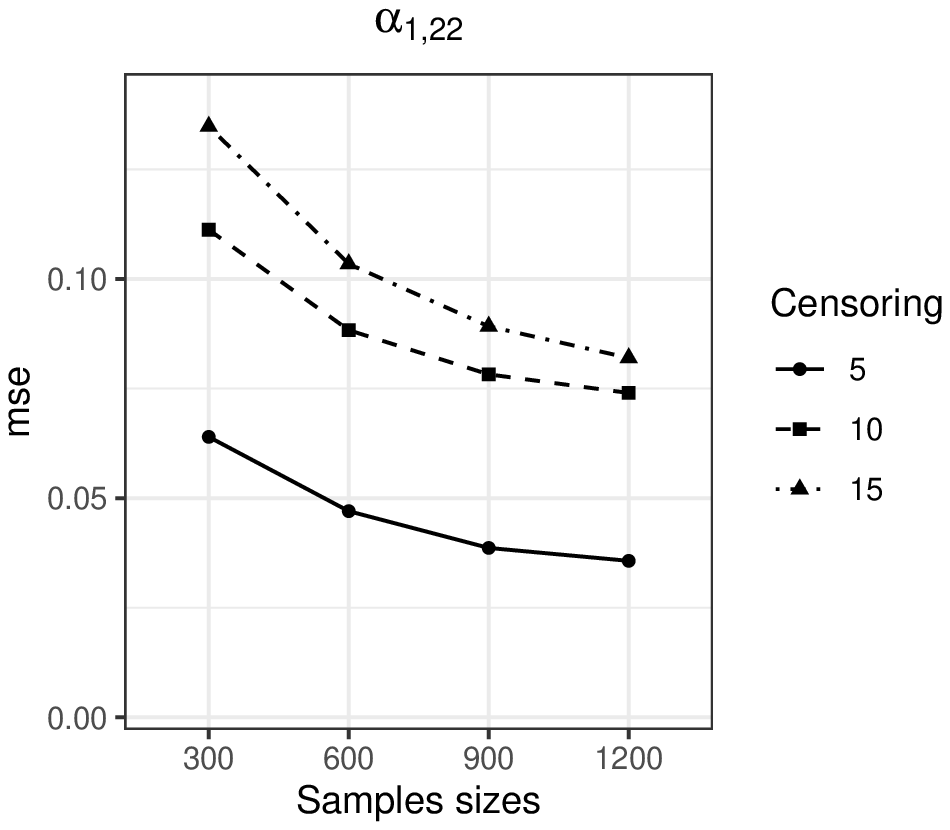}\label{fig:betadensity2}}
\subfigure{\includegraphics[width = 3cm]{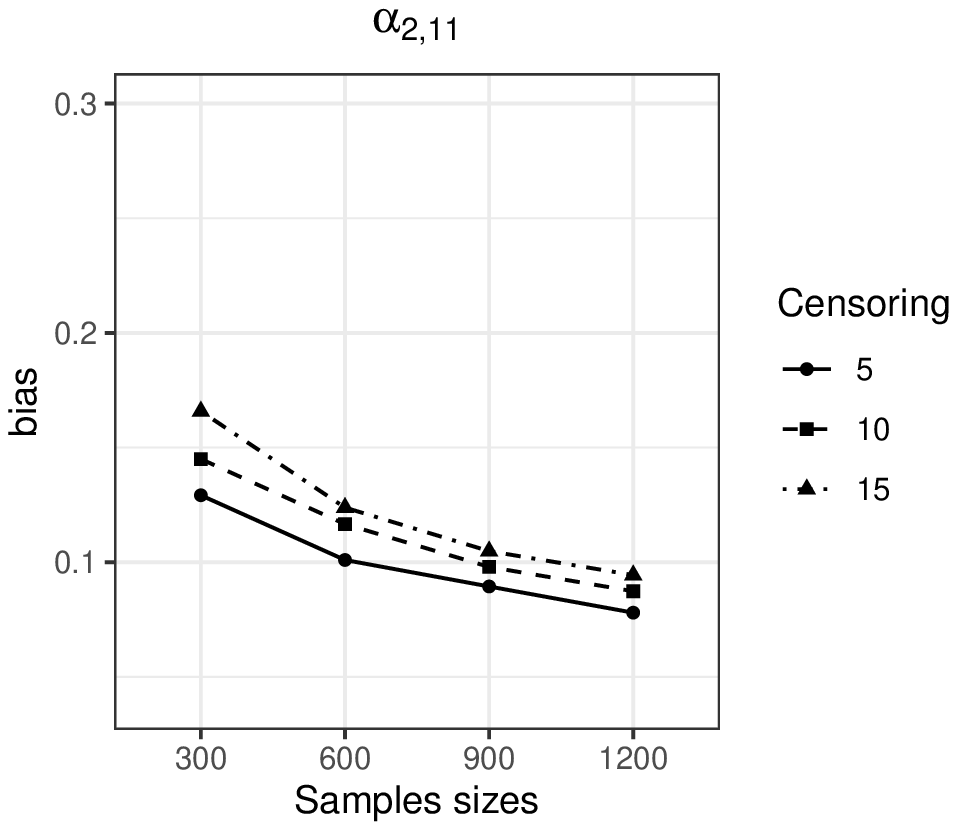}}
\subfigure{\includegraphics[width = 3cm]{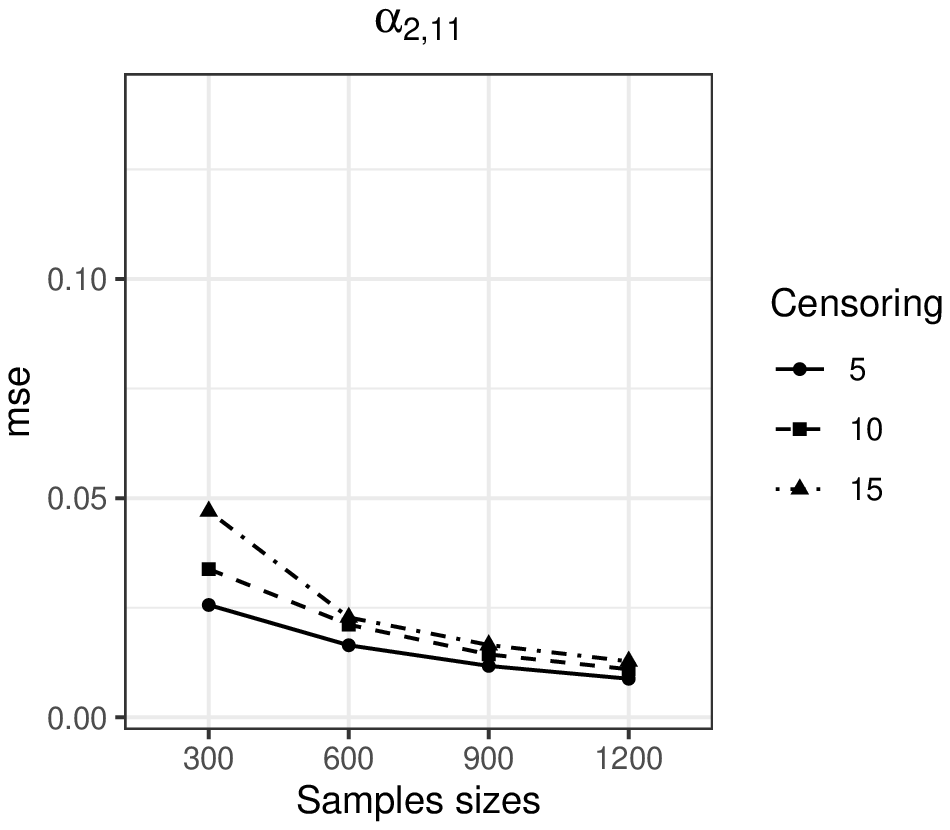}\label{fig:betadensity2}}
\subfigure{\includegraphics[width = 3cm]{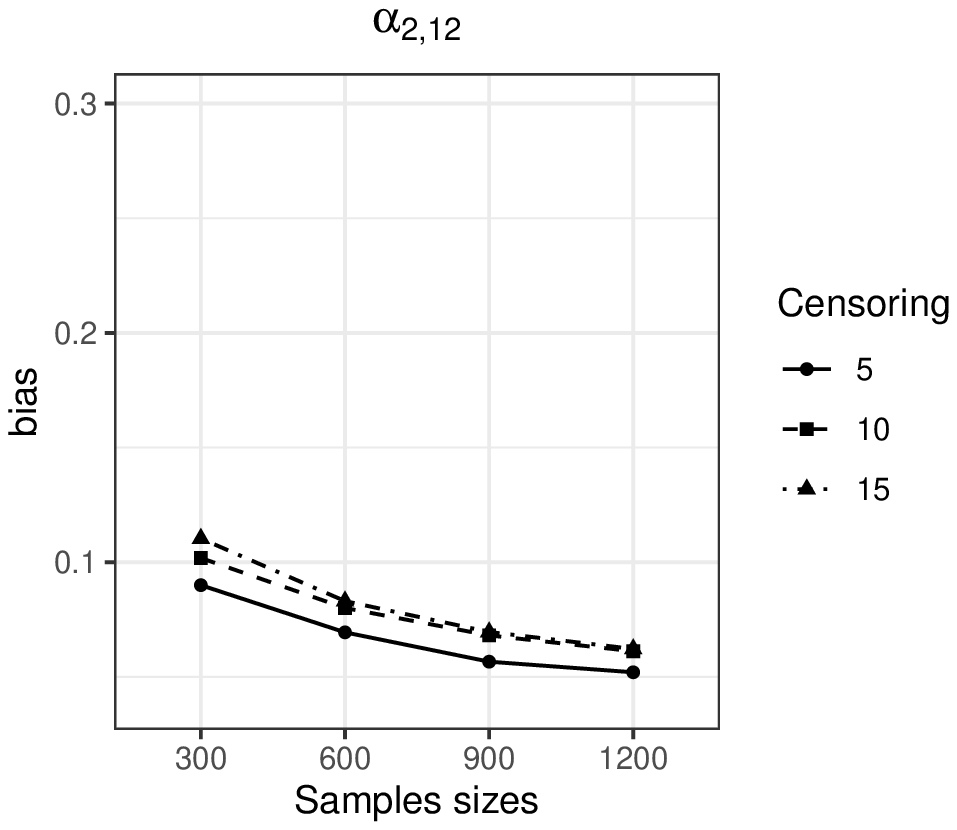}\label{fig:betadensity1}}
\subfigure{\includegraphics[width = 3cm]{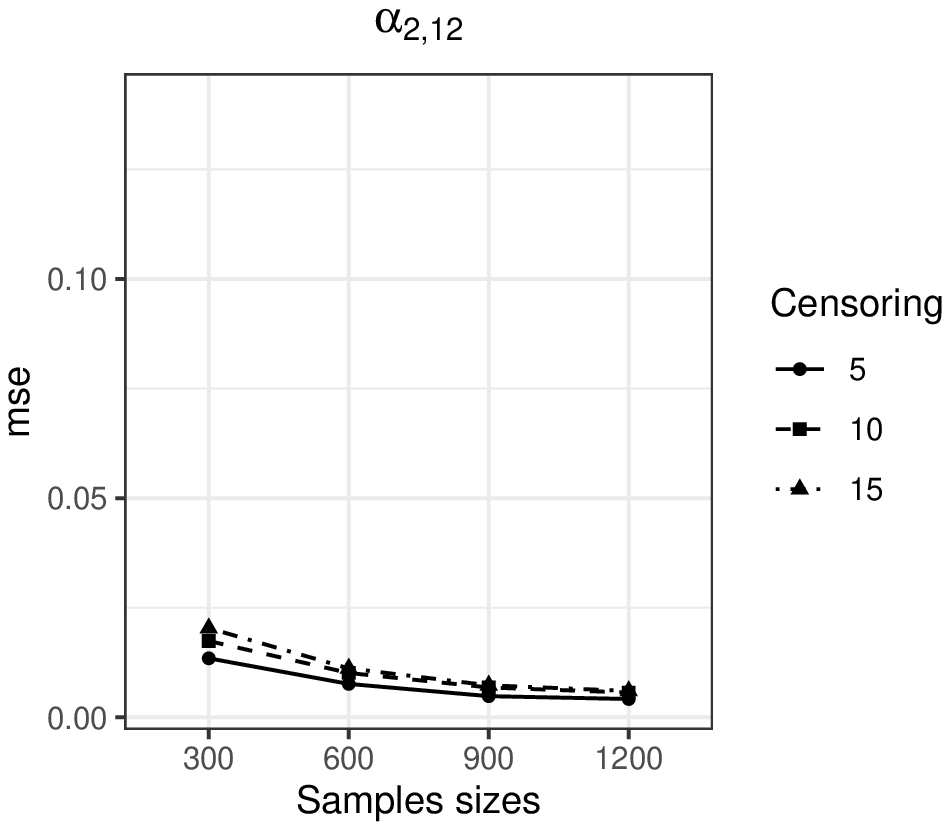}\label{fig:betadensity2}}
\subfigure{\includegraphics[width = 3cm]{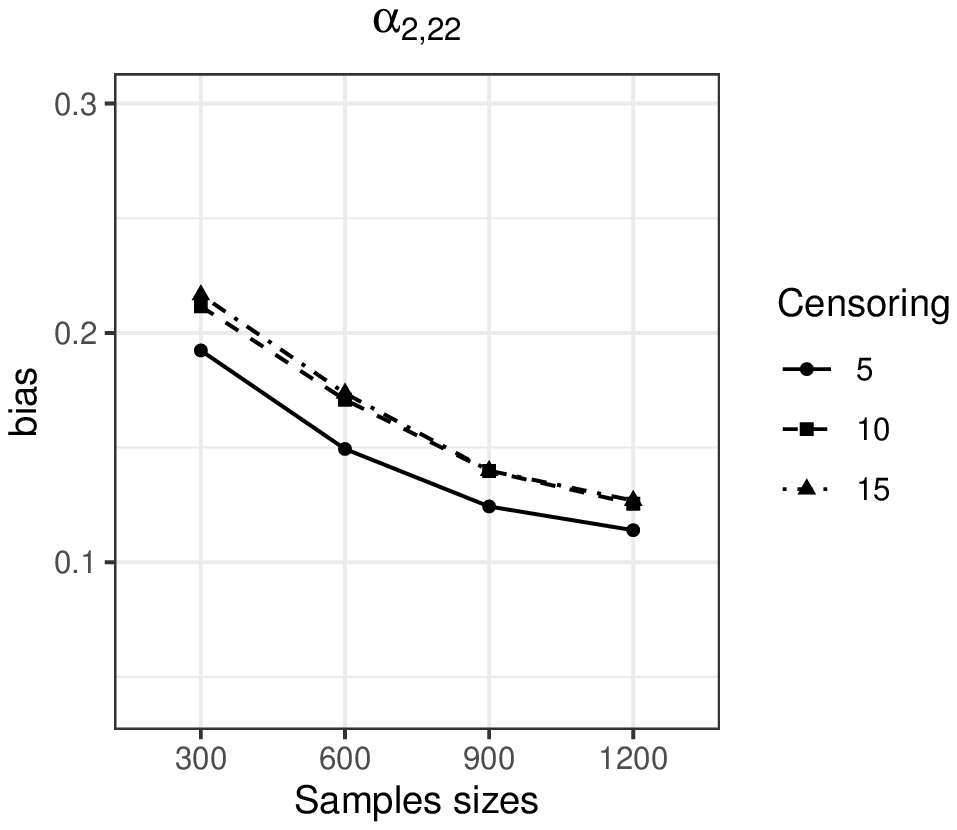}\label{fig:betadensity1}}
\subfigure{\includegraphics[width = 3cm]{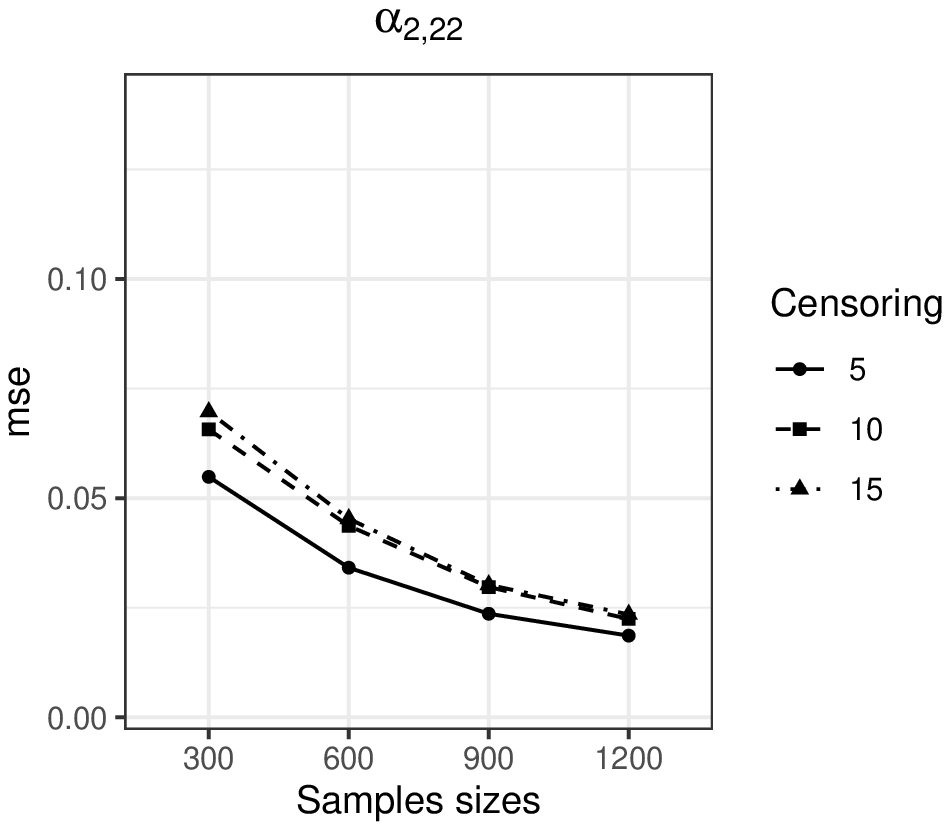}\label{fig:betadensity2}}
\subfigure{\includegraphics[width = 3cm]{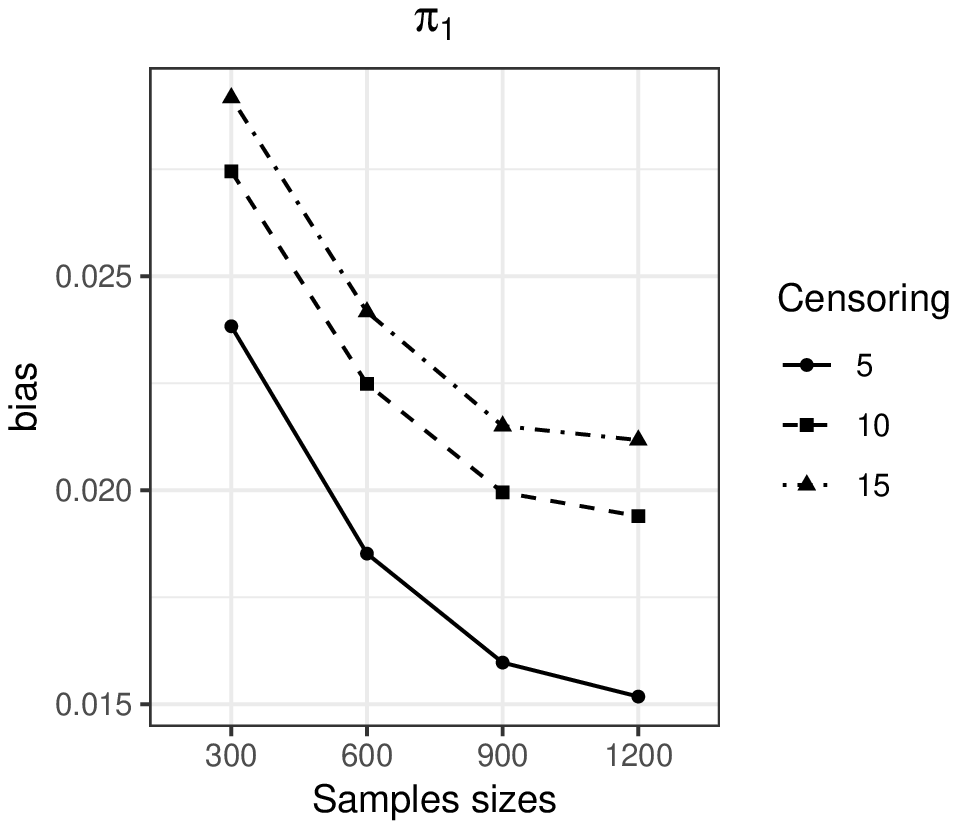}\label{fig:betadensity2}}
\subfigure{\includegraphics[width = 3cm]{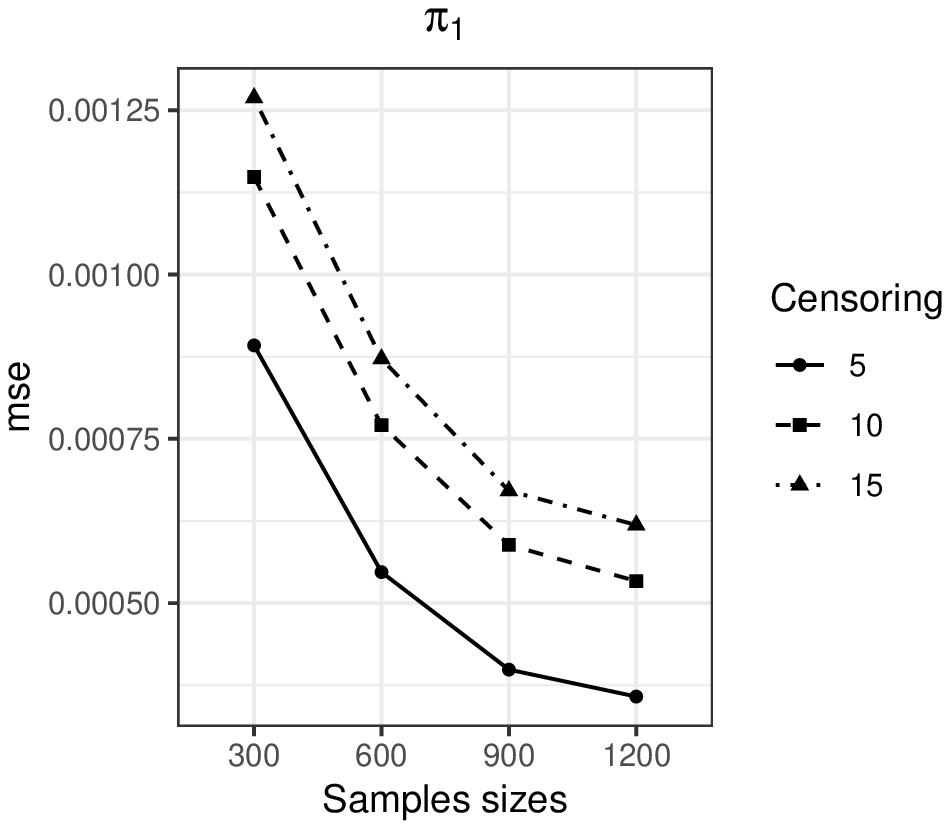}\label{fig:betadensity2}}}
\caption{\small{Simulated data: Asymptotic properties. bias and mse of $\bF_1 = \bSigma_1^{1/2}$, $\bF_2 = \bSigma_2^{1/2}$ and $\pi$ estimates in the FM-MSNC model with
different censoring levels:  $5\%$ (solid line), $10\%$ (dashed line), $15\%$ (dot-dashed line).}}
\label{fig:assinto2}
\end{figure}

The results of the estimates of {\small$\bmu_1, \bF_1, \blambda_1, \bmu_2, \bF_2, \blambda_2$} and {\small$\pi$} are shown in Figures \ref{fig:assinto1}, \ref{fig:assinto3}, and \ref{fig:assinto2}. As a general rule, we can say that the $bias$ and $mse$ tend to approach zero when the sample size increases, indicating that the estimates based on the proposed EM algorithm, under the {\small FM-MSNC} model, provide good asymptotic properties.

\section{Application}\label{app}
To illustrate the performance of our proposed model and algorithm, we consider a dataset of trace metal concentrations collected by the Virginia Department of Environmental Quality (VDEQ) that was previously analyzed by \cite{he2013mixture} and \cite{lachos2017finite} using the normal and Student-t distribution, respectively. 

This dataset consists of {\small$p = 5$} concentration levels of dissolved trace metals in independently selected {\small$n = 184$} freshwater streams across the Commonwealth of Virginia. The five attributes are trace metals: copper (Cu), lead (Pb), zinc (Zn), calcium (Ca) and magnesium (Mg). The Cu, Pb, and Zn concentrations are reported in $\mu$g/L of water, whereas Ca and Mg concentrations are reported in mg/L of water. Since the measurements were taken at different times, the presence of multiple limits of detection values are possible for each trace metal \cite{richmond2003quality}. The limits of detection for Cu and Pb are both the {\small$0.1$} $\mu$g/L, {\small$1.0$} $\mu$g/L for Zn,  while Ca and Mg have limits of {\small$0.5$} mg/L and {\small$1.0$} mg/L, respectively.

\begin{figure}[!h]
\centering
\center{\subfigure{\includegraphics[width = 3.5cm]{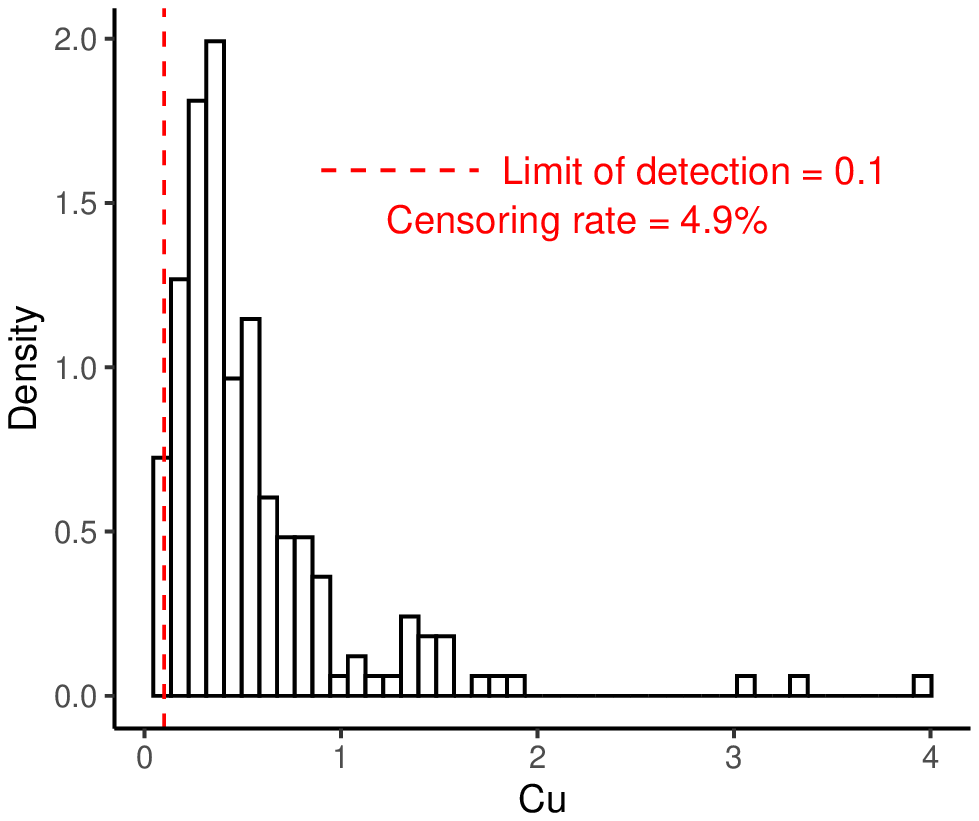}}
\subfigure{\includegraphics[width = 3.5cm]{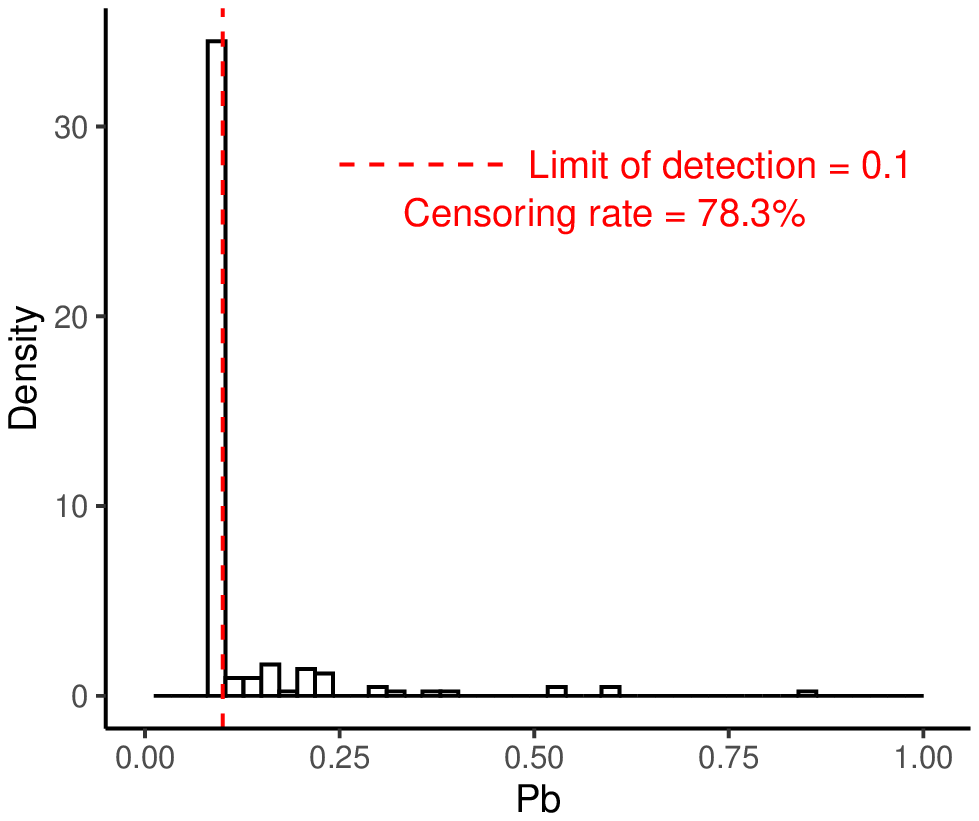}\label{fig:betadensity2}}
\subfigure{\includegraphics[width = 3.5cm]{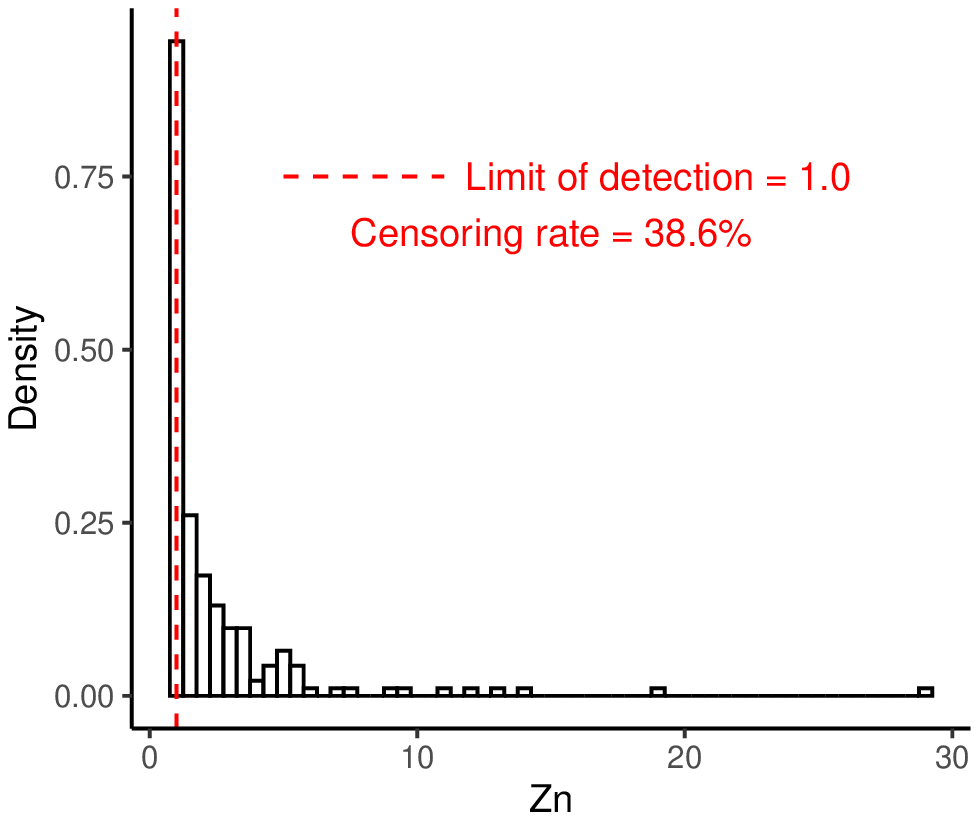}\label{fig:betadensity1}}\\
\subfigure{\includegraphics[width = 3.5cm]{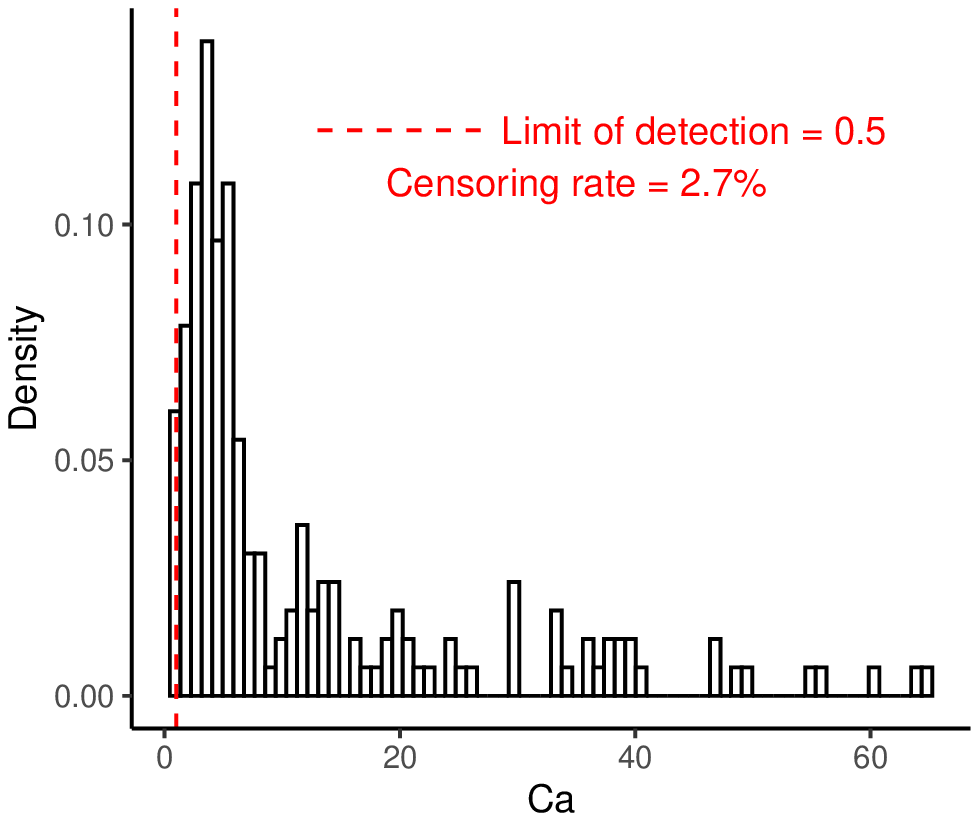}\label{fig:betadensity2}}
\subfigure{\includegraphics[width = 3.5cm]{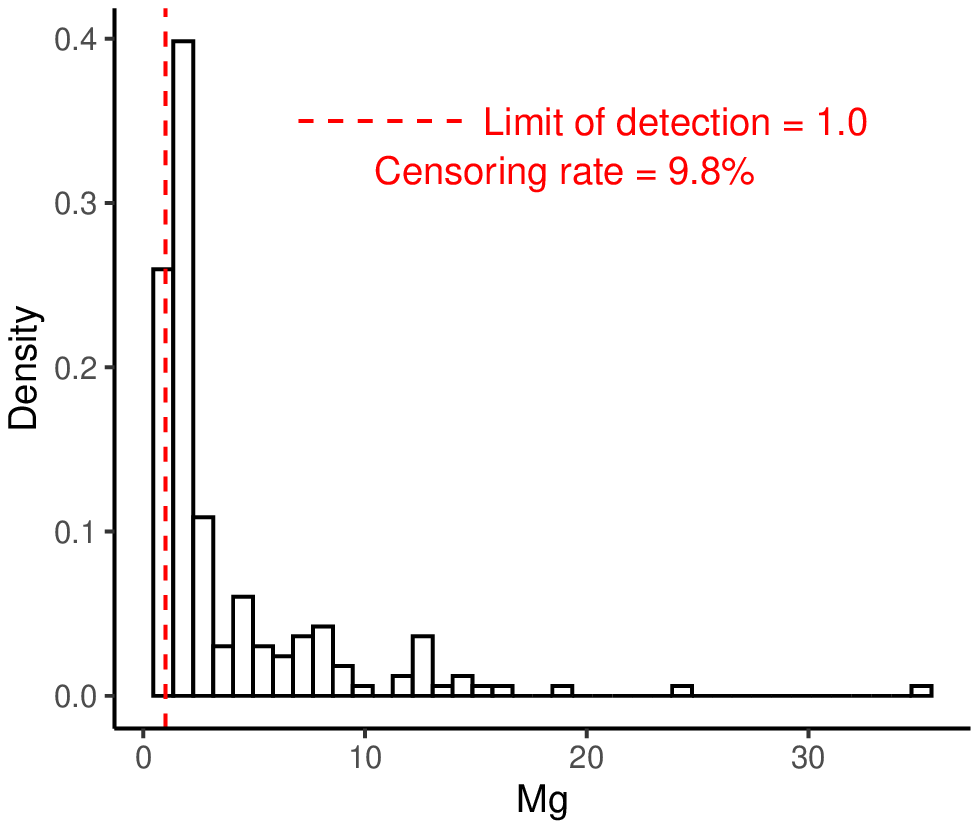}\label{fig:betadensity1}}
\subfigure{\includegraphics[width = 3.5cm]{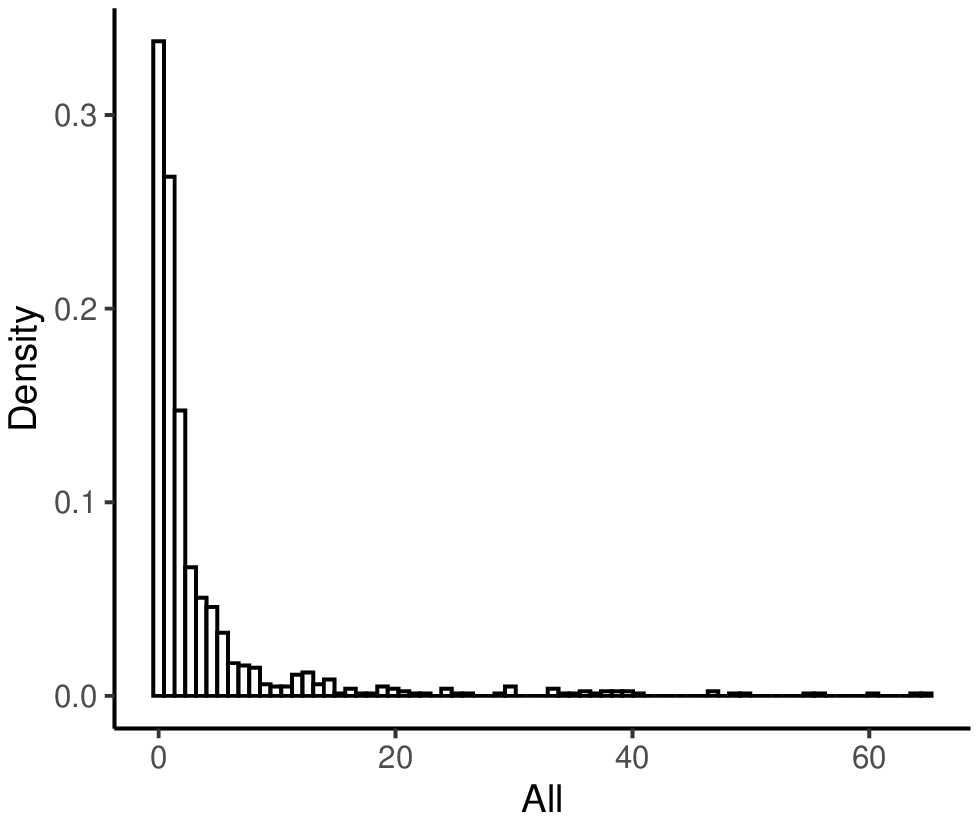}\label{fig:betadensity2}}
}
\caption{\small{VDEQ data. Histograms of all the original five attributes of Virginia trace metal concentration data and all attributes together. The red line means the censoring limit detection for each concentration.}}
\label{fig:hist_concen}
\end{figure}

The percentage of left-censored values of {\small$2.7\%$} for (Ca), {\small$4.9\%$} for (Cu), {\small$9.8\%$} for (Mg) are small in comparison to {\small$78.3\%$} for (Pb) and {\small$38.6\%$} for (Zn). Also note that {\small$17.9\%$} of the streams had {\small$0$} non-detected trace metals, {\small$39.1\%$} had {\small$1$}, {\small$37.0\%$} had {\small$2$}, {\small$3.8\%$} had {\small$3$}, {\small$1.1\%$} had {\small$4$} and {\small$1.1\%$} had {\small$5$}. As the concentration levels are strictly positive measures, to guarantee this, we consider an interval-censoring analysis by setting all lower limits of detection equal to {\small$0$} for all trace metals. Also, due to the different scales for each trace metal, we standardize the dataset to have zero mean and variance equal to one as in \cite{wang2019model}. The work mentioned before considered this dataset to be left censored  without taking into account the possibility of predicting negative concentration levels for the trace metals. For instance, note that Pb censored concentrations take values in the small interval {\small$[0,0.1]$}. Thus, after transforming the data, the new limits of detection are {\small$-0.8776$} (Cu), {\small$-0.3124$} (Pb), {\small$-0.4719$} (Zn), {\small$-0.7894$} (Ca), {\small$-0.6289$} (Mg). Figure \ref{fig:hist_concen} shows the histogram for each {original} trace metal with the detection limits and all of them together.It can be seen that most of the distributions associated with the variables have two or more modes and are right skewed. For this reason we propose to fit a {\small FM-MSNC} model. 

We fit the data with 1, 2 and 3 components considering the {\small FM-MSNC}, {\small FM-MtC}  and {\small FM-MNC} models, for the {\small FM-MtC} model we consider fixed degrees of freedom, as described in \cite{lachos2017finite}. The number of groups of the model is chosen according to the information criteria as shown in Table \ref{tab:selection_crite}. It can be seen that according to all model selection criteria the {\small FM-MSNC} model with three components fits the data best. We considered the variance-covariance {\small $(\bGamma)$} to be equal in order to reduce the number of parameters to be estimated (parsimonious model).

\begin{table}[!h]
\tiny
\begin{center}
\caption{\label{CompGroupDR} VDEQ data. Model selection criteria for various FM-MSNC and FM-MNC models. Values in bold correspond to the best model according to the criteria.}
\begin{tabular}{lccccccc}
\toprule
&  \multicolumn{3}{c}{FM-MSNC} &  & \multicolumn{3}{c}{FM-MNC} \\
\cmidrule{2-4} \cmidrule{6-8}
Criteria & $G=1$ & $G=2$ & $G=3$ & & $G=1$ & $G=2$ & $G=3$  \\
\midrule
Log-likelihood & -1269.302 & -910.3387 & \textbf{-697.6815} & & -1351.596 & -1268.848 &-1210.626\\
AIC & 2588.604 & 1892.677 & \textbf{1489.363} & & 2743.192 & 2589.695 & 2485.253\\
BIC & 2668.977 & 2008.415 & \textbf{1640.465} & & 2807.491 & 2673.284 & 2588.13\\
EDC & 2606.427 & 1918.343 & \textbf{1522.871} & & 2757.451 & 2608.232 & 2508.066\\ \midrule
Time & 1.7023 min.& 4.0229 min.& 9.2552 min. &  & 1.054 sec. & 7.729 sec. & 26.0381 sec. \\\midrule		
&	\multicolumn{7}{c}{FM-MtC}\\ \midrule
&  \multicolumn{3}{c}{$\nu = 3$} &  & \multicolumn{3}{c}{$\nu = 4$} \\
\cmidrule{2-4} \cmidrule{6-8}
Criteria & $G=1$ & $G=2$ & $G=3$ & & $G=1$ & $G=2$ & $G=3$  \\
\midrule
Log-likelihood & -1040.276 & -1018.943 & -1074.852 & & -1061.702 & -1036.393 & -1072.487\\
AIC & 2120.553 & 2089.887 & 2213.705 & & 2163.404 & 2124.786 & 2208.974\\
BIC & 2184.852 & 2173.475 & 2316.583 & & 2227.702 & 2208.375 & 2311.852\\
EDC & 2134.812 & 2108.423 & 2236.519 & & 2177.662 & 2143.323 & 2231.788\\ \midrule
Time & 16.672 sec. & 35.9952 sec. & 2.8734 min. & & 13.1202 sec. & 29.9673 sec. & 2.68 min.\\ 
\bottomrule
\end{tabular}
\label{tab:selection_crite}
\end{center}
\end{table}

The ML estimates of the parameters were obtained using the EM algorithm described in Subsection \ref{sectionem}. The results are shown in Table \ref{tab:estimative_parameters}. As can be seen, {\small$46.99\%$} of the freshwater streams belong to Cluster 1, Cluster 2 contains around {\small $34.40\%$} of them and the remaining {\small $15.78\%$} belong to Cluster 3. {Table  \ref{tab:selection_crite} shows that the best  {\small FM-MNC} model has two components, and in this {\small$96.73\%$} of freshwater streams belong to Cluster 1 and the remaining {\small$3.27\%$} are in Cluster 2. For the {\small FM-MtC} model, the best model  also has two components and three degrees of freedom. In this case, {\small$85.19\%$} of freshwater streams belong to Cluster 1 and the remaining {\small$14.81\%$} belong to Cluster 2.}

\begin{table}[!h]
\scriptsize
\centering
\caption{VDEQ data. ML estimates of parameters from fitting the FM-MSNC model with 3 components to the Virginia trace metal
concentration data.}
\begin{tabular}{cl}
\toprule
Parameter & Estimate\\
\midrule
$(\pi_1, \pi_2, \pi_3)$ & $(0.4699, 0.3440, 0.1861)$\\ [3 pt]
$\bmu_1$ & $(-0.3789, -0.5344, -0.722, -0.5744, -0.4833)$\\[3 pt]
$\bmu_2$ & $(-0.8189, -0.6838, -0.4485, 0.805, 0.7948)$\\[3 pt]	
$\bmu_3$ & $(1.2681, -0.463, -0.503, -0.3272, -0.305)$\\[3 pt]
$\blambda_1$ & $(0.4867, -1.4692, 13.2902, -0.4112, -0.0062)$\\[3 pt]
$\blambda_2$ & $(28.3006, -1.012, 6.0129, 4.4658, -1.6659)$\\[3 pt]
$\blambda_3$ & $(-0.5228, 11.9182, 12.3316, -1.3583, 1.2298)$\\[3 pt]
$\bF_1 = \bSigma_1^{1/2}$ & $\left[ \begin{array}{ccccc}
0.3328 & 0.1229 & 0.0419 & 0.0338 & 0.0573 \\
& 0.3733 & 0.0060 & -0.0105 & 0.0038 \\
&  & 0.5540 & -0.0174 & -0.0061 \\
& & & 0.0952 & 0.0554\\
&  &  &  & 0.1208\\
\end{array}\right]$\\[3 pt]
$\bF_2 = \bSigma_2^{1/2}$ & $\left[ \begin{array}{ccccc}
1.0582 & -0.035 & 0.1628 & 0.1981 &  0.0710 \\
& 0.102 & -0.0074 & 0.0025 & 0.0111 \\
&  & 0.4745 & 0.0157 & 0.1260 \\
& & & 1.0244 & 0.3920\\
&  &  &  & 1.2392\\
\end{array}\right]$\\ [3 pt]
$\bF_3 = \bSigma_3^{1/2}$ & $\left[ \begin{array}{ccccc}
1.4992 & -0.1232 & -0.2405 & 0.1341 & 0.0950\\
& 2.2952 & 0.8411 & -0.0459 & -0.0915 \\
&  & 2.1716 & -0.1003 & -0.0633 \\
& & & 0.2824 & 0.1472\\
&  &  &  & 0.2640\\
\end{array}\right]$\\
\bottomrule
\end{tabular}
\label{tab:estimative_parameters}
\end{table}

\begin{figure}[!htb]
\centering

\center{
\subfigure{\includegraphics[width=\textwidth]{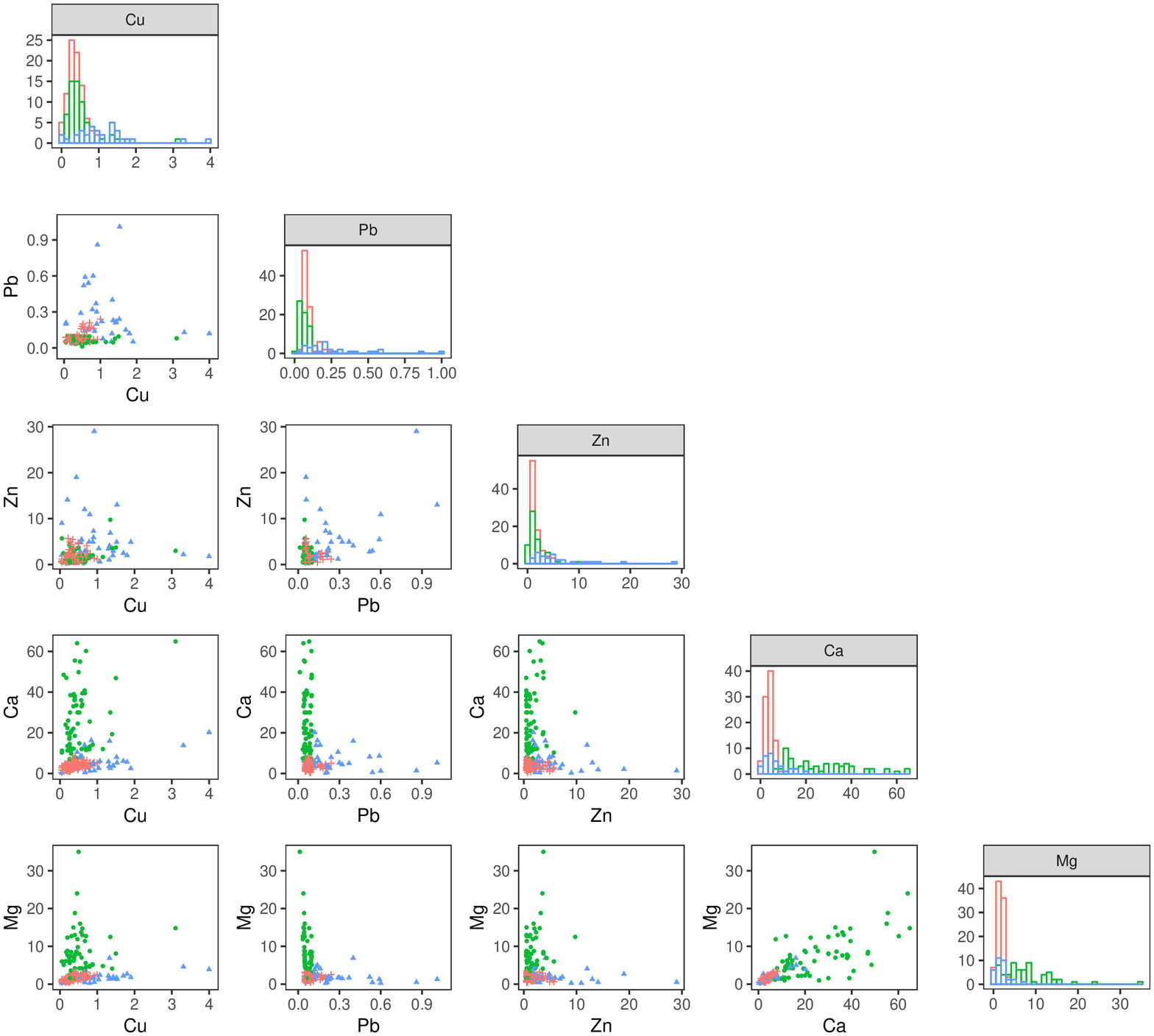}\label{fig:betadensity2}}
}
\caption{Scatter plots and histograms of the fitted values of $\y_j$ for each predicted clusters of the VDEQ data. Lower diagonal
entries: $\color{ForestGreen}{\bullet}$ cluster 1; $\color{cyan}{\blacktriangle}$ cluster 2; $\color{pink}{+}$ cluster 3.}
\label{fig:appli_Metal}
\end{figure}

In Figure \ref{fig:appli_Metal}, we fit the data using the {\small FM-MSNC} with three components. The scatter plots of the observations {\small$y_i$} {\small $(i = 1, \ldots , 184)$} for each pair of trace metals reveal that it is difficult to classify freshwater streams by visualization because these observations almost mix together.

\section{Conclusions}
\label{sec:6}

In this paper, a novel approach to analyze multiply censored and missing data is presented based on the use of finite mixtures of multivariate skew-normal distributions. This approach  generalizes
several previously proposed solutions for censored data, such as, the finite mixture of Gaussian components \citep{karlsson2014finite,caudill2012partially, he2013mixture} and the finite mixture of Student-t components \citep{lachos2017finite}, which are also restricted to a left or right censored problem. A simple and efficient EM-type algorithm was developed, which has closed-form expressions at the
E-step and relies on formulas for the mean vector  and covariance matrix  of the
multivariate truncated skew-normal distribution, for which the the \texttt{R MomTrunc} library is used \citep{GalarzaTrunSN2019}. The proposed EM algorithm was implemented as part of the \verb|R| package \verb"CensMFM" and is
available for download at the CRAN repository. The experimental results and
the analysis of a real dataset provide support for the usefulness and effectiveness of our proposal. 

The method proposed in this paper can be extended to other types of mixture
distributions, for example, the  multivariate scale mixtures of skew-normal distributions \citep{Cabral2012} or generalized hyperbolic mixtures \citep{browne2015mixture}.  It is also
of interest to develop an effective Markov chain Monte Carlo algorithm for the FM-MSNC models in a fully Bayesian treatment. Finally, the proposed methods can also be easily applied to other substantial areas in which the data being analyzed have censored and/or missing observations, for instance, factor analysis models \citep{wang2017robust} and linear mixed models \citep{lin2009LMM, lachos2011linear}.



\bibliographystyle{unsrt}  

\end{document}